\documentclass[a4paper,fleqn,usenatbib]{mnras}

\usepackage[T1]{fontenc}
\usepackage{ae,aecompl}

\usepackage{graphicx}	
\usepackage{amsmath}	
\usepackage{amssymb}	

\newcommand{\gimtwod}{\textsc{gim2d}}

\newcommand{\sextractor}{\textsc{SExtractor}}

\defcitealias{2011ApJS..196...11S}{S11}
\defcitealias{2015MNRAS.447.2753T}{T15}
\defcitealias{2015MNRAS.454.1886S}{S15}
\defcitealias{2000ApJ...539..718C}{CF00}

\newcommand{\Seleven}{\citetalias{2011ApJS..196...11S}}

\title[Illustris galaxies as seen by the SDSS]{Galaxies in the Illustris simulation as seen by the Sloan Digital Sky Survey - I: Bulge+disc decompositions, methods, and biases. }

\author[Bottrell et al.]{
Connor Bottrell,$^{1}$\thanks{E-mail: connor.bottrell@gmail.com}
Paul Torrey,$^{2}$\thanks{Hubble Fellow}
Luc Simard,$^{3}$
and Sara L. Ellison$^{1}$
\\
\scriptsize
$^{1}$University of Victoria, Department of Physics and Astronomy, Victoria, British Columbia, V8P 1A1, Canada\\
\scriptsize
$^{2}$Department of Physics, Kavli Institute for Astrophysics and Space Research, Massachusetts Institute of Technology, Cambridge, MA 02139, USA\\
\scriptsize
$^{3}$National Research Council of Canada, Herzberg Institute of Astrophysics, 5071 West Saanich Road, Victoria, British Columbia, V9E 2E7, Canada
}

\date{Accepted XXX. Received YYY; in original form ZZZ}

\pubyear{2016}

\begin{document}
\label{firstpage}
\pagerange{\pageref{firstpage}--\pageref{lastpage}}
\maketitle

\begin{abstract}
We present an image-based method for comparing the structural properties of
galaxies produced in hydrodynamical simulations to real galaxies in the Sloan
Digital Sky Survey.  The key feature of our work is the introduction of extensive
observational realism, such as object crowding, noise and viewing angle, to the
synthetic images of simulated galaxies, so that they can be fairly compared to real
galaxy catalogs. We apply our methodology to the dust-free synthetic
image catalog of galaxies from the Illustris simulation at $z=0$, which are then fit with bulge+disc
models to obtain morphological parameters. In this first paper in a series, we detail
our methods, quantify observational biases, and present publicly available bulge+disc decomposition catalogs.
We find that our bulge+disc decompositions are largely robust to the observational biases
that affect decompositions of real galaxies. However, we
identify a significant population of galaxies (roughly 30\% of the full sample) in
Illustris that are prone to internal segmentation, leading to systematically reduced
flux estimates by up to a factor of 6, smaller half-light radii by up to a factor of
$\sim$ 2, and generally erroneous bulge-to-total fractions of (B/T)=0.
\end{abstract}




\section{Introduction}\label{intro}

A range of tools has been developed in the past several decades to model the formation and evolution of galaxies with the overarching goal of reproducing the observed properties of galaxies and their populations in nearby and distant epochs of the universe (see \citealt{2015ARA&A..53...51S} for a recent review). Validation of the models is determined through comparisons with observations. (e.g. \citealt{2003ApJ...591..499A,2011ApJ...728...51B,2011MNRAS.410.1391A,2011ApJ...742...76G,2014MNRAS.440L..51C,2015arXiv150900853A,2015arXiv151005645F}). 
The growth in observational data from modern observational large galaxy redshift surveys such as the Sloan Digital Sky Survey (SDSS) \citep{2011AJ....142...72E}, the Two-degree-Field Galaxy Redshift Survey (2dF) \citep{1999MNRAS.308..459F}, the Deep Extragalactic Evolutionary Probe 2 survey (DEEP2) \citep{2013ApJS..208....5N}, the Cosmic Assembly Near-infrared Deep Extragalactic Legacy Survey (CANDELS) \citep{2011ApJS..197...35G} as well as from forthcoming projects such as the Large Synoptic Survey Telescope (LSST) provides an increasingly precise and complete test-bed against which the models must be benchmarked.

The most direct way of deriving observable properties of galaxies from theoretical predictions is to numerically track the coevolution of dark and baryonic matter in hydrodynamical cosmological simulations (e.g. \citealt{1992ApJ...399L.109K,1996ApJS..105...19K,1997ApJ...477....8W,2002ApJ...571....1M,2003MNRAS.339..289S,2005MNRAS.363....2K,2008MNRAS.390.1326O,2009MNRAS.399.1773C,2009MNRAS.400...43C,2010MNRAS.402.1536S,2010MNRAS.406.2325O,2012MNRAS.425.3024V}). Tracking the dynamics of both the dark and baryonic matter self-consistently to small spatial scales allows predictions to be made about the internal structure of galaxies including the distribution of gas \citep{2012MNRAS.425.2027K,2012MNRAS.427.2224T} and the structures formed from the stellar components \citep{2003ApJ...591..499A,2003ApJ...597...21A,2004ApJ...607..688G,2011MNRAS.410.1391A,2012MNRAS.423.1544S,2012EPJWC..1908008M}. Numerical simulations are useful tools for interpreting the observed properties of galaxies because they facilitate controlled explorations of galaxy formation and evolution. In particular, the latest generation of cosmological hydrodynamical simulations (e.g. Illustris \citealt{2014MNRAS.444.1518V}, EAGLE \citealt{2015MNRAS.446..521S}, FIRE \citealt{2014MNRAS.445..581H}, APOSTLE \citealt{2016MNRAS.457.1931S}, BAHAMAS \citealt{2016arXiv160302702M}) are designed specifically to allow straight-forward comparisons between observed and simulated galaxy populations across a range of redshifts.  

Explicit tracking of the birth mass, chemical evolution, mass loss, ages and motions of stellar populations within galaxies in hydrodynamical simulations allows the assignment of a full spectrum to each stellar population at every time-step. Synthetic images can be constructed from this information that provide representations of the simulated galaxies as observed with real photometric or spectroscopic instruments (\citealt{2006MNRAS.372....2J,2008ApJ...678...41L,2011ApJS..196...22B,2011A&A...536A..79R}; see \citealt{2013ARA&A..51...63S} for a recent review). The tools for converting hydrodynamical simulations into synthetic images bridge the gap between theory and observations such that consistent analysis of the data of each camp are possible. However, the crucial requirement of a fair comparison between the properties derived from observed galaxies and simulations is that the same biases affect the interpretation of their results \citep{2010ApJ...708...58C,2012MNRAS.424..951H,2013MNRAS.434.2572H,2013MNRAS.428.2529H}. Furthermore, the apparent disparity between structural estimates, such as bulge- and disc-to-total ratios (e.g. \citealt{2010MNRAS.407L..41S}), derived from photometric decomposition analyses of synthetic images and estimates from the orbital properties obtained directly from the simulations also demonstrates that consistency in methodology is fundamental to the interpretative power of such comparisons. 


In an image-based comparison between the properties of galaxies from a hydrodynamical simulation and observations, \citealt{2015MNRAS.454.1886S} (\citetalias{2015MNRAS.454.1886S}) obtained non-parametric optical morphology estimates for Illustris' galaxy synthetic image catalog (\citealt{2015MNRAS.447.2753T}, \citetalias{2015MNRAS.447.2753T}) and demonstrated the diversity of morphologies produced therein. Two key elements of \citetalias{2015MNRAS.454.1886S} were crucial to their comparison of simulated and observed galaxies: (1) the application of observational realism by inserting the synthetic images directly into real fields; (2) consistent methods of deriving morphological estimates. \citetalias{2015MNRAS.454.1886S} showed that their non-parametric morphologies for simulated galaxies with realism (specifically the Gini and M20 parameters, \citealt{2004AJ....128..163L}), roughly occupied the same space as real galaxies -- an important success in the structural comparison of galaxies from hydrodynamical simulations and the real universe. While several puzzles are highlighted by \citetalias{2015MNRAS.454.1886S}, the morphological similarities shown between simulated and observed galaxies in their analysis motivates a complimentary and orthogonal exploration of galaxy structures that may be obtained through parametric decompositions of simulated galaxies. Furthermore, the realism suite of \citetalias{2015MNRAS.454.1886S} can be built upon by including the statistical biases from crowding, sky brightness, and point-spread function (PSF) resolution to enable even more thorough consistency in comparisons with observations.

In this first paper in our series, we describe our methodology and data products that facilitate consistent, image-based comparisons between structural properties derived from real galaxies and galaxies evolved in hydrodynamical simulations. In particular, we derive structural estimates from SDSS renderings of the synthetic galaxy image catalog from the Illustris simulation \citepalias{2015MNRAS.447.2753T} that are consistent with SDSS analyses using the \gimtwod{} parametric surface-brightness decomposition analysis tool \citep{1998ASPC..145..108S}. Specifically, the analysis of \cite{2011ApJS..196...11S} using the same quantitative morphology pipeline provides an observational reference catalog against which our results can be benchmarked. We clarify that our aim is not to optimize our quantitative morphology pipeline in its capacity to accurately model the simulated galaxies with realism. The experiment described here is rather to apply the identical, calibrated pipeline for deriving morphologies from real galaxy images to realistic images of galaxies from a cosmological simulation. Only in this way do we enable a comparison that is procedurally consistent between observed galaxies and simulated galaxies with realism. We apply an extensive suite of observational realism to the Illustris synthetic images to ensure that the same biases in resolution, signal-to-noise, and crowding affect both model and real galaxies. Furthermore, we provide a detailed characterization of these biases by conducting several experiments that quantify the random and systematic errors associated with these biases and identify their correlations with morphology. We also investigate the biases associated with the projected viewing angle on the morphological parameters. Questioning the adequacy of the synthetic images themselves, we examine the biases associated with post-processing choices of how stellar light is propagated from unresolved stellar populations and their effect on structural estimates. We reserve comparisons of the morphologies to a forthcoming paper \citep{cbottrell2017}.

This paper is organized as follows. Section \ref{ags} provides a basic description of the simulation products and synthetic images and a detailed description of our observational realism suite. Section \ref{morph} describes the application of our 2D parametric quantitative morphologies to mock-observed galaxies and the resulting catalogs. The biases on structural parameters from realism are investigated and discussed in Section \ref{sec:bias}. We adopt cosmological parameters that are consistent with \emph{Wilkinson Microwave Anisotropy Probe} 9-year results in a $\Lambda$CDM cosmogony: $\Omega_{\text{m}} = 0.2726$; $\Omega_{\Lambda} = 0.7274$; $\Omega_{\text{b}} = 0.0456$; $\sigma_8 = 0.809$; $n_s= 0.963$; and $H_0 = 100 h$ km s$^{-1}$Mpc$^{-1}$ where $h=0.704$ \citep{2013ApJS..208...19H}.

\section{Simulated Galaxies}{\label{ags}}

In this section, we describe the Illustris simulation, production of synthetic stellar mock galaxy images, and applied observational realism used to generate the image that form the basis of the analysis in this paper. For details of the simulation models, we refer the reader to \cite{2013MNRAS.436.3031V,2014MNRAS.444.1518V,2014Natur.509..177V}.



\subsection{Illustris Simulation}{\label{illustris}}

The Illustris simulation is a large-volume cosmological hydrodynamical simulation \citep{2014Natur.509..177V,2014MNRAS.444.1518V,2014MNRAS.445..175G}. The simulation employs a broad physical model that includes a sub-resolution interstellar medium (ISM), gas cooling (including primordial and metal-line cooling), star-formation, stellar evolution and enrichment, supernova feedback, black-hole seeding and merging, and active galactic nucleus (AGN) feedback. Parameters of the model are tuned to reproduce the stellar mass function at $z=0$ and global star-formation rate density across cosmic time \citep{2014MNRAS.438.1985T,2014MNRAS.445..175G}.


\citetalias{2015MNRAS.454.1886S} showed that Illustris contains a variety of galaxy morphologies at $z=0$ including spirals, ellipticals, and irregulars whose non-parametric structures broadly agree with observations. In particular, the presence of rotationally-supported disc galaxies in coexistence with elliptical populations is an important result in the current generation of cosmological hydrodynamical simulations. Not only is it relevant to the capacity with which galaxy populations from the simulations can be compared with the observable universe, but also in resolving the long-standing problem of general angular momentum deficit, high central concentrations, and unrealistic rotation curves in disc-formation experiments \citep{2000ApJ...538..477N,2012MNRAS.423.1726S,2012MNRAS.427.2224T}. The implication of the \citetalias{2015MNRAS.454.1886S} analysis is that realistic galaxy morphology can emerge from simulations that couple sophisticated numerical methods to reasonably complete galaxy formation (specifically feedback) modules.

Current large-volume hydrodynamical simulations such as Illustris and EAGLE have been shown to broadly reproduce fundamental relations and galaxy morphologies observed in the real universe (e.g. \citealt{2015MNRAS.454.1886S,2015arXiv151005645F,2015MNRAS.450.1937C,2015MNRAS.452.2879T,2016MNRAS.tmp..895T}; Lange et al. in prep.). Their new levels of fidelity make them well-suited for experiments comparing the observationally accessible properties of simulated galaxies with those of real ones. A quantitative morphology analysis that employs bulge+disc surface brightness decompositions and will yield structural information about the physical components is ideally fashioned for this task. The key to the interpretive power in relating the results with observations is that the theoretical/synthetic images include the same realism found in observational datasets. We describe our methods for creating realistic synthetic images in the following sections.

\subsection{Synthetic Galaxy Images}{\label{stellar mocks}}

We employ the synthetic galaxy image catalog of \citetalias{2015MNRAS.447.2753T} which is comprised of 6891 galaxies with $N_{\star} > 10^4$ stellar particles. The particle resolution cut places a lower limit on the total stellar mass of galaxies in our sample at $\log \text{M}_{\star}/\text{M}_{\odot} \gtrsim10$. Systems below this cut are neglected owing to their poor internal resolution. All galaxies in our sample are taken from redshift $z=0$ and their surface brightness distributions are redshifted to $z=0.05$. The corresponding luminosity distance and angular scale at this redshift, assuming the cosmological parameters stated at the end of Section \ref{intro}, are $d_L=221.3$ Mpc and 0.973 kpc/arcsecond. 

\subsubsection{Stellar Light \& Surface Brightness Smoothing}\label{sec:stellar_light}

The methods for producing synthetic spectral energy distributions (SEDs) and idealized simulated galaxy images are described in detail in \citetalias{2015MNRAS.447.2753T}. In short, stellar particles inherit an initial mass $ \mathrm{M}_{\star}\approx1.3\times10^6\; \mathrm{M}_{\odot}$, time of birth, and a metallicity from the local ISM where the particle is created. Using the \textsc{starburst99} single-age stellar population synthesis (SPS) models \citep{1999ApJS..123....3L,2005ApJ...621..695V,2010ApJS..189..309L} an SED is assigned to each stellar particle based on its mass, metallicity, and age. The \textsc{sunrise} code \citep{2006MNRAS.372....2J,2010MNRAS.403...17J} is used to map stellar light from each particle and generate a synthetic image. The dust absorption, scattering, or emission functionalities of the \textsc{sunrise} code are not used in the radiative transfer (see Section \ref{sec:dust}).

Stellar particles represent discretized unresolved stellar populations in Illustris. 
\citetalias{2015MNRAS.447.2753T} use adaptive smoothing of each star particle using a gaussian kernel with full-width at half maximum equal to the $16^{\text{th}}$ nearest neighbour distance to convert the discrete simulation particle distribution to a smooth distribution of light. However, \citetalias{2015MNRAS.447.2753T} caution that neither this, nor any other adaptive or fixed-length smoothing prescription that they explore is any more or less valid when treating individual star particles as full unresolved stellar populations. While a more physical light assignment procedure may exist, we limit our exploration of methods for distributing stellar light only to the biases on specific parameters in the morphological decompositions (Section \ref{sec:smooth}). 



\subsubsection{Creation of Synthetic Images}
Photon packets from the stellar particles are propagated into pinhole cameras with $256\times256$ pixel resolution providing each pixel with the integrated SED of the photon packets incident upon it - effectively creating a mock integral field unit (IFU) data cube. The field of view from each camera is $10\;rhm_{\star}$ where $rhm_{\star}$ is the stellar half-mass radius measured from the gravitational potential minimum of a galaxy. Cameras are located on the arms of a tetrahedron whose centroid is located at the gravitational potential minimum. The camera locations provide 4 camera angles with effectively random orientation with respect to the galaxy of interest. The mock IFU can then be convolved with a filter transmission function and resampled onto the desired angular resolution. The broadband image data products of \citetalias{2015MNRAS.447.2753T} can be manipulated to create synthetic galaxy images in 36 unique bands using the dedicated python module \textsc{sunPy}.\footnote{http://www.github/ptorrey/sunpy-master/}

\subsubsection{Dust obscuration}\label{sec:dust}

The synthetic images are created without using the dust absorption, scattering, or emission functionalities of the \textsc{sunrise} code. The inclusion of dust effects in the radiative transfer stage is a vital element of a truly comprehensive procedure for observational realism. However, the mass and spatial resolutions afforded by the Illustris simulation are not expected to enable converged radiative transfer results with dust \citep{2010MNRAS.403...17J}. Furthermore, given that the dust is not traced explicitly in current large-volume state-of-the-art cosmological simulations on the scale of Illustris (though, see recent work by \citealt{2016MNRAS.457.3775M}), the properties derived from synthetic images of galaxies that are generated using radiative transfer with dust may depend sensitively on how dust is implemented in post-processing.

In principle, the dust column density and corresponding obscuration along the line of sight to a stellar particle within a cosmological hydrodynamical simulation can be computed by adopting a model for the distribution of dust that follows the gas density, temperature, and spatial metallicity distribution (e.g. \citealt{hopkins2005physical, robertson2007photometric, wuyts2009recovering, wuyts2009color}). However, a proper treatment of dust in generating synthetic images and spectra requires the capacity to resolve the gas and dust distributions on small spatial scales ($\ll$1 kpc). In the Illustris simulation, the complex structure of the interstellar medium (ISM) is modelled with a pressurized, effective equation-of-state \citep{2003MNRAS.339..289S} that is limited to $\sim$ kpc scale resolution. Consequently, gas discs in Illustris galaxies have larger scale-heights and are more chemically and thermodynamically homogeneous than in the ISM of real galaxies. It may be possible to construct a sub-grid dust-obscuration model that accounts for the unresolved properties of the ISM in Illustris and facilitates accurate and convergent radiative transfer. However, applying such a model would require a dedicated exploration of the impact of sub-grid dust properties on the resulting synthetic images in order to understand the model uncertainties. Indeed, recent zoom-in simulations of Milky Way halos and 25 $h^{-1}$ Mpc volumes that employ the Illustris model and explicitly implement dust have challenged the notion that the dust broadly traces the gas density and metallicity averaged over large scales \citep{2016MNRAS.457.3775M,2016arXiv160602714M}. Evidently, dedicated investigations of the sub-grid or explicit modelling of dust in large-volume hydrodynamical simulations are required to resolve the effects and uncertainties associated with dust on measured properties from synthetic galaxy images. Such an exploration is beyond the scope of this paper. Instead, we acknowledge that the lack of dust in the synthetic images may bias our comparison with observations, and focus in this paper on analysis of the unattenuated synthetic stellar light distributions. Nonetheless, our dustless synthetic images will be a valuable standard for comparisons with future image catalogs and realism suites that include treatments for dust.


\subsection{Observational Realism}{\label{realism}}

Observational realism is applied to the synthetic images of the simulated galaxy sample from Illustris to enable direct comparisons between real and simulated galaxy populations. First, we generate synthetic images of simulated galaxies in SDSS bands $g$ and $r$ using \textsc{sunPy} (as described in Section \ref{stellar mocks}) to be consistent with the canonical simultaneous $g-$ and $r-$band morphological decomposition analysis of 1.12 million real galaxies in the SDSS DR7 Legacy Survey (\citealt{2011ApJS..196...11S}, \citetalias{2011ApJS..196...11S}). We then use the following procedure to create an unbiased Illustris simulated galaxy population in SDSS in $g$ and $r$ bands:

\begin{enumerate}
\item[(1)] \textsc{Selection of SDSS Fields and Photometric Quantities}: We randomly select a unique galaxy objID from SDSS DR7 Legacy photometric galaxy catalog. The SDSS atlas (prefix ``fpAtlas"), PSF (prefix ``psFIeld"), and $g-$ and $r-$band corrected images (prefix ``fpC") for the \texttt{run}, \texttt{rerun}, \texttt{camcol} and \texttt{field} containing this galaxy are then obtained from the SDSS Data Archive Server (DAS). The corrected images are field images that have been reduced through bias-subtraction, flat-fielding, and purging of bright stars. The PSF image contains all of the necessary metadata to reconstruct the PSF in any band and location on the corrected image field. The Atlas image is only collected for the CCD gain that is contained in its image header. The photometric zero points, airmasses, extinction coefficients (\texttt{PhotoPrimary} table), and CCD gain (Atlas image) for the SDSS fields are collected to be used to convert the synthetic Illustris galaxy image fluxes from nanomaggies to counts. 

\item[(2)] \textsc{Photometric Segmentation and Location Assignment}: We employ \textsc{Source Extractor} (\sextractor{}) \citep{1996A&AS..117..393B} to create a segmentation map of the SDSS $r-$band corrected image. We select a random location on the SDSS corrected image as the designated position of the centroid of the simulated galaxy images. While upholding our intention of recreating realistic crowding statistics of the simulated galaxy population, we restrict selection of the centroid location to pixels which have not been flagged as belonging to other objects identified in the segmentation map. Our choices of deblending parameters in \sextractor{} are the same as those used in \citetalias{2011ApJS..196...11S}.

\item[(3)] \textsc{SDSS PSF Image Convolution and Flux Conversion}: We reconstruct the $g-$ and $r-$band PSF images specific to the selected location in the corrected image using the \texttt{read\_PSF} software utility from SDSS\footnote{http://classic.sdss.org/dr7/products/images/read\_psf.html}. We remove the softbias added by SDSS, normalize the PSF images, and convolve them with the noiseless synthetic galaxy images to provide realistic SDSS resolution. Using the photometric information obtained in (1) specific to the choice of SDSS corrected images, we convert the fluxes of the convolved synthetic images from nanomaggies (default data product units) to DN counts.

\item[(4)] \textsc{Addition of Signal Shot Noise to Simulated Galaxy}: Photon shot noise is generated and added to the convolved synthetic galaxy images following Poisson statistics. The contribution of Poisson noise to the total variance in each pixel is expected to be small relative to the sky -- which is the dominant noise term for photometry in the SDSS \citep{2011ApJS..196...11S}. In a preliminary analysis of galaxies at $z = 0$ we found that the inclusion of Poisson noise did not affect parameter estimates significantly, but we elect to include it for completeness.

\item[(5)] \textsc{Placement of Simulated Galaxy Image into the SDSS}: The simulated galaxy images with realism are inserted into the SDSS corrected images with image centroids (which are aligned with the gravitational potential minimum of the galaxy) at the designated location from (2). This provides the bias contributions from the sky, crowding, and any other field-specific properties.

\end{enumerate}

\begin{figure*}
	\includegraphics[width=\linewidth]{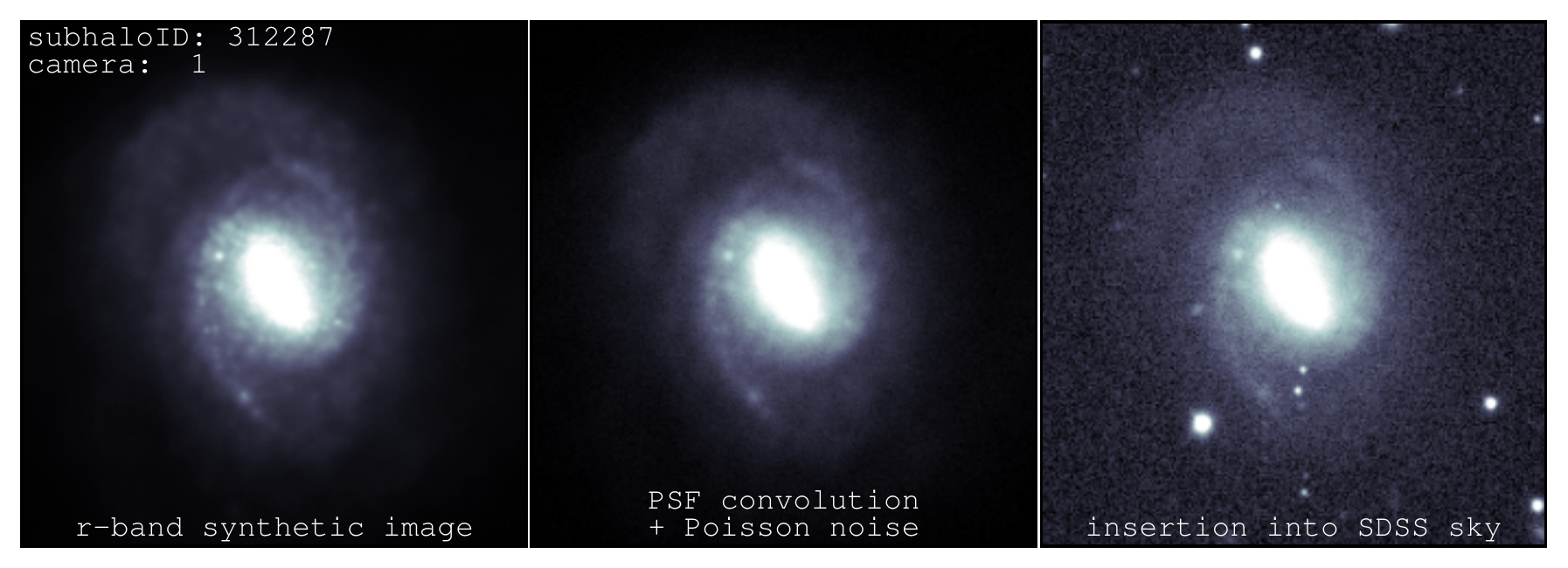}
    \caption[Addition of observational realism]{Addition of observational realism to synthetic images of simulated galaxies. \emph{Left panel}: Synthetic SDSS $r-$band image of Illustris subhaloID 312287 in a nearly face-on projection. \emph{Middle panel}: convolution of the $r-$band synthetic image with the PSF corresponding to its insertion into the SDSS and addition of signal shot noise. \emph{Right panel}: placement of PSF-convolved synthetic image flux into an SDSS corrected image. The logarithmic scale is identical in each image. The flux in each pixel in the right panel has been lifted by 5 counts so that the sky is visible using this scale.}
    \label{fig:i2o}
\end{figure*}

Figure \ref{fig:i2o} shows a demonstration of the steps described above. The rightmost panel of Figure 1 can be analyzed using the same pipeline described in \citetalias{2011ApJS..196...11S}. The only prior is the position of the simulated galaxy in the corrected image -- which is consistent with analysis of real galaxies. One subtlety is that applying steps (1) and (2) to the placement of simulated galaxies achieves effective matching to the statistics of crowding, resolution, sky brightness within the real photometric galaxy catalog. Using this placement procedure, fields containing higher concentrations of galaxies have a naturally higher probability of being selected for placement. While galaxies in isolation will still be represented, this method guarantees that any systematic biases in the recovered parameters from quantitative morphological analysis are (a) consistent with the real galaxy population and (b) statistically quantifiable through analysis of the same simulated galaxy across multiple placements and viewing angles.

\section{Quantitative Morphologies}{\label{morph}}
The observational realism described in the previous section puts the Illustris simulated galaxy population in an observational context that is well suited for quantitative comparisons with real galaxy populations. To facilitate our comparison, we use the same quantitative morphology analysis of \citetalias{2011ApJS..196...11S} which performs 2D photometric surface brightness decomposition with parametric component models. Bulge+disc and single component decompositions contrast the detailed structural properties of simulated and observed galaxies. Furthermore, our realism suite places the morphological decompositions of simulated galaxies on level ground with observations. While \citetalias{2015MNRAS.454.1886S} employ the same quantitative analysis and methods for photometric deblending in the simulated and observed galaxy populations, any biases that are \emph{intrinsic} to the source-delineation or morphological analysis that correlate with resolution, signal-to-noise, and crowding will manifest themselves differently in the observations and simulations if the statistical distributions for these realisms are not the same. Our more extensive treatment of observational biases may have important consequences in comparisons between models and observations. 

In this section, we detail our quantitative morphologies analysis. We describe our methods for delineating photometric boundaries between sources and the sky and between closely projected sources. We then explain our choice of parametric models in the surface brightness decompositions and the structural parameters that are afforded by these choices. Finally, we describe the design of our catalogs -- each of which have associated dedicated experiments for characterization of the biases that our realism considerations have on structural parameter estimates. 

\subsection{Deblending}{\label{deblend}}

The methods that are used to delineate object boundaries (deblending) have been shown to affect morphological parameters -- particularly in crowded images \citep{2011ApJS..196...11S}. The standard SDSS \textsc{photo} pipeline attempts to isolate the flux from an object to reconstruct the image of what the object would have looked like if it were the only source in the image. Therefore, pixels that share flux from multiple sources are attributed to the area of each source with an associated fractional flux contribution based on the reconstruction modelling. However, deblending with the \textsc{photo} pipeline has been shown to produce erroneous photometric and structural estimates such as the production of red-outliers and large scatter in the colour-magnitude diagrams of pair galaxies -- which were previously and erroneously ascribed to a new population of extremely red galaxies in pair systems \citep{2004MNRAS.352.1081A,2009MNRAS.399.1157P,2010MNRAS.401.1043D}. Inaccurate photometric estimates in pairs is an indication that the same inaccuracies are relevant in all objects with closely projected external sources. \citetalias{2011ApJS..196...11S} showed that photometric and structural estimates derived from \sextractor{}\footnote{\sextractor{} deblending uses a multi-threshold flux tree. Starting with the lowest isophotal threshold and moving up, troughs are identified that separate branches which meet the criterion of containing a specified fraction of the total flux. The minima of these troughs delineate the flux boundaries and each pixel is given a flag corresponding to a unique object -- creating a segmentation image. No pixels are shared and therefore no object's segmentation map area extends into the area associated with another object identified through this scheme. Although the fluxes measured directly from the pixels associated with an object would be systematically be underestimated in this scheme in the presence of close neighbours, \citetalias{2011ApJS..196...11S} showed that the missing flux is recovered by fitting a surface brightness profile model -- which integrates the flux of the model (whose form is determined only by the pixels flagged as belonging to the object of interest) out to large radii. In practice, fitting the model recovers the missing flux of \sextractor{} deblended objects in crowded environments.} deblending combined with \gimtwod{} sky measurement and bulge+disc decompositions improved upon other schemes using several sensitive tests: the size-luminosity relation of discs, and the colour-magnitude diagrams and fiber colours of pairs. \Seleven{} showed that the deblending used in the \textsc{photo} pipeline and associated magnitude and colour estimates was the source of the outliers and scatter in the colour-magnitude diagrams of pairs. \citetalias{2011ApJS..196...11S} also demonstrated that deblending using \sextractor{} in tandem with parametric bulge+disc decompositions reduced the scatter and eliminated the outlier populations in the colour magnitude diagrams of pairs -- leaving a tight red sequence and clearly separated blue cloud using SDSS pair catalog of \cite{2011MNRAS.412..591P}. We therefore employ the \sextractor{} source deblending procedure used by \citetalias{2011ApJS..196...11S}. We do not presume that the \citetalias{2011ApJS..196...11S} scheme is optimal or unique in defining object-sky boundaries and separating objects whose constituent pixels may have shared contributions from other sources. However, we note that although any biases from the \citetalias{2011ApJS..196...11S} deblending and sky estimation scheme may affect our morphologies, application of the same scheme to our mock-observed simulated galaxies ensures that the biases are consistent (we explore the impact of alternative deblending on morphological parameters in Appendix \ref{sec:SExpars}). 


\begin{figure}
	\center\includegraphics[width=\linewidth]{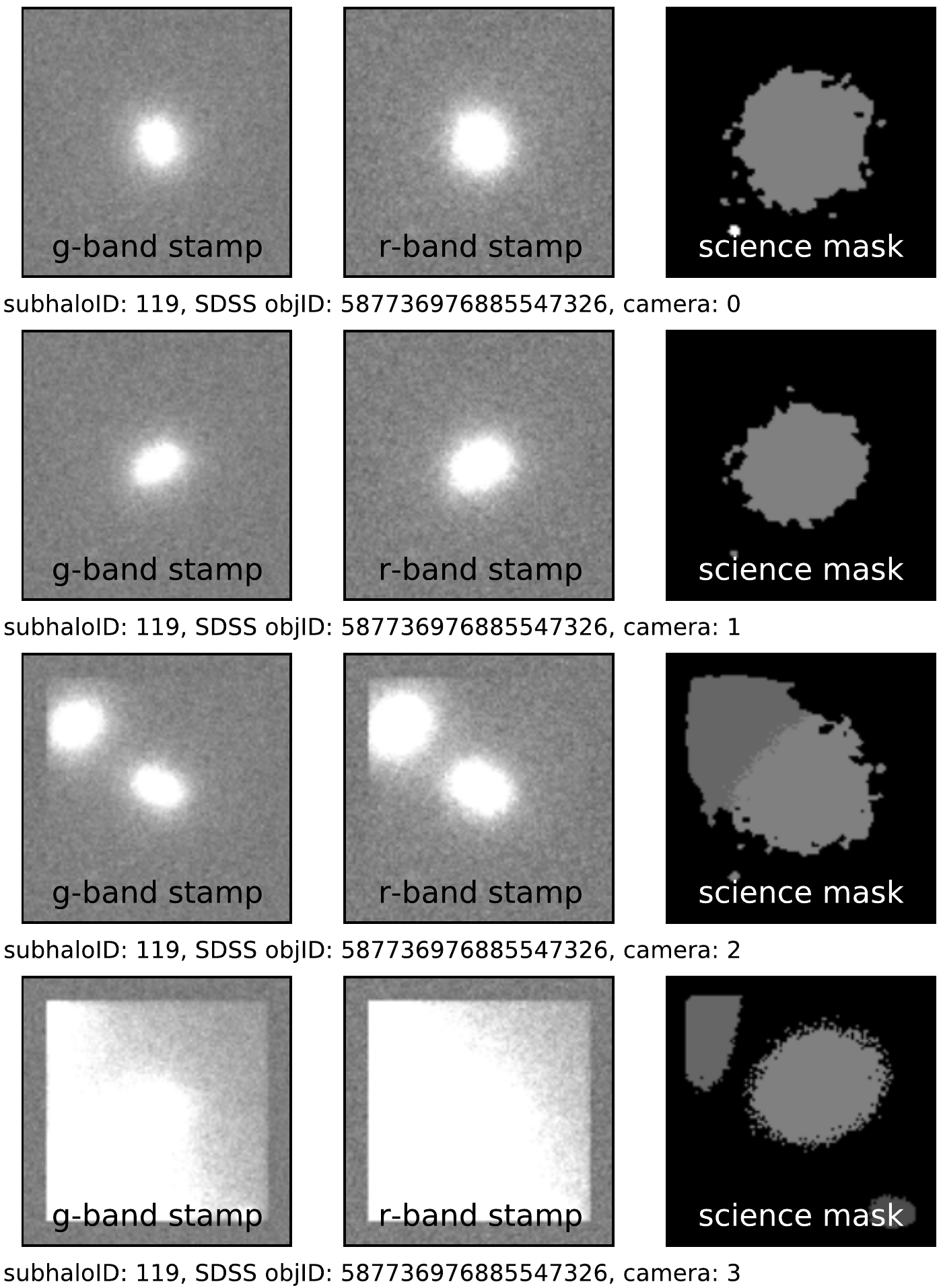}
    \caption[Camera angles and FoF projection]{Four camera angle realizations of the same simulated galaxy with observational realism. The galaxy (subhaloID 119) is inserted into the same location in SDSS for each camera angle -- separated by row in the Figure. The science $r-$band image stamp, $g-$band image stamp, and segmentation stamp are shown from left to right. The third row of panels indicate that the central galaxy has at least one companion that is not within the field of view for the first two rows. Note the delineation of flux for the companion at the boundary of inserted synthetic image. Only pixels for which the central galaxy has a dominant flux contribution are selected as data used in the fitting. In the fourth row, another (more massive) companion's orientation with respect to the central galaxy places it in the line of sight of the cameras field of view. The projection effect results in a large discontinuity between the flux at the boundary of the synthetic image and the rest of the SDSS sky and an erroneous science mask that does not properly delineate between the galaxy's flux and the visible flux from the companion.}
    \label{fig:sciscimsk}
\end{figure}


Before application of our five-step realism suite described in Section \ref{realism}, the synthetic images generated using \textsc{sunPy} have no sky or noise other than a residual noise contribution from the Monte Carlo photon propagation scheme in \textsc{sunrise}. However, since the synthetic images are constructed from the Friends-of-Friends halo finder in Illustris, there may be contributions from other stellar sources within the field of view that give rise to fluxes that do not truncate to zero at the synthetic image boundaries. When added to the SDSS corrected image after the other steps in Section \ref{realism}, the non-zero flux from other sources in the FoF group may result in boxlike flux boundaries between synthetic image and SDSS corrected image into which it is placed. Figure \ref{fig:sciscimsk} shows an example where FoF companions to the galaxy of interest (centred) are in the projected field-of-view (two bottom rows of panels). The undesirable effect is exemplified in the bottom row, where a companion is projected along the line of sight between the galaxy and the camera position.

Galaxy projections with unrealistic artifacts such as shown in Figure \ref{fig:sciscimsk} are easily separable by comparing the total flux at an arbitrary camera angle with any other. Identification of a significant positive systematic bias in the total flux of a synthetic image of a particular galaxy subhaloID with respect to, for example, an estimate of the mode flux for all camera angles is a effective flagging scheme for these situations. These projection effects are also rare. SubhaloID 119 is also a part of the most massive and most crowded FoF in the simulation. The projection effects in Figure \ref{fig:sciscimsk} are only common among galaxies belonging to the most massive groups and clusters. Situations such as seen in the third row of panels in Figure \ref{fig:sciscimsk} are not problematic for our quantitative morphologies pipeline due to our deblending scheme and valid comparisons between the morphological parameters to those from other camera angles and real galaxies can be made. They are not useful, however, in analyses that compare properties of the synthetic images such as total flux and photometric aperture half-light radius with those recovered from the fitting. The aim of our analysis is to decompose the surface brightness profiles of the primary galaxy -- which contains the subhalo's gravitational potential minimum. Therefore, parametric estimates of flux and size will be systematically biased relative to the properties derived directly from the synthetic images for galaxies with prominent projected companions/satellites. However, the bias only occurs commonly amongst massive galaxies belonging to large groups.

\subsection{Parametric Decompositions}
We perform simultaneous parametric 2D surface brightness profile fitting of the $g-$ and $r-$band image data using the \gimtwod{} software package \citep{2002ApJS..142....1S}. \gimtwod{} employs a Metropolis-Hastings Markov-Chain Monte Carlo algorithm for deriving the best-fitting models based on the image data. Only science image pixels identified by \textsc{SExtractor} to belong to the principle object in the mask are used in the likelihood calculations. The initial conditions are determined from a coarse sampling of the very large volume of structural parameter space with generous limits that are computed from the image moments. The procedure for setting the initial conditions and optimization of the models are identical to the \Seleven{} analysis.

Two models are used to fit the images: a bulge+disc (B+D) model and a pure Sersic ($pS$) model. The B+D decompositions employ Sersic profiles with indices $n_b=4$ and $n_d=1$ \citep{1953MNRAS.113..134D,1959HDP....53..311D,1970ApJ...160..811F}, while the $pS$ model allows the Sersic index to vary over $0.5 \leq n_{pS} \leq 8.0$. The fourteen free parameters of the B+D decompositions are the total fluxes in each band $f_g$, $f_r$, bulge-to-total ratios $(B/T)_g$, $(B/T)_r$, semi-major axis bulge effective radius $r_e$, bulge eccentricity $e$, bulge position angle (clockwise y-axis$\equiv0$) $\phi_b$, disc scale length $r_d$, disc inclination $i$, disc position angle $\phi_d$, and centroid positions $(dx)_g$, $(dy)_g$, $(dx)_r$, and $(dy)_r$. The parameters for position angles of the bulge and disc, disc inclination, bulge ellipticity, bulge effective radius, and disc scale length have the added constraint that they must be equivalent in both bands. Furthermore, the centroid positions of the bulge and disc components of the model are constrained to be the same. Similarly, the ten free parameters for the $pS$ model are $f_g$, $f_r$, Sersic index $n_{pS}$, semi-major axis disc effective radius $r_e$, disc eccentricity $e$, disc position angle $\phi_d$, and centroid positions $(dx)_g$, $(dy)_g$, $(dx)_r$, and $(dy)_r$. The position angle of the profile, ellipticity, and effective radius are constrained to be the same in each band in the $pS$ fits. 

\subsection{Simulated Galaxy Population Samples and Catalogs}\label{catalogs}

Several experiments were conducted with increasing completeness to characterize the complexities and biases affecting estimates for the full Illustris galaxy population. Our samples and catalogs are described in this section. The decomposition catalogs are summarized in Table \ref{tab:catalogs} at the end of this section and are made publicly available in the online supplementary information with this paper. Descriptions of catalog parameters are given in Appendix \ref{sec:parameter_descriptions} in Tables \ref{tab:illustris_n4} and \ref{tab:illustris_pS}. 

\subsubsection{Representative Illustris Galaxy (RIG) Sample}
We began by selecting a small, but representative Illustris galaxy (RIG) sample in the parameter space of stellar mass and stellar half-mass radius. The RIG sample was assembled by uniformly sampling 100 galaxies in the $ \mathrm{M}_{\star}-rhm_{\star}$ plane while omitting systems in mergers, with strong tidal features, or problematic projection effects (see Figure \ref{fig:sciscimsk}).
Figures \ref{fig:RIGS_1}-\ref{fig:RIGS_5} show each RIG in the \texttt{CAMERA 0} projection.

The RIG sample is used to perform repeated analyses of galaxies in SDSS fields to obtain parameter estimate distributions for each galaxy and each camera angle -- facilitating a quantitative analysis of the biases from field crowding, camera projection, and stellar light sourcing. Our final catalog includes a single decomposition for all camera angles of every galaxy in the Illustris synthetic image catalog of \citetalias{2015MNRAS.447.2753T}. The characterization of biases with the RIGs preceding the analyses of every galaxy in Illustris enables quantitative assessment of the typical biases associated with specific regions of the size-mass space of Illustris galaxies.
In the following two sub-sections, we examine biases associated with galaxy image generation and camera projection by holding the sky placement fixed.

\subsubsection{\texttt{SMOOTHING} catalog}\label{SMOOTHcat}
The \texttt{SMOOTHING} catalog is constructed to explore the choices for distributing stellar light in creating the synthetic images \citep{2015MNRAS.447.2753T}. Synthetic images are generated for the RIG sample using three alternative stellar light distribution (SLD) schemes to compliment the existing images constructed using the fiducial scheme (Section \ref{sec:stellar_light}). All SLD realizations of galaxies from the RIG sample are placed in the same uncrowded location in SDSS and fitting is performed for a single camera angle: \texttt{CAMERA 0}. The \texttt{SMOOTHING} catalog therefore contains 400 decompositions with four decompositions for each RIG (one for each SLD scheme). The biases on photometric and structural parameters from the SLD schemes may be evaluated with respect to the fiducial scheme and to each other. Four SLD schemes are explored in our analysis:

\begin{enumerate}
\item[(1)] \textsc{Fiducial Smoothing} (\texttt{fn16}): Light from each stellar particle within the FoF halo is projected from an SPH kernel\footnote{A cubic B-spline profile.} with a characteristic scale set by the distance to the 16th nearest stellar particle. Adaptive smoothing allows for smoother distributions of light while avoiding unrealistic compactness around largely isolated stellar particles at large distances from the galactic centre.

\item[(2)]  \textsc{Constant Smoothing} (\texttt{fc1kpc}): Same as (1), but with the characteristic scale length set to a constant 1 kpc for all stellar particles. Total stellar light is conserved with respect to the fiducial scheme, but compact surface brightness features with projected spatial distributions less than the characteristic scale are distributed more broadly. 

\item[(3)] \textsc{Resampled Adaptive Smoothing} (\texttt{rn16}): Young, bright stellar populations associated with the $\sim10^6$ solar mass simulation stellar particles can result in artificially distinct, circular light features in the synthetic images. To mitigate this effect, we resample the light associated with these young, bright star particles into 100 particles (child particles). The child particles contain 1/100th the mass of the parent particle, are spatially distributed within the parent particles original light kernel, and are assigned Gaussian age and metallicity distributions with $1\sigma$ values of 10\% of the original age and metallicity. The resulting stellar population flux is roughly conserved, but the sharp light profile edges are somewhat reduced. The light from all child particles and remaining particles is then propagated in the same way as for (1).

\item[(4)] \textsc{Resampled Constant Smoothing} (\texttt{rc1kpc}): Same resampling as in (3) but but a constant characteristic kernel scale of 1 kpc is employed.

\end{enumerate}

\subsubsection{\texttt{CAMERAS} catalog}\label{CAMScat}
The \texttt{CAMERAS} catalog is constructed to evaluate the biases on parameter estimates from projection. Each galaxy in the RIG sample is placed in a single, uncrowded location in SDSS and fitting is performed for all four camera angles. This guarantees consistent and controlled environment, resolution, and sky in each decomposition and across analyses of each RIG. The \texttt{CAMERAS} catalog contains 400 decompositions with four decompositions for each RIG (one for each camera angle). The variation in best-fitting photometric and structural estimates in each RIG are obtained from the \texttt{CAMERAS} catalog.

\subsubsection{\texttt{ASKA} catalog}\label{ASKAcat}
An \textbf{A}ll \textbf{SK}y \textbf{A}nalysis was performed on the RIG sample to examine insertion effects. The RIGs are placed all over the SDSS sky with an average of $\sim100$ unique fields for each galaxy following the steps in Section \ref{realism}. All four camera angles of a galaxy are modelled in each placement -- providing a distribution of $\sim100$ unique sets of best-fitting morphological parameters for each camera angle. The full \texttt{ASKA} catalog contains $\sim40,000$ decompositions of galaxies from the RIG sample. The distribution of best-fitting parameters for each camera angle of a galaxy samples the real statistics for crowding, signal-to-noise, and resolution that exists for SDSS galaxies as a result of our placement criteria and realism procedures. We use the \texttt{ASKA} catalog to quantify the random and systematic effects of crowding, signal-to-noise, and resolution. The distributions of best-fitting parameters for a galaxy in each fixed camera angle facilitate the statistical quantification of the scatter and systematics from biases associated with placement. 

\subsubsection{\texttt{DISTINCT} catalog}
\begin{figure*}
	\center\includegraphics[width=0.8\linewidth]{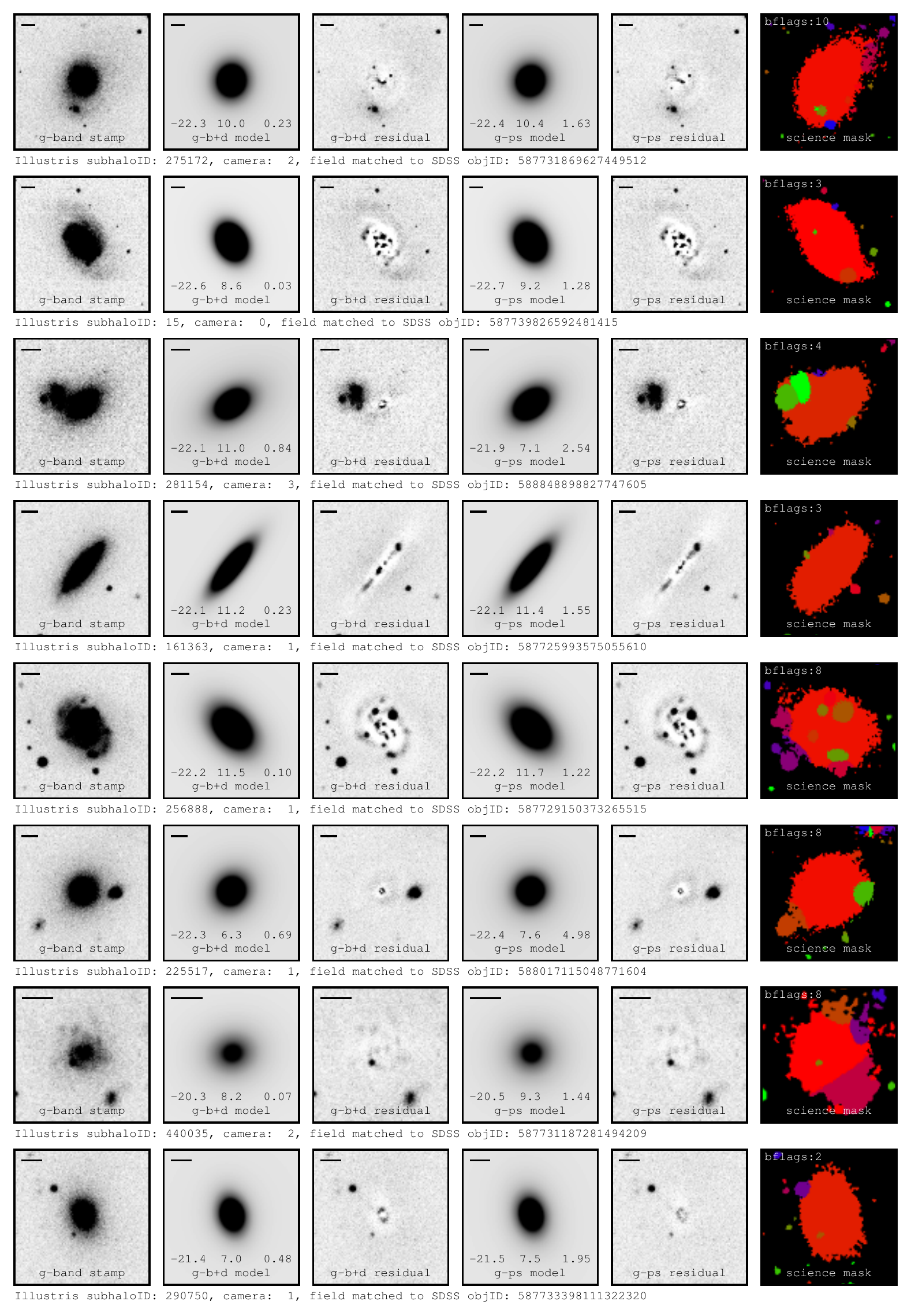}
    \caption[Mosaic of fits to simulated galaxies with realism]{Mosaic of g-band images and best-fitting models for simulated galaxies with realism. \emph{Left to Right:} (1) science cut-out (stamp) of simulated galaxy with realism; (2) B+D model absolute magnitude, circular half-light radius, and bulge-to-total ratio parameters listed from left to right; (3) B+D residual image (model subtracted from science cut-out); (4) $pS$ model with absolute magnitude, circular half-light radius, and Sersic index parameters listed from left to right; (5) $pS$ residual image; (6) source delineation mask (science mask). The science masks are colour-coded by flags associated with distinct objects. Red corresponds to the pixels that are used in the fitting. \emph{bflags} is the number of sources that directly neighbour pixels used in the fitting. A scale in the top left of each panel denotes 10 kpc at z=0.05.}
    \label{fig:mosaic}
\end{figure*}

The \texttt{DISTINCT} catalog contains decompositions for all galaxies in synthetic image catalog of \citetalias{2015MNRAS.447.2753T}. Each camera angle for a given galaxy is assigned probabilistically to a location in the SDSS following item (2) of Section \ref{realism}. The catalog contains $27,564$ decompositions from the 4 camera angles for each of 6891 galaxies from Illustris. Figure \ref{fig:mosaic} shows examples of B+D and $pS$ decompositions of galaxies taken from the \texttt{DISTINCT} catalog. The catalog is designed to investigate the global observational properties of the full Illustris galaxy population. The \texttt{DISTINCT} catalog also forms the basis for comparisons between simulated galaxies and populations of real galaxies in SDSS. Galaxies in the \texttt{DISTINCT} catalog can be sampled to match the luminosity or stellar mass distributions of real galaxies around $z=0.05$. 

\begin{table}
	\centering
	\caption[Decomposition Catalogs]{Decomposition catalogs. Each catalog was fitted with both a pure Sersic and bulge+disc model. SLD denotes which stellar light distribution methods are used to create the synthetic images of each galaxy (Section \ref{SMOOTHcat}). Cameras denotes which camera angle projections of each galaxy are used in a catalog. Whether the insertion into the SDSS sky is fixed to a single location for all decompositions (Fixed) or follows the randomized procedure described in Section \ref{realism} (Random) is indicated by Insertion. $N_{\mathrm{gal}}$ is the number of galaxies used in the decompositions. ($^\star$) Decompositions of all camera angles of galaxies are performed in each insertion location.}
	\vspace{5pt}
	\label{tab:catalogs}
	\begin{tabular}{lccccr} 
		\hline
		Catalog & SLD & Cameras & Insertion & $N_{\mathrm{gal}}$ & $N_{\mathrm{decomp}}$ \\
		\hline
		\texttt{ASKA} & \texttt{fn16} & 0-3 & Random$^{\star}$ & 100 & 42319 \\
		\texttt{SMOOTHING} & \texttt{all} & 0 & Fixed & 100 & 400 \\
		\texttt{CAMERAS} & \texttt{fn16} & 0-3 & Fixed & 100 & 400 \\
		\texttt{DISTINCT} & \texttt{fn16} & 0-3 & Random & 6891 & 27564  \\
		\hline
	\end{tabular}
\end{table}

\section{Characterization of Biases}\label{sec:bias}

In this section we explore estimates for total flux, circular half-light radii, and bulge-to-total light ratios in our decomposition catalogs. We explore the consequences of post-processing choices in how stellar light is distributed spatially from particles that embody full unresolved stellar populations in the simulation. We then assess the sensitivity of photometric estimates to the observational realism of projection and crowding, respectively. In each analysis of a potential bias, we take precautions to control all other potential biases such that the variation in parameters will be sensitive exclusively to the bias under examination. While we focus on demonstrating the B+D decomposition results in this paper, we have verified that the biases reported for galaxy sizes and magnitudes are broadly consistent in the characterization of biases for the $pS$ decompositions (see Section \ref{size_flux} and Appendix \ref{sec:SExpars}).

\subsection{Distribution of Stellar Light}\label{sec:smooth}

\begin{figure*}
	\includegraphics[width=\linewidth]{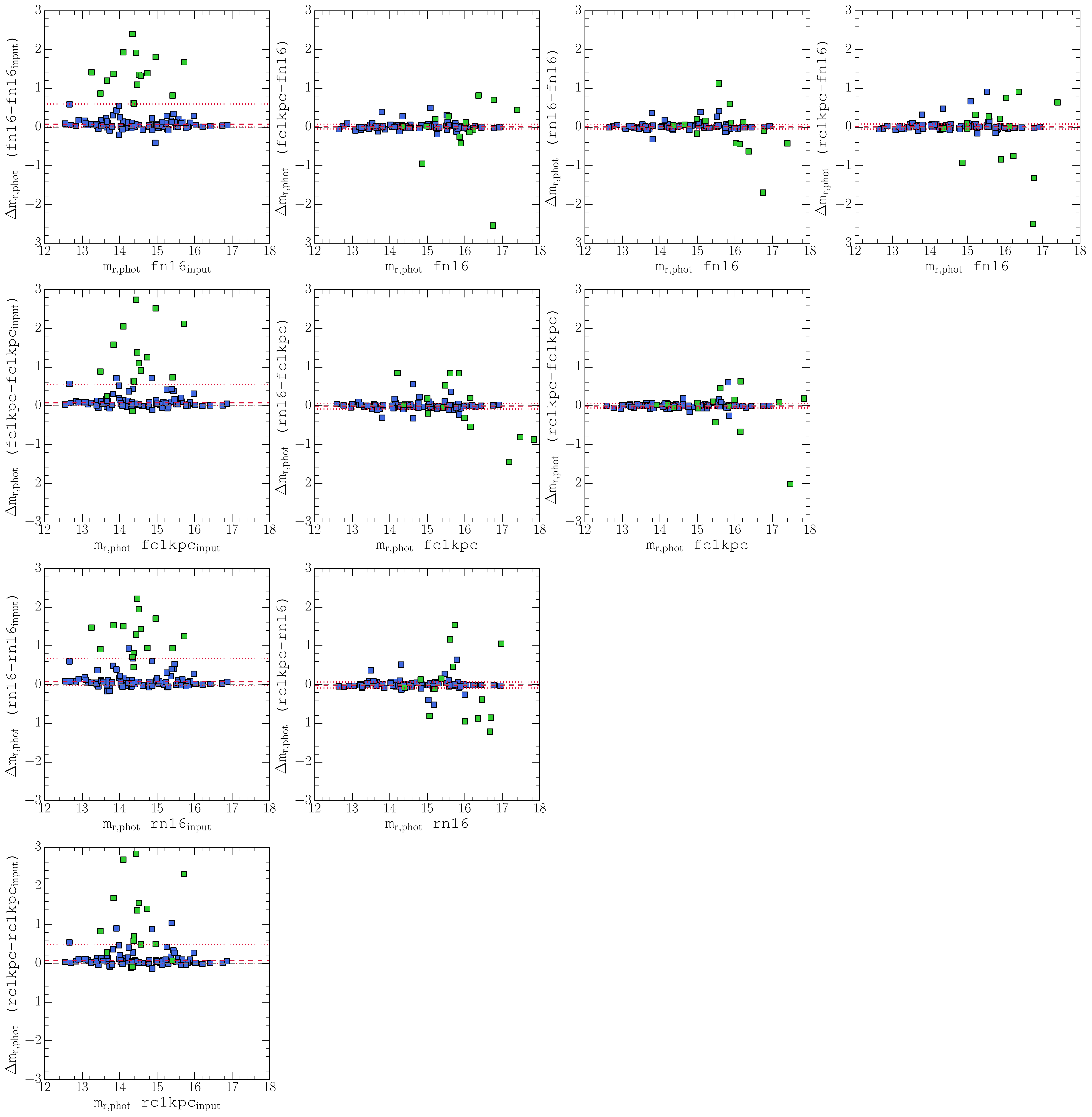}
    \caption[Magnitude bias from the choice of SLD scheme]{Variation and systematic biases in integrated magnitude for alternative SLD schemes. The panels compare the $r$-band integrated magnitude measurements, $m_{\mathrm{r,phot}}$, for the RIG sample taken from the \texttt{SMOOTING} catalog. The first panel of each row shows the systematic bias on $r$-band integrated magnitude computed from the difference between the magnitude of the best-fitting B+D model and the synthetic image for the specified SLD scheme, $\Delta m_{\mathrm{r,phot}}$. All other panels compare the magnitude estimates from one SLD scheme with another. The grey dashed line in each panel indicates $\Delta m_{\mathrm{r,phot}}=0$. The red dashed line and dotted lines show the median and 16$^{\mathrm{th}}$ to 84$^{\mathrm{th}}$ percentile range in each comparison, respectively. Green markers highlight the RIGs with systematic offsets in the fiducial scheme that exceed the $84^{\mathrm{th}}$ percentile in $\Delta m_{\mathrm{r,phot}}$ (i.e., those lying above the upper dotted red line in the top left panel). Panels showing the systematic biases in other schemes show that the choice of scheme does not improve agreement in the magnitude estimates for the RIGs identified as outliers in the fiducial scheme. The systematic outliers in $\Delta m_{\mathrm{r,phot}}$ for the fiducial scheme also consistently generate the majority of the scatter in comparisons of each SLD scheme with each other.}
    \label{fig:mag_smoothing}
\end{figure*}

\begin{figure*}
	\includegraphics[width=\linewidth]{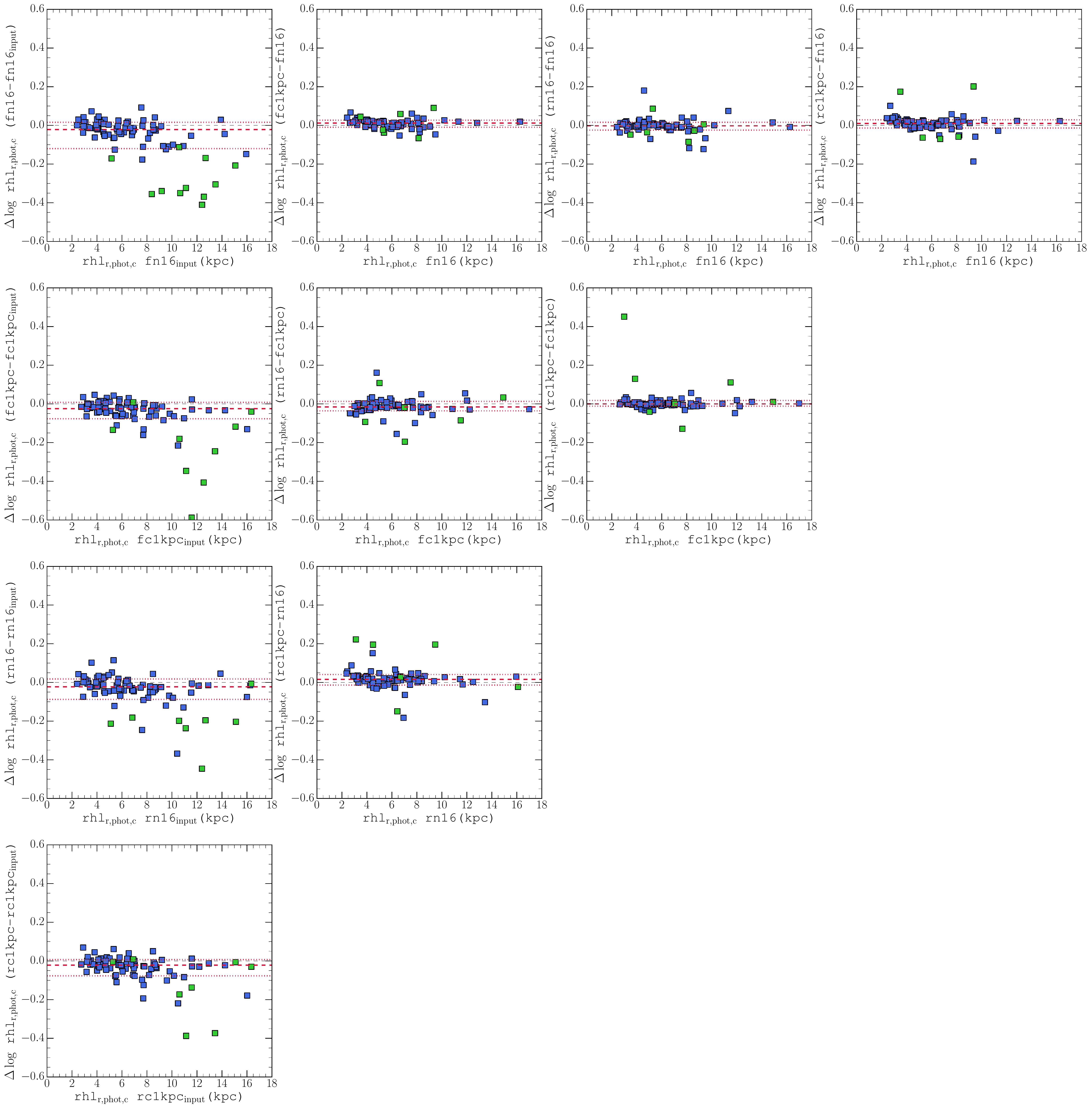}
    \caption[HLR bias from the choice of SLD scheme]{Variation and systematic biases in half-light radius measurements for alternative SLD schemes. The panels compare the $r$-band circular aperture half-light radius measurements, $rhl_{\mathrm{r,phot,c}}$, for the RIG sample taken from the \texttt{SMOOTHING} catalog. The first panel of each row shows the systematic bias on $r$-band half-light radius computed from the ratio between the half-light radius of the best-fitting B+D model and the synthetic image for the specified SLD scheme, $\Delta\log rhl_{\mathrm{r,phot,c}}$. All other panels compare the half-light radius estimates for the RIGs from one SLD scheme with another. The grey dashed line in each panel indicates $\Delta\log rhl_{\mathrm{r,phot,c}}=0$. The red dashed line and dotted lines show the median and 16$^{\mathrm{th}}$ to 84$^{\mathrm{th}}$ percentile range in each comparison, respectively. Green markers highlight the RIGs with large systematic offsets in $\Delta m_{\mathrm{r,phot}}$ in the fiducial scheme identified in Figure \ref{fig:mag_smoothing}.}
    \label{fig:rhl_smoothing}
\end{figure*}

\begin{figure*}
	\includegraphics[width=\linewidth]{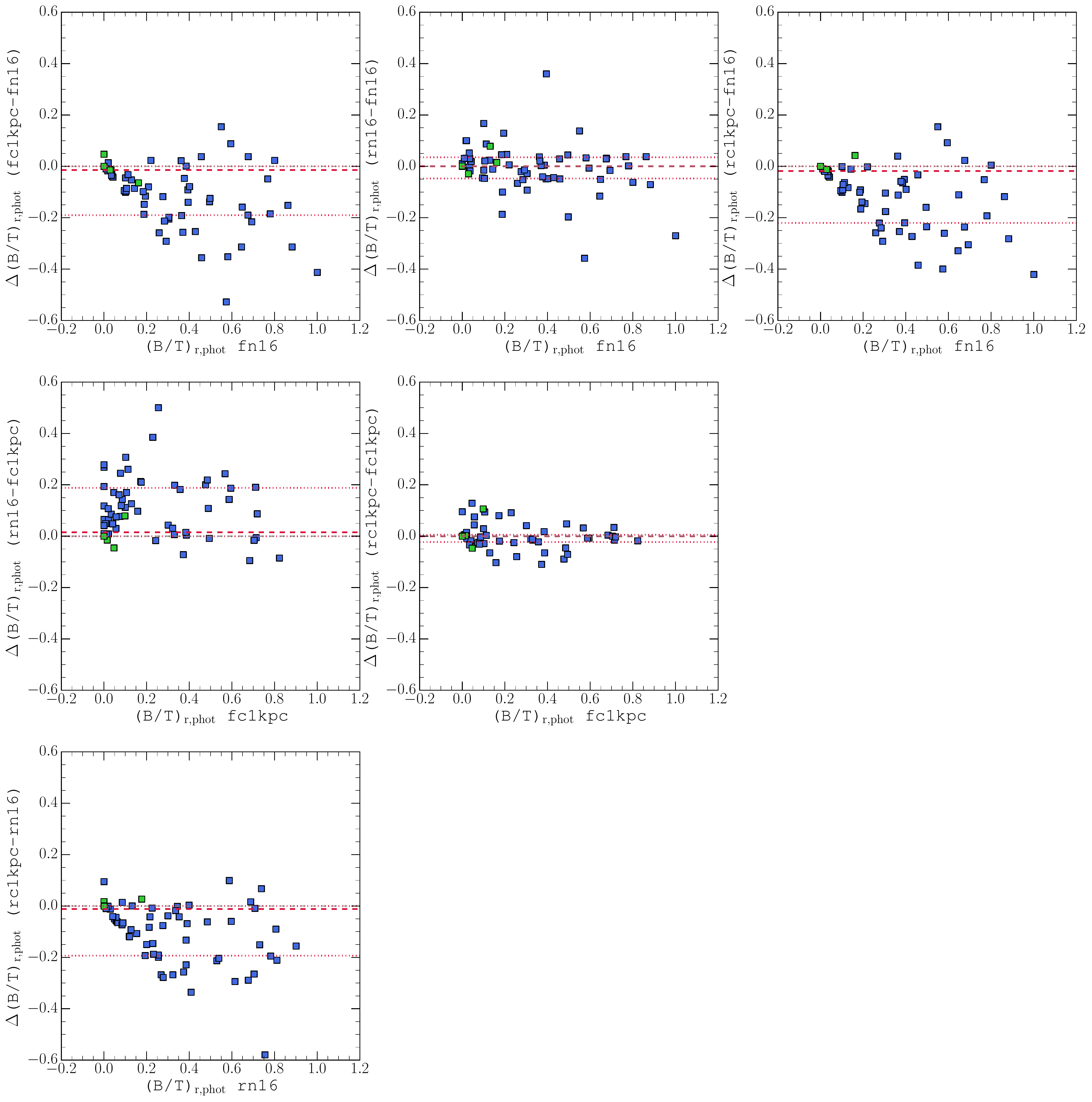}
    \caption[(B/T) bias from the choice of SLD scheme]{Variation in the photometric bulge-to-total light ratio for alternative SLD schemes. The panels compare the $r$-band (B/T) measurements for the RIG sample taken from the \texttt{SMOOTHING} catalog. Each panel compares the (B/T) measurements for the RIGs from one SLD scheme with another. The grey dashed line in each panel indicates $\Delta$(B/T)=0. The red dashed line and dotted lines show the median and 16$^{\mathrm{th}}$ to 84$^{\mathrm{th}}$ percentile range in each comparison, respectively. Green markers highlight the RIGs with large systematic offsets in $\Delta m_{\mathrm{r,phot}}$ in the fiducial scheme identified in Figure \ref{fig:mag_smoothing} -- many of which overlap at ((B/T), $\Delta$(B/T))=(0, 0).}
    \label{fig:btr_smoothing}
\end{figure*}


We produce synthetic images for each of the 100 galaxies in the RIG sample for a single projection using 4 unique stellar light distribution (SLD) schemes to characterize the biases of our fiducial and alternative SLD schemes (see Section \ref{SMOOTHcat}). Our control of placement and projection for decompositions in the \texttt{SMOOTHING} catalog ensures that the variations in parameter estimates for each galaxy is exclusively due to SLD schemes. 


The decompositions from the \texttt{SMOOTHING} catalog are used to quantify the biases from the SLD schemes on integrated magnitude, circular aperture half-light radius, and photometric $(B/T)$. Integrated magnitude and half-light radius can be computed from the synthetic images as well as the best-fitting (B+D) models for each galaxy and SLD scheme. The systematic biases on integrated magnitude and half-light radius are determined by comparing properties of the best-fitting models $m_{b+d}$ and $rhl_{b+d,c}$ with the corresponding properties of the synthetic images $m_{\mathrm{synth}}$ and $rhl_{\mathrm{synth,c}}$ -- which exclude realism. The integrated magnitude and circular half-light radii of the galaxies without realism can be computed from the total flux and aperture photometry of the synthetic images without any assumption about the form of their surface brightness profiles. The properties derived from galaxies using each SLD scheme are also compared among themselves as shown in Figures \ref{fig:mag_smoothing} and \ref{fig:rhl_smoothing}. 


Figure \ref{fig:mag_smoothing} compares measurements of integrated magnitude in the B+D decompositions from the \texttt{SMOOTHING} catalog. Each panel compares the decomposition results for RIGs in two SLD schemes or the decomposition results from a specific SLD scheme with the magnitudes derived from the synthetic images. The systematic biases in model magnitudes for each SLD scheme with respect to their corresponding synthetic images are shown in the first panel of each row. $\Delta m_{\mathrm{r,phot}}$ is the magnitude difference for the RIGs in each comparison. The majority of our integrated magnitude estimates for the representative sample of galaxies are consistent with those computed directly from the synthetic images in all SLD schemes. Meanwhile, a handful of galaxies have systematically larger integrated magnitude estimates than those computed from their synthetic images in the fiducial scheme. No SLD scheme that we employ significantly reduces the number of galaxies with large systematic offsets identified in the fiducial scheme. Visual inspection of the images and masks in each SLD scheme shows us that the outliers are strongly internally segmented due to the prevalence of substructure in their surface brightness distributions. However, each scheme provides unique variations in the segmentation map. Therefore, each best-fitting model to the surface brightness profile of the galaxy is determined from a unique set of science pixels and the best-fitting parameter estimates vary correspondingly. Such changes are reflected in the varying degree by which the outliers are offset in each SLD scheme.


The first row of Figure \ref{fig:mag_smoothing} shows a comparison of the integrated magnitude estimates for the input images and 3 SLD schemes against the fiducial SLD magnitude estimate. The majority of the magnitude estimates are consistent and are correspondingly concentrated around $\Delta m_{\mathrm{r,phot}}=0$ -- showing little sensitivity to SLD scheme. The outliers in each comparison appear randomly distributed about zero and are the same galaxies that demonstrated large positive systematic offsets in all SLD schemes. The random distribution of the outliers indicates that no SLD scheme alleviates the internal segmentation. Instead, the choice of SLD scheme simply provides unique changes to the segmentation maps. 


The biases on photometric sizes of galaxies should be sensitive to the accuracy with which the flux is recovered. Figure \ref{fig:rhl_smoothing} shows the systematic offsets and comparisons between SLD schemes for estimates of circular aperture half-light radius. The outliers in the panels showing the systematic offsets for each SLD scheme are the same galaxies that were systematic outliers in integrated magnitude. Again, no SLD scheme significantly reduces the population of outliers. The negative correlation between the offsets and half-light radii computed from the synthetic images hints that size estimates tend to be less robust for galaxies that are extended. However, several galaxies with large sizes computed from the synthetic images have consistent sizes derived from the models. The extended galaxies with consistent half-light radius measurements between the models and synthetic images tended to be more massive and had less internal segmentation by substructure in their science masks.

As in Figure \ref{fig:mag_smoothing}, the panels apart from the first column of Figure \ref{fig:rhl_smoothing} compare half-light radius measurements from each scheme with each other. All panels show a large population of RIGs with consistent half-light radius estimates between SLD schemes -- particularly in comparisons that both use adaptive or constant smoothing kernels. Interestingly, a systematic offset is present at small half-light radii in comparisons between the adaptive and constant smoothing SLD schemes. The half-light radii computed from the best-fitting models of galaxies with small intrinsic sizes are systematically larger in constant SLD schemes when compared to adaptive schemes. However, the choice of SLD scheme does not strongly affect estimates of half-light radius and full galaxy size estimates do not demonstrate a strong systematic offset based on the SLD method.


Finally, we investigate the variation in photometric (B/T) in comparisons between alternative SLD schemes in Figure \ref{fig:btr_smoothing}. We omit comparisons between the derived and intrinsic (B/T) values owing to ambiguities when defining the intrinsic (B/T). (B/T) estimates show greater sensitivity to the choice of SLD scheme than integrated magnitude and half-light radius. The comparison between (B/T) measurements that are both derived from decompositions of galaxies produced using adaptive schemes (\texttt{rn16 - fn16}) shows that the estimates of (B/T) for the majority of galaxies are consistent with a scatter about $\Delta$(B/T)$_{\mathrm{r}}=0$ of order $\sigma_{\mathrm{(B/T),r}}\approx0.05$. However, several galaxies have discrepant (B/T) between adaptive SLD schemes, and these are different from the delinquent cases in the size and magnitude comparison. A similar outlier population exists in the comparison of the two constant smoothing SLD schemes but with reduced scatter. Interestingly, there is a significant reduction of the (B/T) values derived from the constant schemes compared to adaptive schemes (\texttt{fc1kpc-fn16}, \texttt{rn16-fc1kpc}, \texttt{rc1kpc-fn16}, and \texttt{rc1kpc-rn16}). Many galaxies with large (B/T) using adaptive schemes become virtually bulgeless in constant schemes. One explanation for the reduction in (B/T) in constant schemes is that the light from stellar particles near the centre of the bulge (on which the identification of the bulge relies sensitively) is distributed too broadly relative to adaptive schemes that enable more spatially concentrated light profiles for tight clusters of stellar particles.

The systematic discrepancy in (B/T) estimates using alternative SLD schemes indicates that caution should be exercised in the choice of how stellar light is distributed to facilitate \emph{realistic} light distributions from discrete particles representing unresolved stellar populations. While the total fluxes and sizes are generally robust for all smoothing types (apart from the handful of consistent outliers with strong internal segmentation), the choice of SLD scheme can clearly bias estimates of fundamental structural properties of galaxies. (B/T) estimates for SLD schemes with constant 1 kpc smoothing tend to be low compared with adaptive schemes.

\subsection{Camera Angle}\label{sec:projection}

In this section we address biases introduced in the derived best-fitting parameters for galaxies from variations in the galaxy viewing angle. We employ the \texttt{CAMERAS} catalog described in Section \ref{CAMScat} which contains a single decomposition of each camera angle projection of the RIGs. Synthetic images are constructed using the fiducial SLD scheme \texttt{fn16}. Placement of each image is restricted to the same uncrowded location in the SDSS described in the previous section to ensure that the resulting parameter variation is in response to projection alone.

\begin{figure*}
  \includegraphics[width=0.325\linewidth]{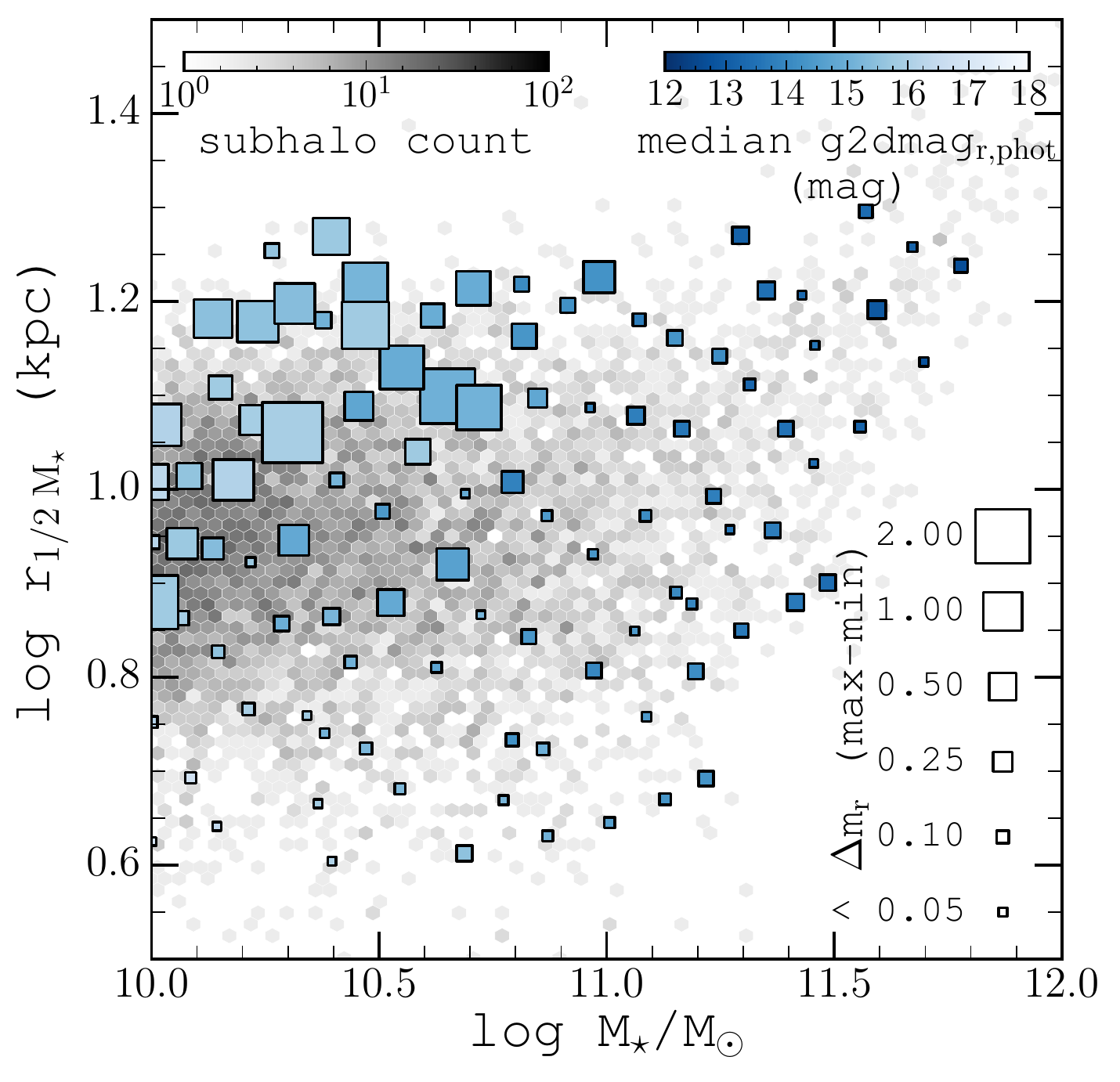}
  \includegraphics[width=0.325\linewidth]{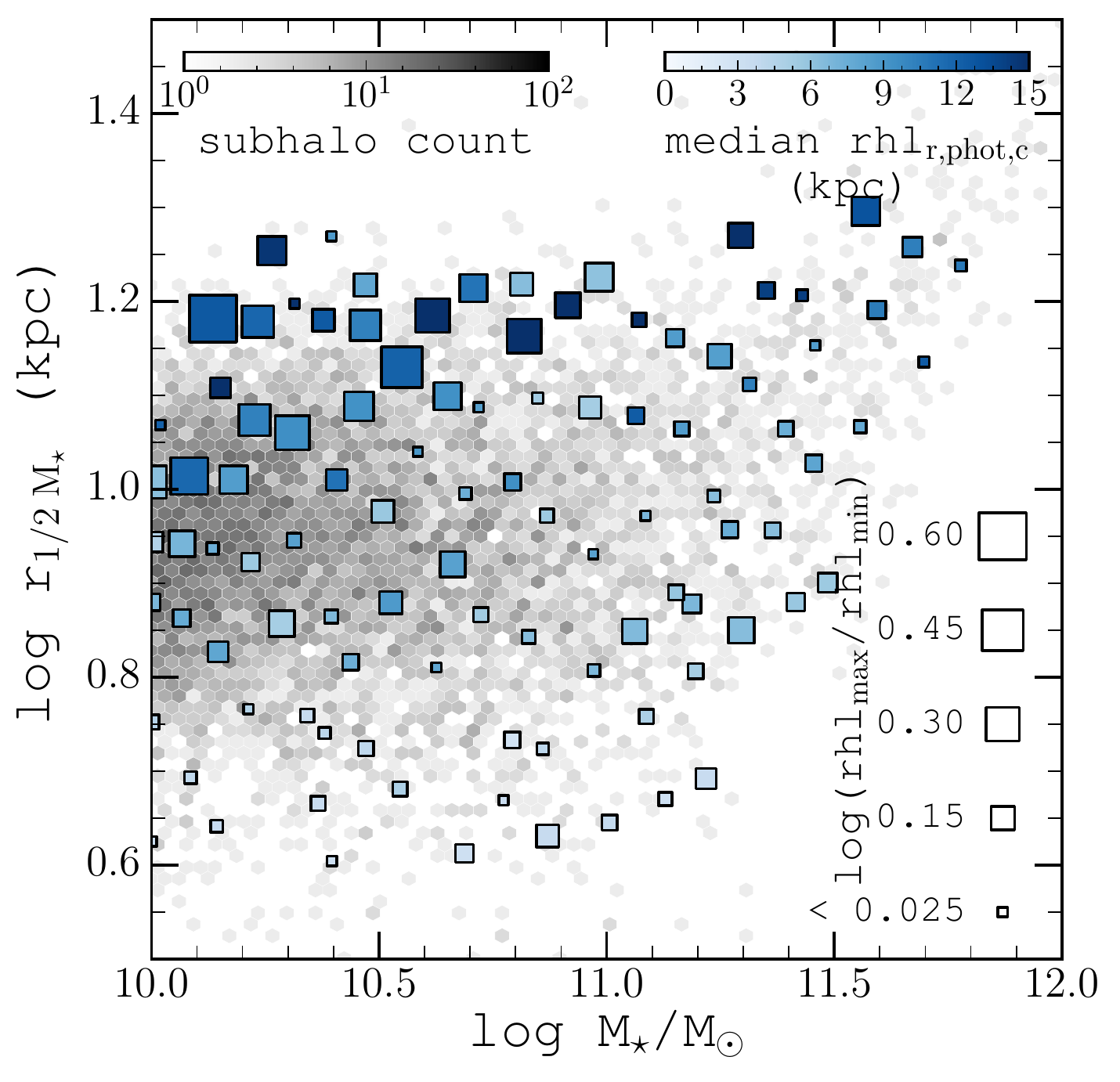}
  \includegraphics[width=0.325\linewidth]{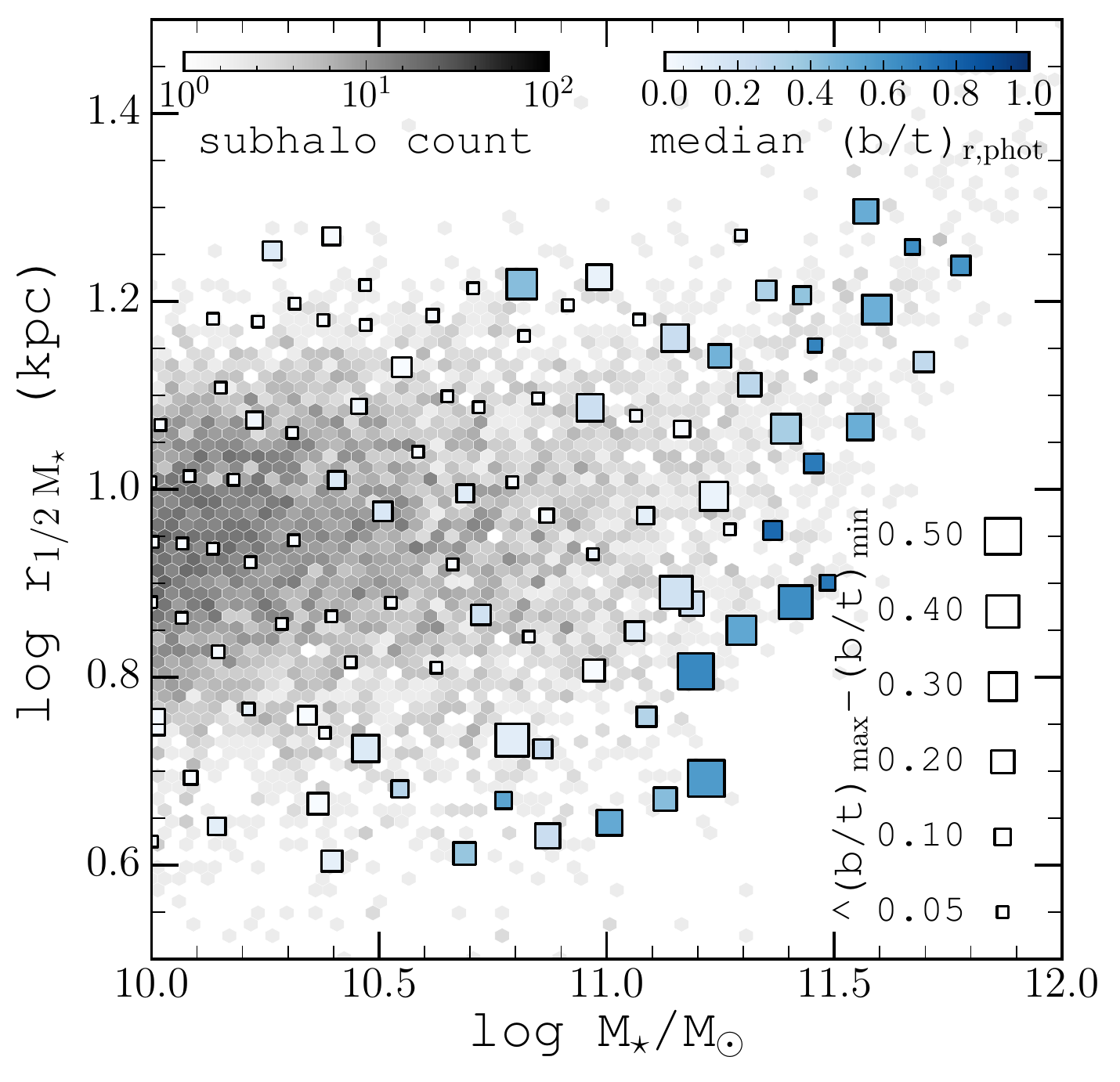}
\caption[Statistical bias from projection]{Characterization of random error from camera angle on $r$-band photometric magnitude (\emph{left panel}), half-light radius (\emph{middle panel}), and photometric bulge-to-total ratio (\emph{right panel}). The \emph{2D greyscale histogram} in each panel shows the distribution of the full \citetalias{2015MNRAS.447.2753T} catalog of Illustris subhalos in the plane of total stellar mass, $\mathrm{M}_{\star}$, and stellar half-mass radius, $r_{1/2 \mathrm{M}_{\star}}$ with colours on a logarithmic scale and colourbar located at the top left of each panel. Coloured boxes are positioned in this size-mass plane according to the corresponding subhalo properties of each RIG. The colour of each box indicates the median of the best-fitting parameters for the 4 projections of each galaxy in a manually selected, uncrowded field. The corresponding colourbar is located at the top right of each panel. The size of each box is calculated from the full range in the recovered parameters. A key of random errors and sizes, along with the exact computation method, is located along bottom right of each panel. 
}
\label{fig:camera_random_errors}
\end{figure*}

Figure \ref{fig:camera_random_errors} shows the sensitivity of bulge+disc model parameters to projection for the RIGs. The greyscale shows the distribution of the full \citetalias{2015MNRAS.447.2753T} catalog of Illustris subhalos in the size-mass plane of stellar half-mass radius and total stellar mass. The locations of the coloured boxes indicate the sizes and masses of the RIGs. The size of each box indicates the variation of the model parameters from projection for each RIG. We quantify the variation using the maximum and minimum parameter estimates from the four decompositions of each RIG (one for each camera angle) due to our small number of estimates. \footnote{The differences between maximum and minimum parameter estimates will be more akin to the $2\sigma$ than $1\sigma$ statistical estimates.} The max-min variations are denoted by $\Delta_{\mathrm{max-min}}$. The exact computation for the range in each parameter is given alongside the key at the bottom right of each panel. 

The left panel of Figure \ref{fig:camera_random_errors} shows the variation in B+D photometric magnitude from projection with respect to position on the size-mass plane. The magnitudes from the B+D decompositions of each projection are consistent for the majority of the RIGs ($\Delta m_{r,\mathrm{max-min}}\lesssim0.1$ mag). However, a handful of galaxies with relatively large sizes and low total stellar mass ($\log(rhm_{\star}$/kpc$) \gtrsim 0.8$ and $\log  \mathrm{M}_{\star}/ \mathrm{M}_{\odot} \lesssim 10.8$) have large variations in magnitude. Galaxies with large variation, $\Delta m_{r,\mathrm{max-min}}\gtrsim0.5$, are identified as the same galaxies that showed significant systematically positive magnitude offsets for all SLD schemes in the previous section. The outliers in the SLD comparison were found to be caused by segmentation issues, so the same factor may also be the cause of the camera angle variations.


We confirmed that internal segmentation is the source of the large variations in magnitude between camera angles by visual inspection of the images and masks for our galaxy sample. Nearly all of the galaxies with large variations in apparent magnitude are extended discs (and some irregulars) -- for which alternative camera angles allow the largest variation in projected surface brightness distribution. For example, it is less likely for substructure in edge-on discs to be deblended from the rest of the galaxy because the general condition for deblending is that the source must be locally discrete in surface brightness. Face-on discs are more prone to segmentation because substructure is more likely to dominate the local surface brightness distribution where that local flux is integrated along the thinnest axis of the disc. The deblending is strongly affected by changes in how these features are distributed with respect to the orientation of the galaxy. The largest variations occurred when a galaxy is strongly segmented in particular projections and not others -- such as when a galaxy is face-on in one projection and edge-on in another, respectively. Though we only show random variation in Figure \ref{fig:camera_random_errors}, we inspected the systematic offsets of our integrated magnitude estimates relative to the synthetic images. All magnitude estimates computed from the models are positively offset in all projections of diffuse galaxies with large amounts of substructure. The middle panel of Figure \ref{fig:camera_random_errors} shows variations in half-light radius for the RIGs in the \texttt{CAMERAS} catalog. As expected, the variations in half-light radius roughly mirror the variations in the flux for each galaxy. 

The right panel of Figure \ref{fig:camera_random_errors} shows the random errors on bulge-to-total light ratio from camera angle. Since many galaxies have estimated (B/T) values of $\sim$0, quoted errors are measured using linear differences. There is an apparent relationship between the variation in $(B/T)$ and the median $(B/T)$ of the four projections, as colour coded on the scale bar. Galaxies with intermediate median $(B/T)$ estimates show the largest variations in (B/T) estimates in each projection while high and small $(B/T)$ estimates have smaller variations. The implication is that $(B/T)$ estimates can vary significantly with projection when significant bulge and disc components both exist in a galaxy, but vary weakly when one component dominates.


\subsection{Environment and Crowding}\label{sec:crowding}


In this section, we remove our placement constraint in order to investigate the random and systematic variation in parameter estimates that are associated with location. For this experiment, we employ the decompositions from the \texttt{ASKA} catalog. The \texttt{ASKA} catalog uses synthetic images of galaxies from the \texttt{RIG} sample that are generated with the fiducial SLD scheme. Decompositions of each RIG were performed in approximately 100 unique locations in the SDSS and for all four camera angles \footnote{We state that we wish to remove any biases from projection, so the use of all 4 projections naïvely seems at odds with this goal. However, it is always possible to extract and compare results only for a specified projection of a galaxy -- as we do. We also wish to perform a sanity check that any trends for the variation in parameters for galaxies holding each camera angle fixed, respectively, are qualitatively similar. Analysis in all four camera angles at each location facilitates such a check.}
at each location to build the \texttt{ASKA} catalog (see Section \ref{ASKAcat}). We analyze the $\sim100$ decompositions for each simulated galaxy and projection to explore the effects of external crowding. 

As stated in Section \ref{ASKAcat}, the galaxies in the RIG sample have been visually inspected to ensure that there are no projection effects from other objects within the subhalo for any camera angle and that there are no \emph{obvious} structural disturbances from merger activity or pair proximity. The absence of projection effects from other sources in the subhalo's FoF group ensures that the statistics for crowding are uniform and the sources of crowding are external for all galaxies in all camera angles. Galaxies that are prone to internal segmentation are not excluded, since they represent a \emph{bona fide} population of galaxies in Illustris. 

We compute the random variation and median systematic offsets using the distribution of best-fitting (B+D) model parameters for each camera angle of the RIGs. We employ the same metrics from previous sections: integrated magnitude, circular aperture half-light radius, and photometric $(B/T)$. The random variation in parameter estimates is computed using the $16^{\mathrm{th}}-84^{\mathrm{th}}$ percentile range in the distribution of estimates for each galaxy. We also compute the systematic errors by comparing the median in the distribution of parameter estimates for each galaxy with the respective values computed from the synthetic images. The median systematic offset in $r$-band integrated magnitude, $\Delta m_{r,50\%}$, for example, is computed from from the difference between the median integrated magnitude in the distribution of estimates from the models to the integrated magnitude of the synthetic image. Median systematic offsets in half-light radius $\Delta \mathrm{rhl}_{r,50\%}$ are computed similarly -- taking the ratio of the median half-light radius from the distribution of best-fitting models parameters and the half-light radius computed from photometry of the synthetic image.

\subsubsection{Random Variation}\label{sec:crowd_random}

\begin{figure*}
  \includegraphics[width=.325\linewidth]{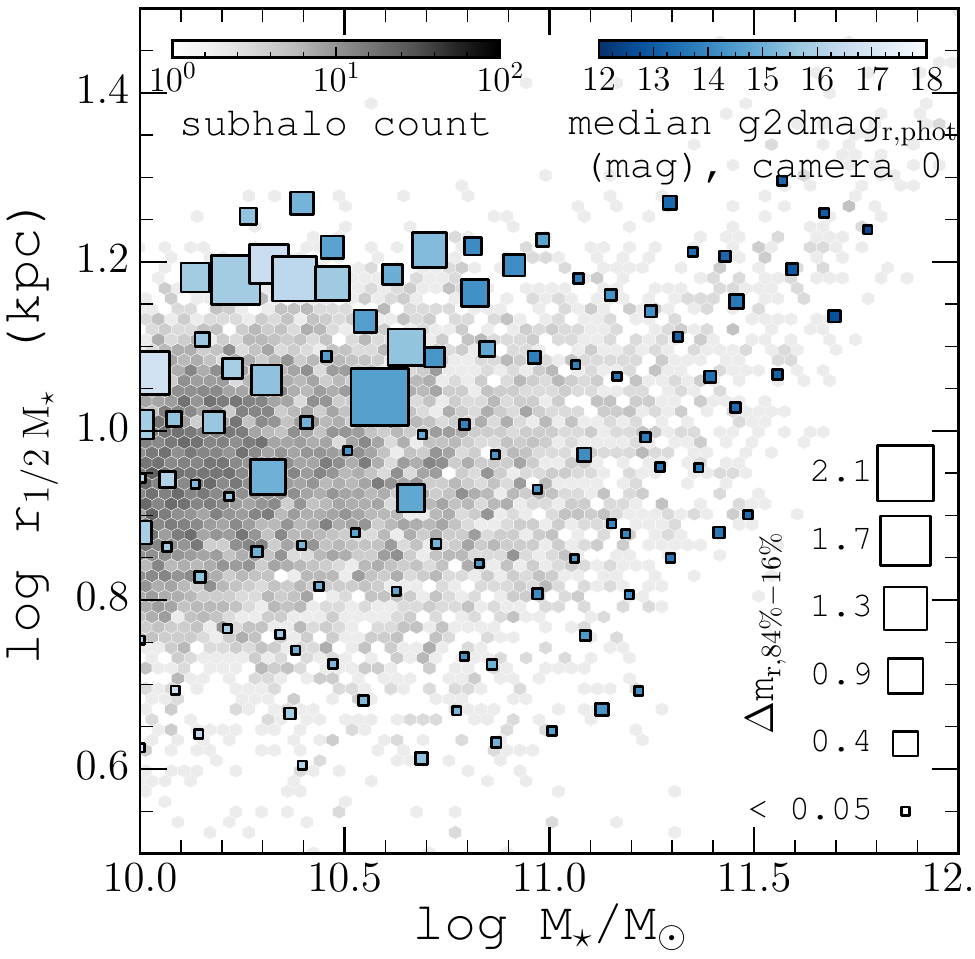}
  \includegraphics[width=.325\linewidth]{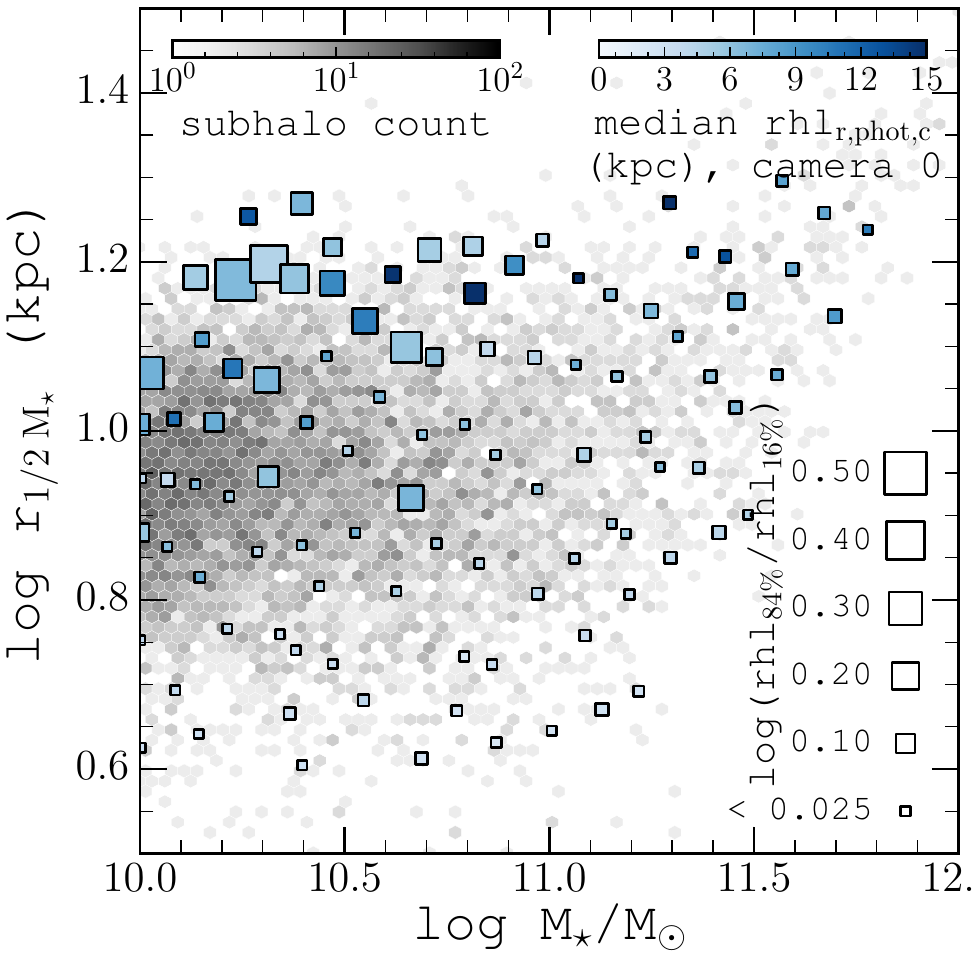}
  \includegraphics[width=.325\linewidth]{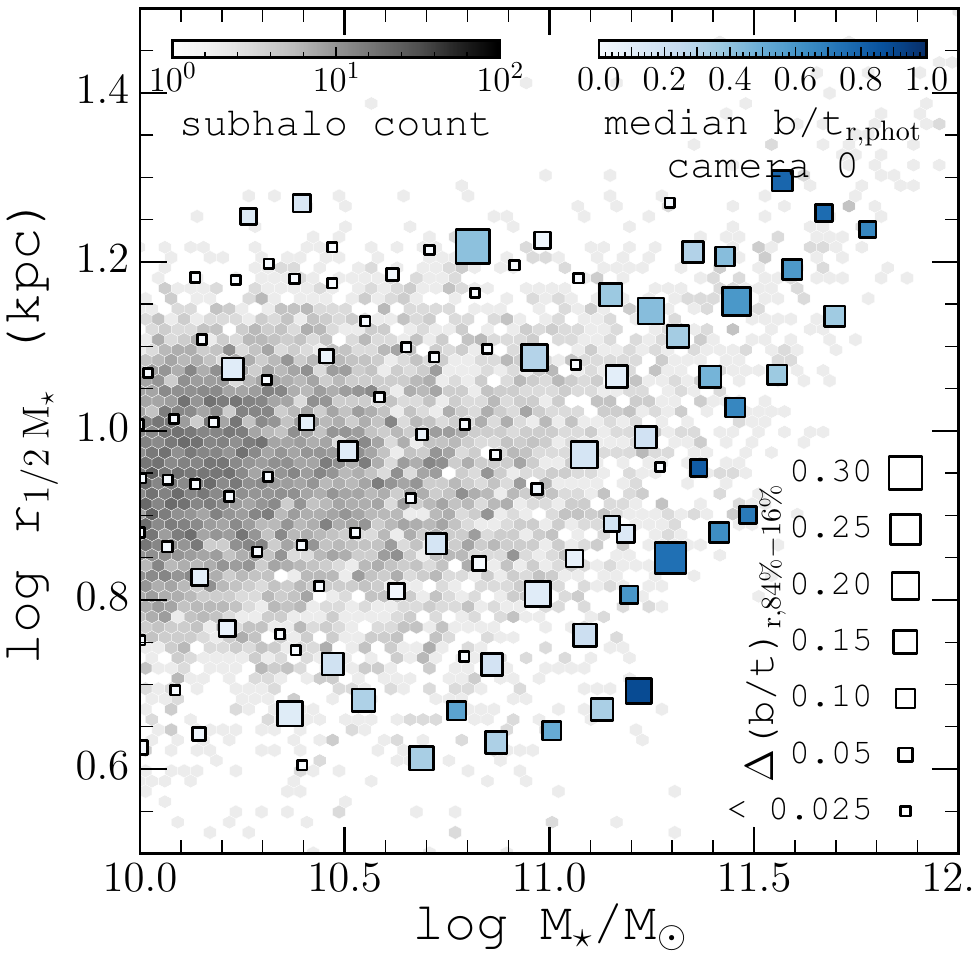}
\caption[Statistical bias from placement]{Characterization of random error from crowding on $r$-band photometric magnitude (\emph{left panel}), half-light radius (\emph{middle panel}), and photometric bulge-to-total ratio (\emph{right panel}) where each galaxy from the \texttt{ASKA} catalog is projected onto its position in the plane of stellar half-mass radius, $r_{1/2 \mathrm{M}_{\star}}$, and total stellar mass, $ \mathrm{M}_{\star}$. The \emph{2D greyscale histogram} in each panel shows the distribution of the full \citetalias{2015MNRAS.447.2753T} catalog of Illustris subhalos in on this parameter surface with colours on a logarithmic scale and corresponding colourbar on the top left in each panel. Coloured boxes are positioned in this plane according to the properties of each RIG subhalo. The colour of each square indicates the median in the distribution of estimators obtained from placing each galaxy in an average of 100 randomly selected SDSS locations with fixed camera angle (\texttt{CAMERA 0} in all cases here). The corresponding colourbar is located on the top right of each panel. The size of each square shows the random error that can be expected for a parameter estimate for an arbitrary SDSS location that is computed by comparing the $16^{\mathrm{th}}$ and $84^{\mathrm{th}}$ percentiles of the distribution of best-fitting estimates for each galaxy. A key of random errors and sizes, along with the exact computation method, is located in the bottom right of each panel.}
\label{fig:crowd_random_errors}
\end{figure*}

Figure \ref{fig:crowd_random_errors} shows the random variation in parameter estimates from crowding effects for the RIG sample. As in Figure \ref{fig:camera_random_errors}, the greyscale shows the distribution of the full \citetalias{2015MNRAS.447.2753T} catalog in the size-mass. The coloured boxes are positioned corresponding to the location of each RIG according to their intrinsic total stellar masses and half-mass radii. From left to right, the panels show the random error that one can expect on an estimate of $r$-band integrated magnitude, rest-frame half-light radius in kpc, and photometric bulge-to-total fraction for an arbitrary placement in the SDSS. The colour of each square denotes the median of the distribution of parameter estimates for the galaxy. The size of each square is set by the $16^{\mathrm{th}}-84^{\mathrm{th}}$ percentile range. The exact computation for the comparison is given to the left of the key for random error and square size, located at the lower right of each panel. We show only the results for \texttt{camera 0} decompositions of each galaxy in the Figure because the results are qualitatively similar for all camera angles. 

The left panel of Figure \ref{fig:crowd_random_errors} shows the variation of integrated magnitude estimates derived from the \gimtwod{} (B+D) models. We have verified that the $\pm1\sigma$ interval about the mean assuming normally distributed parameter estimates provides very similar results. As with the camera angle tests, many RIGs with total stellar mass $\log \mathrm{M}_{\star}/\mathrm{M}_{\odot}\lesssim11$ and half-mass radius $\log(rhm_{\star}/\mathrm{kpc})\gtrsim0.8$ have an unusually high sensitivity to placement inferred from the range in their parameter estimates. Meanwhile, the remainder of our sample is largely robust to biases associated with location in the SDSS. Typical random errors, excluding galaxies in the upper left corner, are $\Delta m_{r,84\%-16\%}\la0.05$ mags which corresponds to flux variations of approximately $\pm5\%$ in flux about the median. For the relatively low mass, highly extended galaxies at the upper left, random errors are as large as $\Delta m_{r,84\%-16\%}\approx2$ mags, which roughly correspond to flux variations by factors of 6. 

The variation in estimates of circular aperture half-light radius shown in the middle panel of Figure \ref{fig:crowd_random_errors} mirrors the variation in the integrated magnitudes, i.e galaxies that demonstrate weak sensitivity to placement also have small variation in their size estimates. Galaxies with large random variation on the magnitude estimates have correspondingly large variation in the size estimates -- which are again confined to a particular region of the size-mass space. Precise size estimates, as with flux, for a given galaxy has two requirements: accurate evaluation of the sky level and proper delineation of boundaries between multiple sources. The fact that every galaxy in our sample statistically experiences the same variations in crowding and sky level indicates that the large sensitivity to placement for extended but relatively low mass galaxies is driven by their morphologies. Note also that the median estimates for half-light radius in each galaxy, denoted by the colours in the boxes, do not demonstrate a clear correlation with half-mass radius among galaxies with large random variation. Further statement about the driving source of the large random errors requires inspection of the the images and \emph{systematic} errors on these estimates. But the fact that the highly sensitive galaxies we see here are the same group with high sensitivity to camera angle and large systematic offsets in all SLD schemes gives a strong case for internal segmentation as the source of the problem.

The characterization of the random error on $(B/T)_r$ is shown in the right panel of Figure \ref{fig:crowd_random_errors}. A similar relationship between the variation in (B/T) and the median (B/T) that was seen in the camera angle analysis is also manifested here. Galaxies with intermediate median values of (B/T) have the largest random errors about these medians. This is expected because the separation of the bulge and disc components is the most challenging in decompositions of galaxies with simultaneously significant bulge and disc components. Furthermore, the significance of the bulge component is sensitive to the accuracy with which the location and light profile of the peak in the bulge surface brightness distribution can be inferred -- which may be affected by placement specific biases such as crowding. Accurate modelling of the surface brightness profile at large radii where the S/N becomes small will also provide variation in both components' flux estimates. Galaxies with median $(B/T)_r>0.8$ and $(B/T)_r<0.2$ have relatively small random errors. We also note that the variation in estimates for the low-mass, extended galaxies in the upper left of the distribution does not mirror the variations in size and magnitude estimates shown in the other panels, and all have median (B/T) estimates that are close to zero. In general, it appears that the variation in (B/T) that arises from biases associated with placement are not correlated with variations in integrated magnitude and half-light radius estimates. 

\subsubsection{Systematic Offsets}\label{sec:crowd_system}

In this subsection we inspect the median systematic offsets for integrated magnitude and half-light radius relative to the synthetic images to examine accuracy of our measurements. 

\begin{figure*}
  \includegraphics[width=0.49\linewidth]{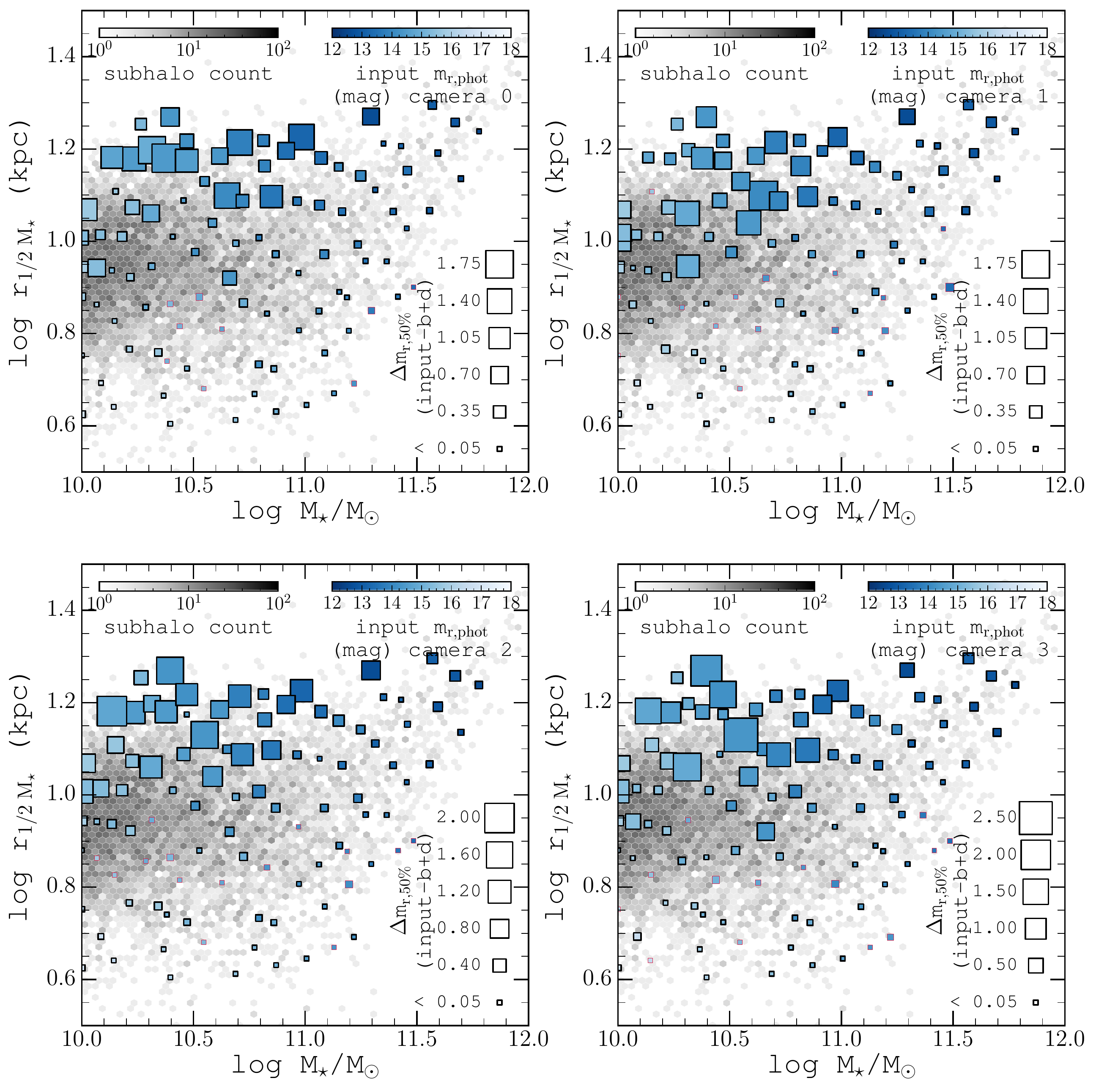}
  \includegraphics[width=0.49\linewidth]{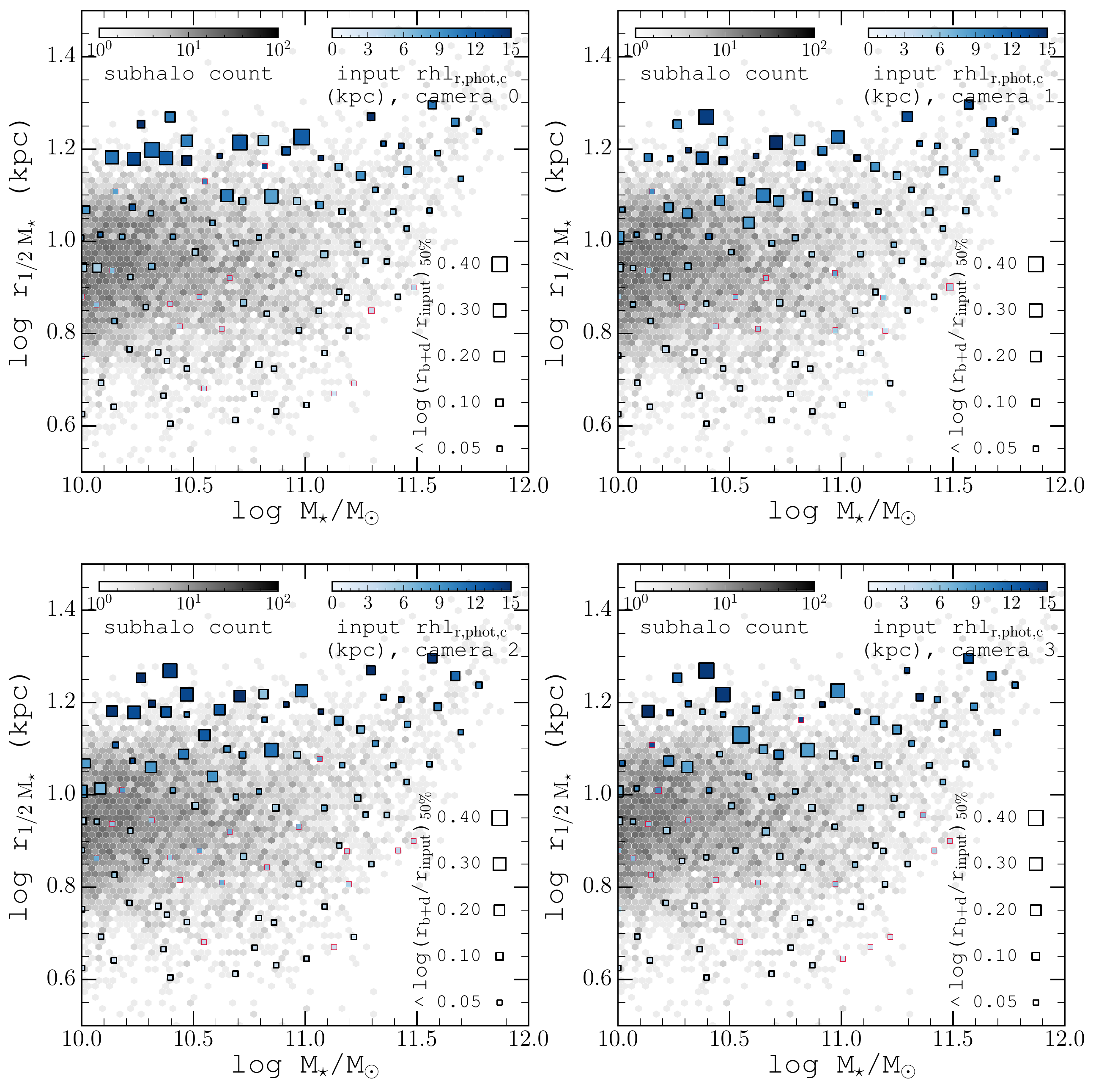}
\caption[Systematic biases on magnitude and HLR from placement]{Characterization of systematic errors on best-fitting estimates of magnitude (\emph{left panel}) and half-light radius in kpc (\emph{right panel}). Figure properties are the same as for the first two panels of Figure \ref{fig:crowd_random_errors} with the exceptions that the size of each coloured square denotes systematic error relative to the properties computed directly from the synthetic images and the border colour denotes whether the systematic indicated along the key is positive (\emph{red}) or negative (\emph{black}). Note that a negative systematic error in magnitude corresponds to a higher magnitude estimate from the model relative to the synthetic image and therefore a relative decrease in flux. Median systematic errors for the magnitudes in each galaxy are computed from difference between the integrated B+D analytic model magnitudes (output) and the magnitude of the galaxy as calculated from the sum of the flux in the synthetic images before any realism is added (input). Similarly, output and input half-light radii are computed from circular aperture photometry of the best-fitting models and the synthetic images, respectively. Note that positive systematic errors are all very small are randomly distributed amongst galaxies with small negative systematic errors -- indicating that crowding in these galaxies provides weak variations about 1:1 flux recovery for these galaxies, but does not systematically drive the best-fitting model estimates in a particular direction.
}
\label{fig:crowd_systematic_errors}
\end{figure*}

Figure \ref{fig:crowd_systematic_errors} shows the systematic errors on our integrated magnitude estimates and half-light radii from the bulge+disc fitting of the RIG sample. The properties of each figure are similar to those described for Figure \ref{fig:crowd_random_errors}, with the difference that we now look at median statistics of each galaxy relative to the the respective properties of the synthetic images before realism is added. $\Delta m_{r,50\%}$ denotes the difference between the integrated magnitude computed from the total flux in the synthetic images and the median in the distribution on estimates for integrated magnitude from the (B+D) decompositions (input-B+D).\footnote{The unconventional calculation of the difference using (input-B+D) rather than (B+D-input) is solely for consistency in the colours of the left and middle panels of the figure (i.e. boxes with black borders correspond to negative systematic offsets in \emph{flux} and size, respectively).} Similarly, $\Delta\mathrm{rhl}_{r,50\%}$ is the logarithm of the ratio between the median in the distribution of half-light radii over all placements and the half-light radius computed from the synthetic image. The colour of the borders on each square denote negative (\emph{red}) and positive (\emph{black}) systematics according to their computation, shown along the keys in each panel. The colour of each square shows the respective input integrated magnitudes and half-light radii from the synthetic images against which each median statistic is compared.

The left panel of Figure \ref{fig:crowd_systematic_errors} shows the median systematic offsets for the distributions of best-fitting B+D model magnitudes of the RIGs over all placements for a single camera angle. We show only the results for each galaxy in a single camera angle because we find that the general trends are the same when holding each other camera angle fixed. Systematic offsets in magnitude are small apart from the low mass, extended galaxies and a few of the higher mass, extended galaxies. Furthermore the systematics are randomly distributed about $\Delta m_{r,50\%}=0$. The random distribution of small systematic offsets indicates that the effects of crowding and other placement-sensitive biases do not systematically affect estimates of the flux. Crowding and other positional biases only provide scatter about $\Delta m_{r,50\%}=0$ for the majority of our sample of galaxies and morphologies. The second panel shows the systematic offsets of half-light radius estimates. The half-light radius offsets mirror of the systematics on integrated magnitude -- both for the majority of the sample and for the diffuse galaxies with large negative systematic offsets. 

In summary, our bulge+disc decomposition results for the majority of the RIGs show weak sensitivity to the the biases associated with placement: crowding, sky background, and PSF resolution. However, a recurring handful of diffuse galaxies have large systematically negative median offsets in both integrated magnitude and half-light radius -- consistent with the systematics seen in the smoothing analysis in Section \ref{sec:smooth}. The corresponding random variations for these galaxies are also large and show strong sensitivity to both placement and camera angle (Section \ref{sec:projection}). The discrepancies that we have highlighted for these diffuse galaxies are the following: systematic under-estimations of flux and size over all environments; significantly greater random sensitivity to placement, projection, and SLD scheme than other galaxies on the size-mass plane; and that errors of this magnitude do not exist in our analyses of real galaxies. We explore the source that drives the unusually high sensitivity to observational biases in Illustris' diffuse galaxies in the next section.


\subsection{Internal Segmentation by Artificially Discrete Substructure}\label{sec:segmentation}

The population of low-mass, diffuse galaxies in the RIG sample have consistently demonstrated high sensitivity to each observational bias. In this section, we investigate the reasons that observational biases on the surface brightness distributions of the diffuse RIGs generate large random and systematic errors in their best-fitting model estimates.

\begin{figure*}
	\includegraphics[width=\linewidth]{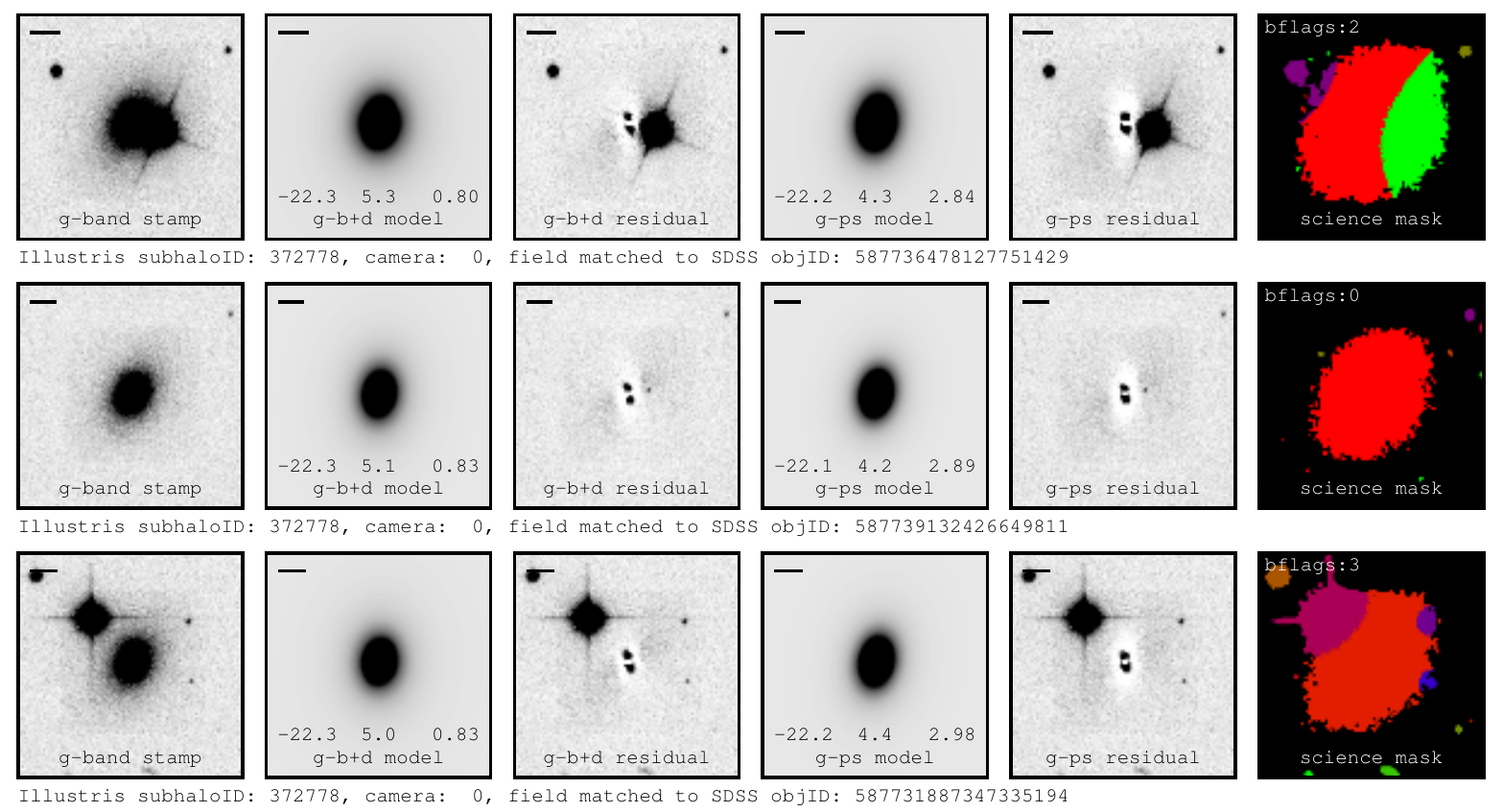}
    \caption[Examples showing low sensitivity of parameter estimates to crowding]{Examples of structural and photometric parameters varying weakly as a function of crowding. Each row of panels corresponds to the same camera angle of the same galaxy inserted in a unique SDSS field location. The middle row shows the galaxy in a relatively uncrowded environment. Note that this galaxy is not intrinsically prone to internal segmentation. The top and bottom rows show the galaxy where it happens to have been placed in environments where it is strongly crowded by bright field stars. The model parameters are nearly identical despite the significant variation in crowding effects and proximity of the additional source. A scale in the top left of each panel denotes 10 kpc at z=0.05. }
    \label{fig:crowdgal}
\end{figure*}

We begin by comparing the effects of placement on the diffuse RIGs with an example taken from the rest of the population. Figure \ref{fig:crowdgal} shows an example of the robustness of our structural and photometric estimates to crowding effects that is representative among the majority of the RIGs. Each row shows the decompositions for the same galaxy in a different location in the SDSS. We use the \emph{bflags} parameter to quantify the segmentation of a galaxy from internal and external sources. As outlined in Section \ref{deblend}, \sextractor{} allocates each pixel in an image to a source by assigning it a flag in the segmentation map. The pixels that have the same flag as the central pixel of the synthetic image (i.e. belong to the same source) are always used in the fitting. \emph{bflags} is computed as the number of uniquely flagged sources that are directly contiguous to the source pixels being used in the fitting. For example, in the middle row of Figure \ref{fig:crowdgal}, the galaxy is located in a relatively uncrowded field with no external sources directly bordering the pixels used in the fitting. The top and bottom rows show the same galaxy with the same camera angle in locations where bright stars contaminate the line of sight and significantly crowd the galaxy (see that \emph{bflags}$ = 0$). The delineation of boundaries between the galaxy and the external sources along with the corresponding \emph{bflags} number are shown in the science masks. The best-fitting B+D parameter estimates listed in the second panel of each row show maximum variations on the order of a few percent in magnitude, half-light radius, and bulge-to-total fraction, respectively. The corresponding best-fitting $pS$ results show remarkable consistency in integrated, half-light radius, and Sersic index. The quality of the residual images in each case aside, the residuals for the galaxy light profile are consistent in each case. The example in Figure \ref{fig:crowdgal} illustrates why the decomposition results for the majority of the RIGs show weak sensitivity to crowding -- despite possibly significant variations in crowding by external sources.

\begin{figure*}
	\includegraphics[width=\linewidth]{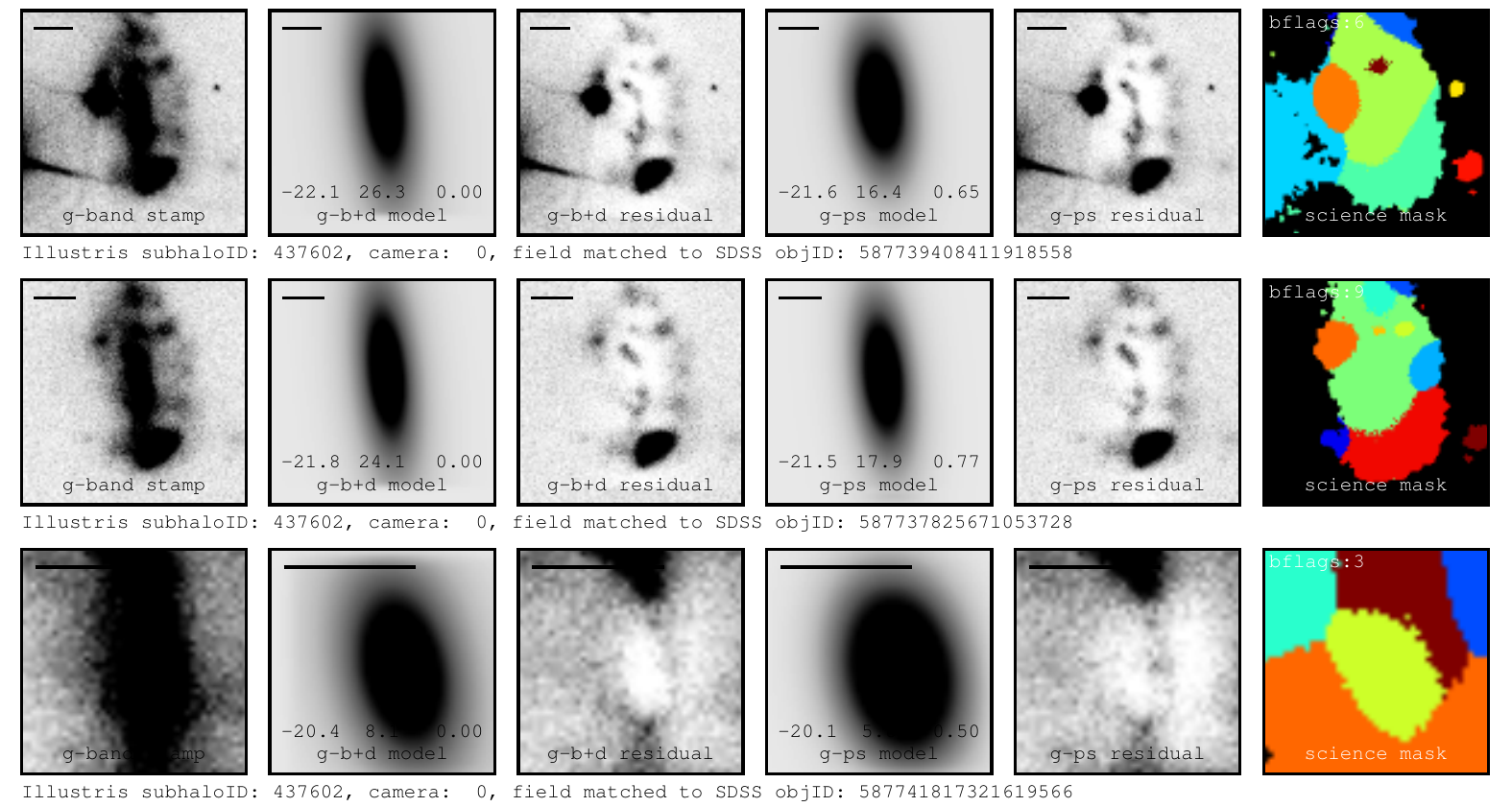}
    \caption[Examples showing high sensitivity of parameter estimates to internal segmentation]{Examples of structural and photometric parameters varying strongly as a function of internal segmentation. Each row of panels corresponds to the same camera angle of the same galaxy inserted in a unique SDSS field location. Variations in the crowding by external sources and sky variance provide unique variations in the segmentation map for each placement. When there are intrinsically large amounts of internal segmentation due to high surface-brightness substructure within the galaxy, further changes to the segmentation map through crowding can provide significant random error on the best-fitting estimates. Changes to the colourmap are made in the science masks of this mosaic to aid in visual distinction between zones in highly segmented systems. A scale in the top left of each panel denotes 10 kpc at z=0.05.}   
    \label{fig:segmentgal}
\end{figure*}

However, the best-fitting estimates for the diffuse galaxies in the RIG sample show high sensitivity to the observation biases from placement. Figure \ref{fig:segmentgal} shows an example of the variations that arise for the diffuse RIGs. As for Figure \ref{fig:crowdgal}, each row of Figure \ref{fig:segmentgal} shows the same galaxy in a different location in the SDSS. In every case, the galaxy in Figure \ref{fig:segmentgal} is strongly internally segmented, even in uncrowded fields; as shown in the image incarnation in the middle row along with the corresponding \emph{bflags} number in the science mask. The top and bottom rows of Figure \ref{fig:segmentgal} show that the observational biases associated with placement (crowding, sky background, and PSF resolution) drive significant variations in the segmentation maps of diffuse galaxies such that decomposition results are not consistent. In the top row of Figure \ref{fig:segmentgal}, the presence of external sources in the form of two stars leads to a significantly higher estimate of the integrated absolute magnitude, $\Delta M_{r,(b+d)}=0.3$ mag, relative to the uncrowded placement in the middle row. In the bottom row of Figure \ref{fig:segmentgal}, the science mask is sufficiently segmented that a locally bright feature in the surface brightness distribution of the galaxy is identified as a distinct source. The collection of pixels that share the flag of the locally distinct ``source'' includes the central pixel in the image (which is aligned with the gravitational potential minimum). Only pixels that have this flag (colour-coded yellow-green in the segmentation image) are used in the fitting. The segmentation leads to a factor of 5 reduction in the flux and a factor of 3 in the half-light radius compared to estimates where the galaxy is not so brutally shredded by segmentation (middle row of \ref{fig:segmentgal}). In addition, each segmentation image (even in the uncrowded field) excludes a significant component of the galaxy's flux that is bound in locally bright substructure. We inspected the fields into which the galaxies from the middle and bottom rows of Figure \ref{fig:segmentgal} were inserted to find no significant sources of crowding. Therefore, crowding, sky background variations, PSF resolution, or combinations thereof can all provide sufficient modification to the surface brightness distribution to cause additional \emph{internal} segmentation of galaxies but only when a galaxy has locally discrete substructures in its surface brightness distribution that make it prone to internal segmentation.

\begin{figure*}
	\includegraphics[width=\linewidth]{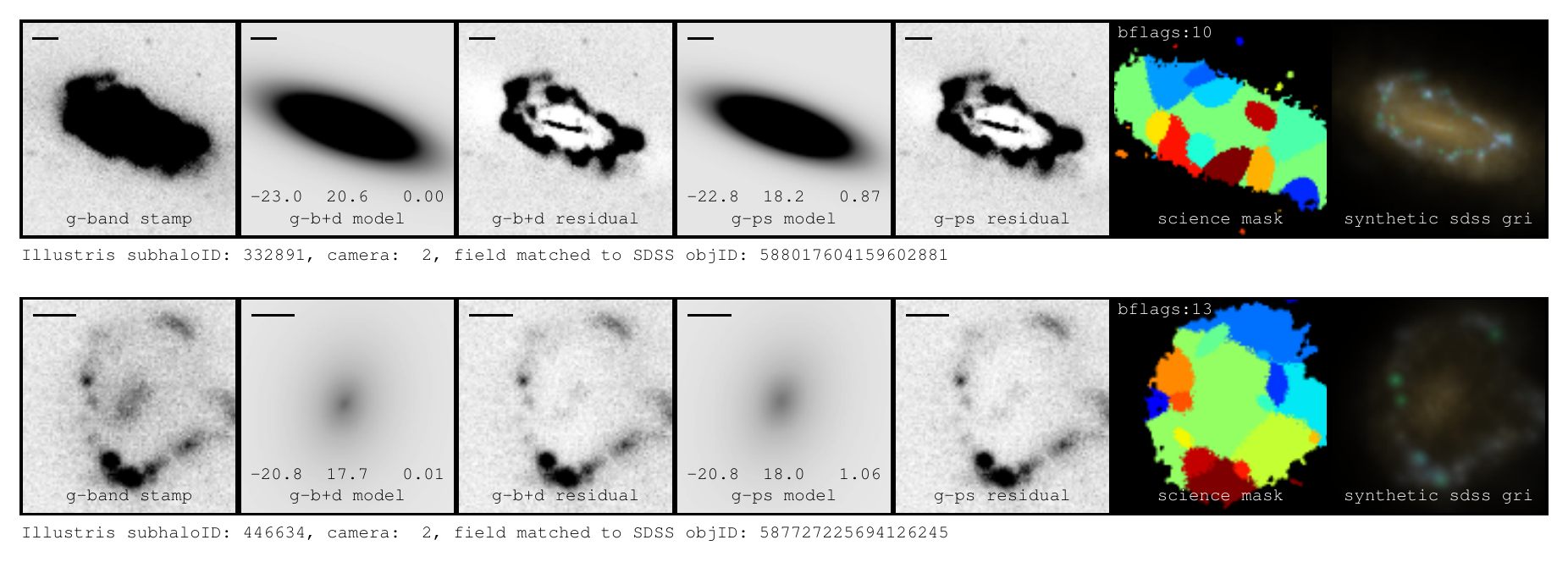}
    \caption[Examples of internal segmentation in diffuse galaxies]{Examples of highly extended galaxies where luminous and localized substructure creates a highly segmented science mask for the galaxy, despite no significant presence of external crowding. The last panel in each row shows the synthetic SDSS \emph{gri} image of each galaxy. These two galaxies (subhaloIDs 332891 in the \emph{upper row} and 446634 in the \emph{lower row}) have half mass radii of 18.6 kpc and 15.8 kpc and total stellar masses, $\log  \mathrm{M}_{\star}/ \mathrm{M}_{\odot}$, of 11.29 and 10.31, respectively. Their best-fitting model r-band half-light radii are 13.3 kpc and 15.0 kpc. The galaxy in the lower row is significantly more diffuse, with a total stellar mass that is an order of magnitude lower than the galaxy in the upper row but similar size when size is quantified by both mass and photometry. Bright green in the science mask denotes the pixels allocated to fitting. A scale in the top left of each panel denotes 10 kpc at z=0.05.}
    \label{fig:shreddedgals}
\end{figure*}

The most extreme scenarios where we see differences as large as $\Delta m_{r,50\%}\approx2$ mag in Figure \ref{fig:crowd_systematic_errors} (a factor of 6.3 decrease in total flux) arise from two possible situations: (1) a large fraction of the galaxy's light is bound in substructure components that are not included in the fitting; (2) discrete substructure is projected relative to the galaxy such that it is identified as the source on which to perform the fitting (which occurs when the substructure is discrete in a particular projection and overlaps with the image centre). In either scenario, the galaxy is not identified as a single source, but is a contiguous collection of deblended features that each have a unique flag in the mask. Figure \ref{fig:shreddedgals} shows two examples of galaxies whose science masks are shredded by segmentation from discrete substructure that contains appreciable fractions of the galaxies' total fluxes. The galaxies in the top and bottom rows of Figure \ref{fig:shreddedgals} have stellar masses $\log  \mathrm{M}_{\star}/ \mathrm{M}_{\odot}=11.3$ and $\log  \mathrm{M}_{\star}/ \mathrm{M}_{\odot}=10.3$, respectively. The supplementary synthetic colour images in right-most panel of each row show blue rings of containing knots of substructure in the form of young and highly luminous populations of stars that seemingly orbit at a fixed radius from the galactic centre. While it is \emph{visually} apparent that both B+D and \emph{pS} models reproduce the fraction of the surface brightness distributions of the galaxies that excludes the rings, the residuals and masks demonstrate that a significant fraction of the total flux is lost because the bright substructure is masked out. For the galaxy in the first row of Figure \ref{fig:shreddedgals}, the median ratio of the fluxes determined by the B+D decompositions to those of the synthetic images is $f_r/f_{r,\mathrm{synth}}=0.35$ across all environments, meaning that roughly 2/3 of the total flux is in the substructure. The galaxy in the second row is 10 times less massive, but has roughly the same half-mass radius -- making it significantly more diffuse. The median flux fraction for this galaxy is $f_r/f_{r,\mathrm{synth}}=0.56$ -- so roughly half of the light is locked up in masked substructure. In each case, the substructure systematically represents a significant fraction of the total stellar light. While variation in sky and crowding will produce variations in what substructure is masked, the large negative systematic errors on flux and photometric size arise from the exclusion of bright substructure from the fitting.

\begin{figure}
	\includegraphics[width=\linewidth]{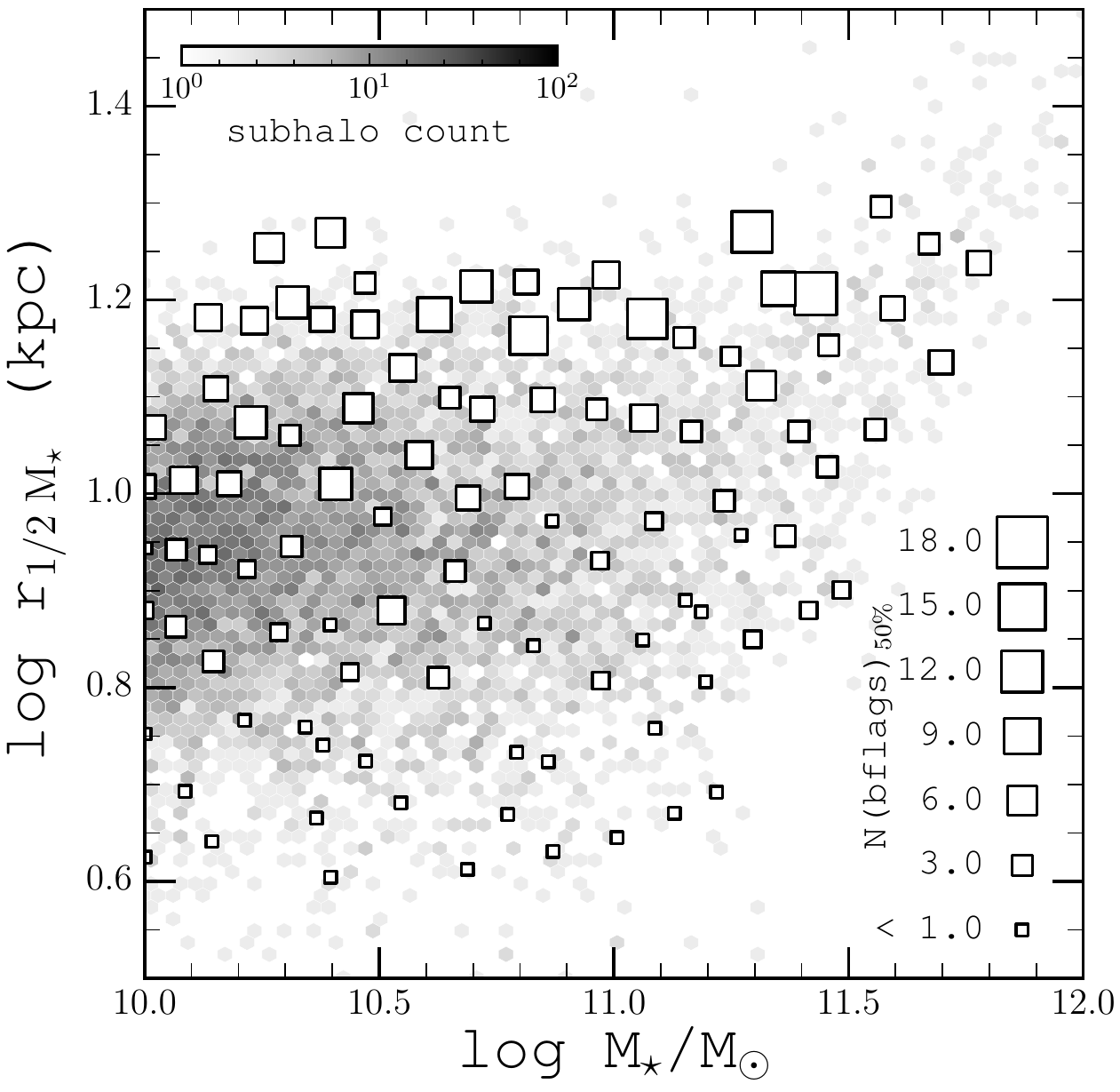}
    \caption[Median \emph{bflags} for each galaxy in the \texttt{ASKA} catalog]{The median number of \emph{bflags} across all placements and camera angles for each RIG. \emph{bflags} is computed as the number of flagged sources that directly border pixels that are flagged as science pixels (pixels that are used to determine the likelihood of the models). It is a parameter that is sensitive to external crowding and internal segmentation. The locations of each square denotes the position of a RIG on the plane of total stellar mass and half-mass radius. The size of the square indicates the median \emph{bflags} parameter over all placements and camera angles in the \texttt{ASKA} catalog. Despite uniform crowding statistics between galaxies, the extended and diffuse population of Illustris have intrinsically large amounts of segmentation.}
    \label{fig:msk_bflags_errors}
\end{figure}

The final piece of evidence that internal segmentation is driving the large random and systematic biases among diffuse galaxies is shown in the median \emph{bflags} estimates over all placements and projections of each galaxy on the half-mass radius and total stellar mass plane, shown in Figure \ref{fig:msk_bflags_errors}. The \emph{bflags} parameter is not sensitive to whether the source of segmentation is internal from discrete substructure or external from crowding. However, our procedure for creating the \texttt{ASKA} catalog dictates that the placement of any particular galaxy is random. Therefore, the distribution of best-fitting model parameter estimates for each galaxy over all placements is affected by the same crowding \emph{statistics} as any other and \emph{bflags} should be roughly uniform for each galaxy. Although the \emph{bflags} value should not depend on galaxy properties, Figure \ref{fig:msk_bflags_errors} shows that the diffuse galaxies (i.e. large radii for their stellar mass) have significantly greater median \emph{bflags} estimates than the rest of the population. This indicates that the intrinsically large median \emph{bflags} estimates are driven by internal segmentation. The fact that the galaxies with large amounts of internal segmentation seen here directly coincide with ones having high sensitivity to observational biases confirms that internal segmentation by discrete substructure is the source of the sensitivity.

The properties of the diffuse RIGs explain their high sensitivity to observational biases. Figure \ref{fig:crowd_systematic_errors} showed that large median systematic offsets in integrated magnitude and half-light radius are seen for morphologically diffuse galaxies. The source of the systematic offsets is the deblending of locally discrete substructure that contain appreciable fractions of the flux. The sensitivity of the best-fitting model estimates to internal segmentation in diffuse galaxies is exacerbated by observational biases. The biases provide unique variations in the the segmentation that generate the large random variation and inconsistency between estimates in different environments, projections, and SLD schemes seen in Figures \ref{fig:crowd_random_errors}, \ref{fig:camera_random_errors}, and \ref{fig:mag_smoothing}, respectively. Galaxies for which (1) the majority of the galaxy's total stellar light is contained in masked substructure or (2) a locally discrete feature of the galaxy is systematically identified as the source on which to perform fitting have the largest negative systematic offsets in flux and half-light radius. Conversely, galaxies with compact stellar light distributions are less likely to have their substructure deblended and therefore show little to no variation with observational biases and have best-fitting model properties that are consistent with those of the synthetic images.

In principle, it is possible to optimize the deblending procedure for our simulated galaxies to remove internal segmentation. In practice, this would require a detailed exploration into how choices of \textsc{SExtractor} deblending parameters affect the recovered fluxes and sizes of the simulated galaxies with realism. For example, it is relatively simple to choose a set of deblending parameters that produce \emph{bflags}=0 for every galaxy in the same uncrowded field. However, determining whether a given set of deblending parameters will enable consistent decomposition results across all sky backgrounds and spatial resolutions is more challenging. For example, although changing the deblending parameters could reduce segmentation, it could simultaneously limit the ability of \textsc{SExtractor} to distinguish neighbouring objects which \emph{should} be deblended. Furthermore, our goal is not to obtain the best or most accurate models for the simulated galaxy images. Our aim is to examine the result of applying the identical, calibrated methodology that is used to quantify structure in real galaxies to simulated galaxies with extensive observational realism. Only by using the same deblending and decomposition tools is a consistent and fair comparison guaranteed. Therefore, in the main body of this paper, we do not investigate alternative or optimized deblending as a means of eliminating the biases that are identifiably intrinsic to the simulated galaxy images. Nonetheless, the curious reader is directed to Appendix \ref{sec:SExpars} where a comparison is drawn between the B+D and $pS$ model properties of the RIGs with an alternative set of \textsc{SExtractor} parameters that gives relatively little segmentation.


In the next section, we investigate the systematic offsets on estimates of apparent integrated magnitude and circular aperture half-light radius for the full Illustris galaxy population to characterize the influence of these biases in comparisons with real galaxies.

\subsection{Systematics of Size and Flux Estimates in the Full Illustris Galaxy Population} \label{size_flux}

In this section, we investigate the systematics on total flux and half-light radius estimates in the full Illustris population using the \texttt{DISTINCT} catalog. The catalog contains the best-fitting model estimates from a single B+D and $pS$ decomposition of each camera angle for every galaxy in the  Illustris synthetic image catalog of \citetalias{2015MNRAS.447.2753T}. Placement of each camera angle projection the galaxies is randomized following the procedure described in Section \ref{realism}. Any systematic offset between the decomposition results and the derived properties of the synthetic images for a particular galaxy or morphology generally arises from the inadequacy or inflexibility of our quantitative morphology analysis in handling the particular surface brightness distribution of the galaxy in its environment. Our analysis will allow us to characterize the collective systematic biases of that arise from morphology and observational realism in the full Illustris population including the prevalence of bias from internal segmentation. For simplicity, in this section we will refer to every \emph{insertion of a galaxy} simply as a galaxy because there are four decompositions of each Illustris galaxy in the catalog (one for each camera angle).

\subsubsection{Recovery of the Integrated Flux}

\begin{figure*}
	\center\includegraphics[width=\linewidth]{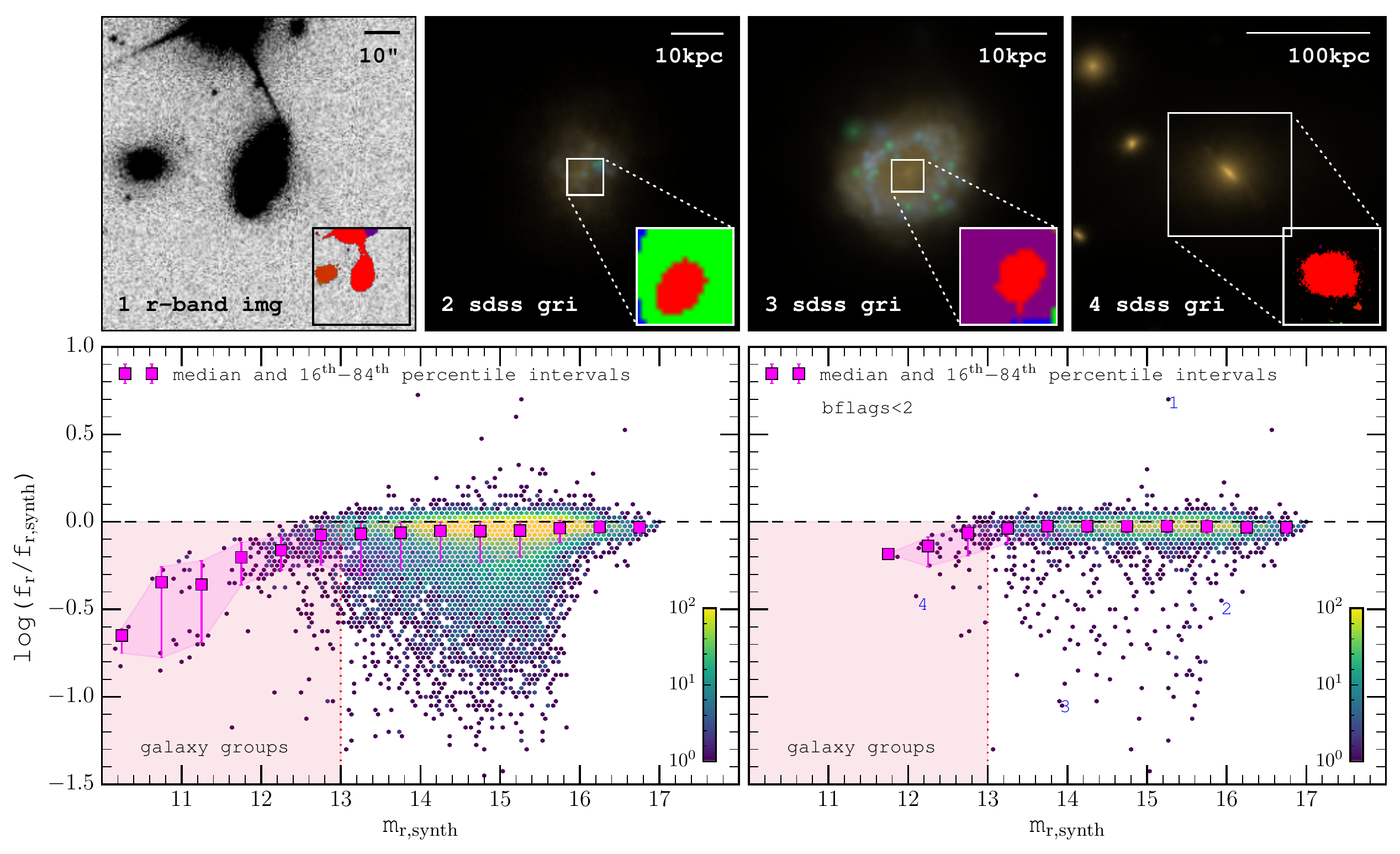}
    \caption[Recovery of flux in the \texttt{DISTINCT} catalog]{\emph{Lower Panels:} The distributions of recovered, $f_{\mathrm{r}}$, and synthetic image, $f_{\mathrm{r,synth}}$, total $r$-band fluxes plotted against the apparent magnitudes derived from the sum of the flux in the synthetic images, m$_{\mathrm{r,sim}}$ for the \texttt{DISTINCT} catalog. The lower left panel shows the distribution of flux offsets for the full \texttt{DISTINCT} catalog. The lower right panel shows only galaxies for which the number of flagged sources that directly border the pixels used in the fitting is \emph{bflags}$<2$. The \emph{black dashed line} in each of the lower panels denotes a 1:1 relation between model and synthetic image fluxes. \emph{Magenta symbols} show the 1-dimensional medians of recovered flux fraction with binning along$m_{\mathrm{r,synth}}$. Errorbars denote the 16-84$\%$ percentile range about the median. The \emph{salmon region} that is left of the \emph{red dotted line} at $m_{\mathrm{r,synth}}=13$ denotes the domain of $m_{\mathrm{r,synth}}$ computed from subhalos that predominantly contain galaxy groups. \emph{Upper Panels:} characterization of outliers in the \texttt{DISTINCT} catalog where \emph{bflags}$<2$, numbered 1-4 in the lower right panel.}
    \label{fig:flux}
\end{figure*}

First, we inspect the systematics on $r$-band integrated flux, $f_r$, by comparing our estimates derived from the best-fitting B+D models of galaxies with observational realism to the sum of the flux in the corresponding $r$-band synthetic images, $f_{r,\mathrm{synth}}$ in Figure \ref{fig:flux}. The apparent integrated magnitude estimates, $m_{r}$ and $m_{r,\mathrm{synth}}$, used in previous sections are computed directly from the respective integrated fluxes, $f_r$ and $f_{r,\mathrm{synth}}$, of the model and synthetic images. Plotted in the lower panels of Figure \ref{fig:flux} is the ratio of recovered to input fluxes against input apparent magnitudes. In the lower left panel of Figure, we show the distribution of flux recovery for the \texttt{DISTINCT} catalog -- which includes a single fit for every galaxy in the synthetic image catalog of \citetalias{2015MNRAS.447.2753T}. In the lower right panel, we show the same plot for galaxies that have \emph{bflags}$<2$. The images in the upper panels correspond to numbered outliers in the lower right panel and will visually aid in characterizing these outliers.

The lower left panel of Figure \ref{fig:flux} shows the distribution of systematic flux offsets for all galaxies from the \texttt{DISTINCT} catalog. The distribution demonstrates that the majority of flux estimates that are consistent with the fluxes computed from the synthetic images. However, roughly $30\%$ of galaxies across the full range of magnitude have systematically negative flux offsets. In previous sections, we have demonstrated that the only bias that is compatible with these large systematic offsets in flux is internal segmentation by discrete substructure in relatively diffuse galaxies. In the lower right panel of Figure \ref{fig:flux}, we attempt to visualize the systematics on flux estimates in the absence of the population of internally segmented galaxies in the \texttt{DISTINCT} catalog by performing a \emph{bflags}$<2$ cut. We find that the cut largely eliminates the population of galaxies with systematically poor flux recovery -- with only a few remaining outliers. However, the cut also significantly reduces the density of galaxies that reside around $\log (f_r/f_{r,\mathrm{synth}})=0$ for the full magnitude range. The loss of density among galaxies with accurate flux recovery is not surprising, however, when we recall that genuine external crowding (which \emph{bflags} is also sensitive to) in the absence of internal segmentation provides no systematic bias on the integrated fluxes and sizes. Therefore, the cut is also eliminating galaxies with accurate flux recovery but experienced crowding by external sources in their placements. 

We showed in Section \ref{sec:crowd_random} and \ref{sec:crowd_system} that our estimates of flux and size in placements with various degrees of crowding are consistent for the majority of the RIGs with no discernible systematic biases. In an effort to roughly quantify the number of galaxies with accurate flux estimates but that were culled because of the \emph{bflags} parameter's insensitivity to the source of segmentation (whether internally from discrete substructure or externally from crowding), we compare the number of galaxies in the range $-0.1<\log (f_r/f_{r,\mathrm{synth}})<0.1$ before and after the \emph{bflags}$<2$ cut. The total number of galaxies in the left and right panels of Figure \ref{fig:flux} are $\sim28,000$ and $\sim7,700$, respectively, so $72\%$ of galaxies in the full population do not satisfy the cut. We find that the number of galaxies that satisfy $-0.1<\log (f_r/f_{r,\mathrm{synth}})<0.1$ are 19,300 (70\%) in the \texttt{DISTINCT} catalog (left panel) and 7,100 (92\%) after the \emph{bflags}$<2$ cut (right panel). Therefore, approximately 12,200 galaxies have \emph{bflags}$\geq2$ due to genuine crowding effects or weak internal segmentation that does not strongly affect the overall surface brightness distribution of the galaxy. The contribution to the density of galaxies around $\log (f_r/f_{r,\mathrm{synth}})=0$ provided by these galaxies are \emph{accidentally} removed in the \emph{bflags} cut as a result of the sensitivity of \emph{bflags} to external crowding. As an example, note that all of the systems in Figure \ref{fig:mosaic} would be excluded under the \emph{bflags}$<2$, yet in most cases have visually satisfying residuals and prove to have accurate flux estimates relative to their synthetic images (not shown in the Figures). Their \emph{bflags} number is indicated in the top left corner of each science image mask. Taken together, our analysis indicates that roughly 8,000 (30\%) galaxies in the \texttt{DISTINCT} catalog are affected by strong internal segmentation or other bias that has so far not been identified. Meanwhile, no significant systematics from internal segmentation are identified in the \Seleven{} analysis (as expected, since the deblending was calibrated to accurately model the surface brightness distributions of real galaxies in the SDSS in isolation and in crowded images).

The upper panels of Figure \ref{fig:flux} show four example galaxies that are significant outliers from $\log (f_r/f_{r,\mathrm{synth}})=0$ that survive the \emph{bflags}$<2$ cut. Panel (1) shows the r-band science image of a galaxy that has been placed in the close vicinity of a bright star such that the galaxy's surface brightness distribution overlaps partially with a diffraction spike. Inset in this panel is the associated science mask where red corresponds to the flag for pixels that are used in the fitting, other colours correspond to additional sources that do not, and sky pixels are transparent. In this case the galaxy's flux is insufficient to warrant branching from the flux tree of the star as detailed in Section \ref{deblend}. Cases such as this are rare. We find that this scenario is common to each of the few large systematically positive outliers. 

In panels (2) and (3) of Figure \ref{fig:flux} we illustrate the scenarios that we found to be common amongst the systematically negative outliers (apart from those in the salmon region left of $m_{r,\mathrm{synth}}=13$). A discrete component at or near the centre of the galaxies' surface brightness distribution has been separated from the larger structure due to internal segmentation. Galaxies such as these survive the cull \emph{bflags}$<2$ cull because the pixels allocated for fitting are bordered by only a single non-sky source -- but still have the undesirable properties of discrete substructure that lead to negative systematic bias in flux estimates. Shown are the SDSS $gri$ false-colour images of these galaxies without any realism with a zoomed inset of the post-realism science image mask. In panel (2) a small young stellar population has been delineated from the rest of the galaxy but contains the pixel that is centred on the gravitational potential minimum of the galaxy. The substructure is therefore allocated as the sole source on which the fitting is performed, but contains only a fraction of the galaxy's total flux. 

Similarly, in panel (3) of Figure \ref{fig:flux} we demonstrate that a discrete nuclear component that is embedded within a ring of star formation has been identified as the primary source. The ring-like morphologies such as this one are described in \citetalias{2015MNRAS.454.1886S} with possible origins arising from choices of ISM and feedback models within Illustris.


We investigated the source of the negative systematic flux offsets for galaxies with $m_{r,\mathrm{synth}}<13$ shown in the salmon coloured region in each of the lower panels of Figure \ref{fig:flux}. The $gri$ false-colour synthetic image of the galaxy subhalo in panel (4) shows that the galaxy that is decomposed in the fitting is nested at the gravitational potential minimum of a group containing several galaxies. Therefore, the apparent systematic error in the flux does not necessarily arise because of segmentation or problematic fitting. Instead, it is often a natural consequence of the total flux in the input image being the sum of the flux from all galaxies that belong to the subhalo. The fitting is performed only on the galaxy containing the minimum of the gravitational potential -- which will contain a fraction of the total flux from the group. Only a handful of the groups on the bottom left panel satisfy the \emph{bflags}$<2$ criterion shown in the right panel, because there may be several projections for which the surface brightness distribution of a galaxy that belongs to a group is contaminated by other group members. The systematic descent with magnitude below $m_{r,\mathrm{synth}}=13$ is associated with groups that contain increasingly larger number of constituent galaxies and therefore increased total fluxes relative to the flux that can be obtained from the central. Such discrepancies should also be evident in estimates of the half-light radius of the best-fitting models and the synthetic images. 

\begin{figure*}
	\includegraphics[width=0.492\linewidth]{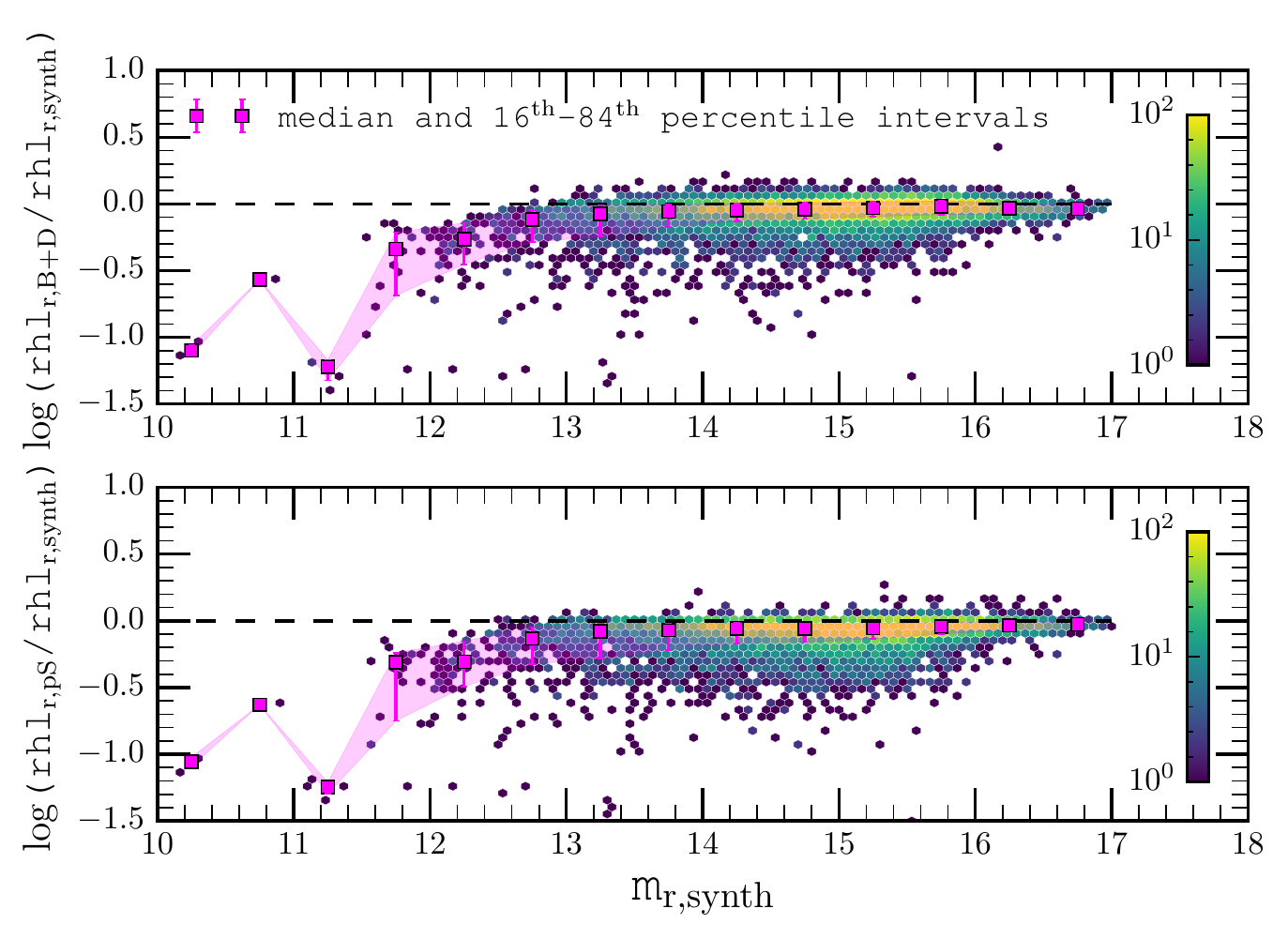}
	\includegraphics[width=0.48\linewidth]{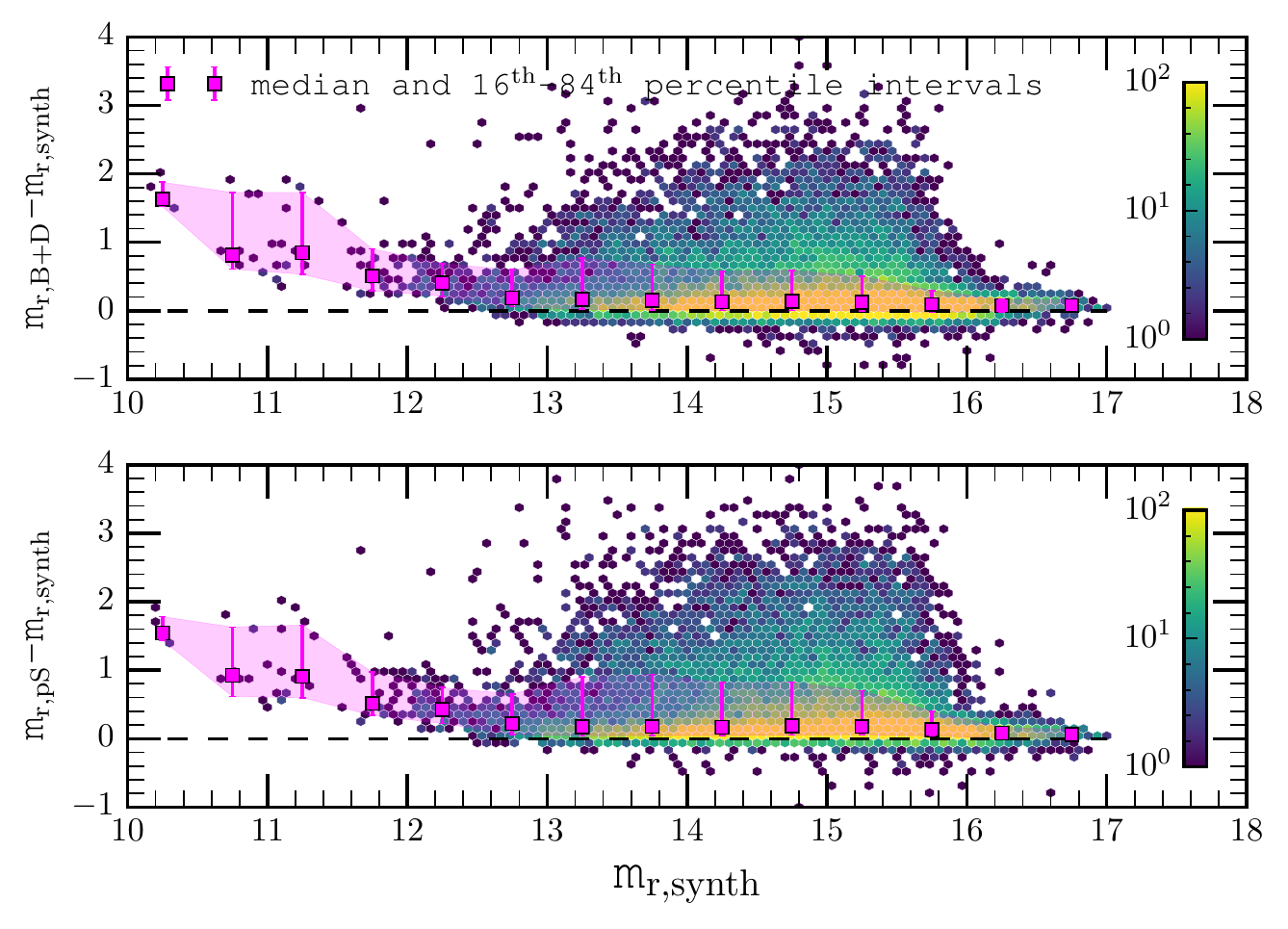}
    \caption[Accuracy of model HLR estimates in the \texttt{DISTINCT} catalog]{Comparisons of circular aperture half-light radii and integrated magnitudes computed from the models and input synthetic images from the \texttt{DISTINCT} catalog. \emph{Left panels}: systematic offsets between the half-light radii computed from the B+D models (upper panel) and $pS$ models (lower panel) and the respective synthetic images without realism. \emph{Right panels}: systematic offsets between the integrated apparent magnitudes computed from the B+D models (upper panel) and $pS$ models (lower panel) and the respective synthetic images without realism. The median and $16^{\mathrm{th}}-84^{\mathrm{th}}$ percentile intervals are shown by the magenta points and filled region.}
    \label{fig:ocrhalf}
\end{figure*}

\begin{figure}
	\includegraphics[width=\linewidth]{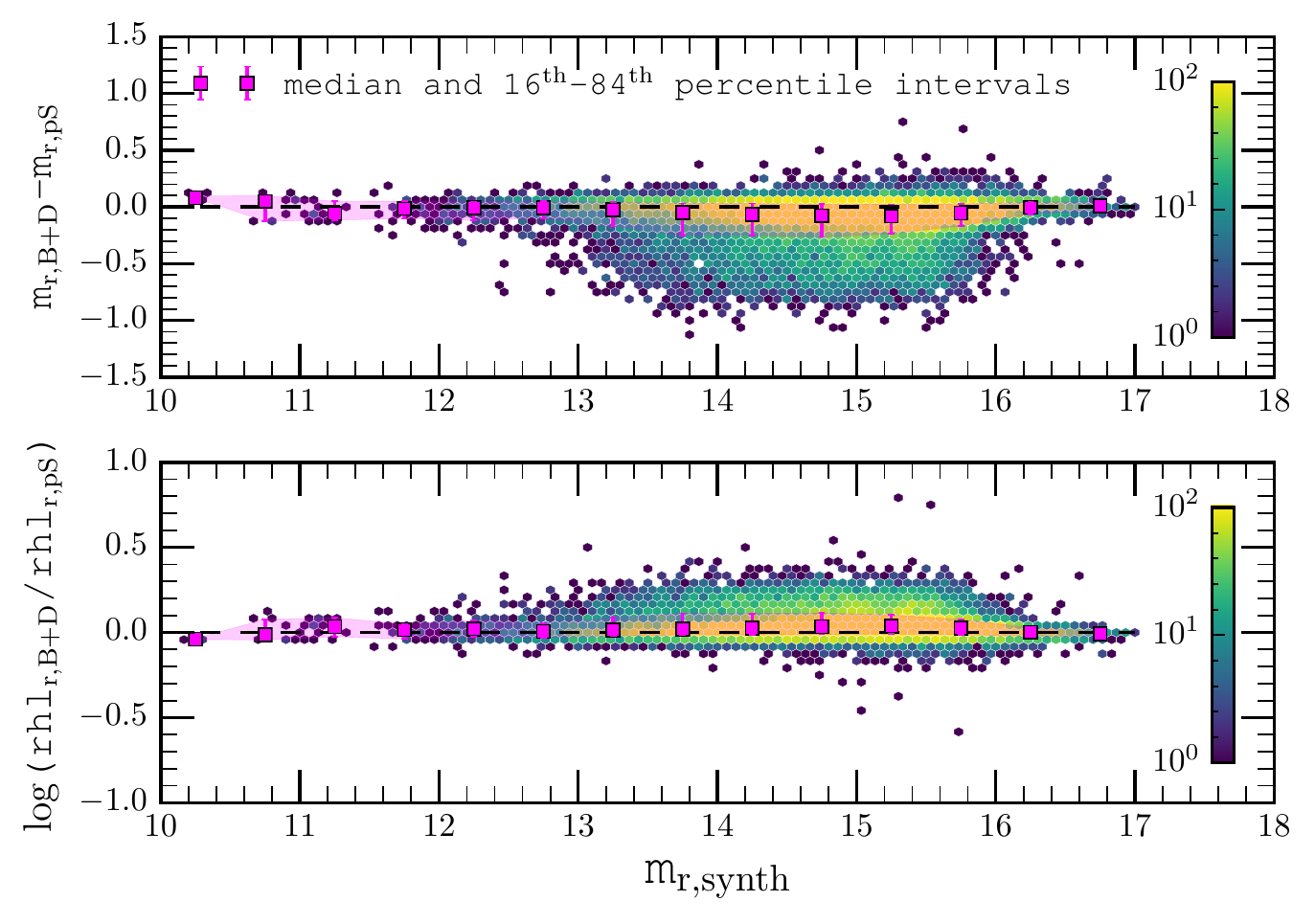}
    \caption[Bulge+Disc and Pure Sersic Magnitudes and Sizes]{Comparisons of B+D and $pS$ model integrated magnitudes (upper panel) and half-light radii (lower panel) from the \texttt{DISTINCT} catalog. Note the difference in scale from Figure \ref{fig:ocrhalf}. The median and $16^{\mathrm{th}}-84^{\mathrm{th}}$ percentile intervals are shown by the magenta points and filled region.}
    \label{fig:BDandpS}
\end{figure}

\subsubsection{Recovery of the Half-Light Radius}

In the last section, we characterized the systematics on estimates of the flux in comparisons between the integrated fluxes derived from the models and the synthetic images for the full population of Illustris galaxies in the \texttt{DISTINCT} catalog. These systematics should be mirrored in the estimates of half-light radius due to the dependence of accurate half-light radius estimates on accurate flux recovery over the full surface brightness distribution of the galaxy. In this section, we perform comparisons of circular aperture half-light radii that are computed from the best-fitting models and synthetic images. Furthermore, we test the robustness of our size and brightness estimates to the choice of surface brightness profile model by comparing the half-light radii and integrated apparent magnitudes derived from the best fitting B+D models (bulge: Sersic $n$=4, disc: Sersic $n$=1) to those from the $pS$ models (Sersic $n$=free) models for each galaxy. 

Figure \ref{fig:ocrhalf} shows the systematic offsets in the photometrically derived size and magnitude estimates in the \texttt{DISTINCT} catalog as a function of $r$-band input integrated apparent magnitude. In the left panels, the $r$-band half-light radii that are computed from photometry of the B+D (upper panel) and $pS$ (lower panel) models, $rhl_{r,b+d}$ and $rhl_{r,pS}$, are compared with those of the raw synthetic images $rhl_{r,\mathrm{synth}}$. In the right panels, we show that the offsets seen for the galaxy half-light radii mirror the integrated magnitude offsets (note that the upper right panel is a scaled reflection of the lower left panel from Figure \ref{fig:flux}).

The upper left and lower left panels of Figure \ref{fig:ocrhalf} show the same systematic negative offsets at $m_{r,\mathrm{synth}}<13$ that reflect the downturn in the flux recovery for groups in the \texttt{DISTINCT} catalog. The majority of our estimates at $m_{r,\mathrm{synth}}>13$ are consistent with the derived properties of the galaxies from the synthetic images without observational realism. However, image segmentation leads to a population of galaxies with negative systematic errors in their size estimates. The similarity of the distributions indicates that results from the best fitting surface brightness profiles are affected similarly.

\subsubsection{Comparison of B+D and $pS$ models in the \texttt{DISTINCT} catalog}

The analyses shown in this paper have been focused primarily on the B+D decomposition results. We have verified that the biases on size and magnitude estimates for the $pS$ models broadly agree quantitatively. We also look at trends in (B/T) from the B+D models and Sersic index from the $pS$ models, and their respective errors (though we reserve this comparison for a future paper -- in which we compare with observed trends). Figure \ref{fig:BDandpS} compares estimates of integrated magnitude and half-light radius from the $B+D$ and $pS$ models directly. Note the difference in scale in each panel with Figure \ref{fig:ocrhalf}. The large reduction in the scatter relative to Figure \ref{fig:ocrhalf} shows the differences between the properties of the B+D or $pS$ models is significantly smaller than either model's systematic offset from the input properties for a given galaxy. In other words, the systematics on the model parameters in the either B+D or $pS$ decompositions broadly agree -- which is expected when internal segmentation is the dominant systematic and both the $B+D$ and $pS$ models use the same segmentation maps. In Appendix \ref{sec:SExpars}, we explore agreement between B+D and $pS$ models in greater detail. But in general, the systematic offsets and random errors we report for the $B+D$ decompositions in the next section are consistent for the $pS$ decompositions. Other work (e.g., \citealt{2013MNRAS.436..697B}) showed that the bright end of the luminosity function inferred from surface-brightness decompositions of galaxies in the SDSS can be sensitive to the assumed light profile model. Although there appears to be no difference in our $pS$ and ($n_b=4$, $n_d=1$) bulge+disc decompositions at bright intrinsic magnitudes, this may be due to comparing different pairs of models. \citet{2013MNRAS.436..697B} compare magnitudes from their pure Sersic decompositions with results that use a ($n_b=$ free, $n_d=1$) Sersic+exponential model whereas we are using a fixed bulge Sersic index light profile model. Future investigation of the sensitivity of measured magnitudes to certain assumed light profiles and sky estimation for realistic synthetic galaxy images may provide insight into the sensitivities observed in analyses of real galaxies.

\section{Summary}\label{summary}

In this first paper in a series, we have described a new procedure for deriving image-based quantitative morphologies of simulated galaxies that enables fair comparisons with observations. We employ the new procedure in an analysis of galaxies from the Illustris simulation in a way that facilitates comparison with observed galaxies from the SDSS. The central tenet of a meaningful comparison between the observational properties of real and simulated galaxies is that the observational biases and methodology for measuring galaxy properties are consistent. Our methods are unique from previous attempts at comparisons between simulations and observations by combining three factors: (1) Using mock-observations of simulated galaxies to enable image-based comparisons with observations; (2) Applying an extensive suite of observational realism to the simulated galaxy images that facilitates unbiased comparisons with observations (3) Consistent methodology for derivation of parametric structural and photometric properties of observed and simulated galaxies.

\subsection{Observational Realism and Morphological Decompositions}

As described in Section \ref{realism}, we first ensure consistent observational biases for measuring galaxy properties by applying an extensive suite of observational realism to dust-free synthetic images of galaxies from the Illustris simulation. Next, parametric quantitative morphologies of the simulated galaxies are derived from bulge+disc decompositions. In brief, the following procedure was employed in the analyses of simulated galaxies:

\begin{itemize}
\item Selection of the SDSS fields into which simulated galaxies are inserted is determined by the projected positions of real galaxies in the SDSS -- which ensures that positional biases on decomposition results from crowding, sky brightness, and PSF resolution are statistically the same as for observed galaxies. 
\item The flux from the simulated galaxies is convolved with the SDSS PSF corresponding to the specific location at which the synthetic images are inserted and signal shot noise is added to the synthetic image flux. Convolution with the reconstructed SDSS PSF for each placement ensures the same statistics for PSF resolution given that placement is assigned quasi-randomly from the positional distribution of real galaxies.
\item \sextractor{} is used to perform deblending of galaxy flux with other sources and the sky. The parameters used in the deblending thresholds are the same as were used by \cite{2011ApJS..196...11S} in their analysis of galaxies from the SDSS. Consistency in the deblending is crucial to consistency in the decompositions (see Section \ref{deblend}).
\item 2-D parametric surface-brightness decompositions are performed with \textsc{gim2d} -- making the procedure for deriving structural and photometric properties of simulated galaxies completely consistent with observed galaxies. Furthermore, the parametric decompositions enable measurement of the structural properties of the physical components from the surface brightness distributions. Both bulge+disc decompositions and pure Sersic decompositions are performed for each galaxy for consistency with the catalogs of \cite{2011ApJS..196...11S} -- though we primarily focus on the bulge+disc decompositions in our characterization of biases.
\end{itemize}

\subsection{Characterization of Biases}
Several experiments were designed to characterize the biases in the decomposition results. The decomposition catalogs used in each experiment are described in Section \ref{catalogs} and summarized in Table \ref{tab:catalogs}. The results enabled quantification of statistical and systematic errors from observational biases and our decomposition pipeline (summarized in Table \ref{table_biases}). Many of the experiments were aimed at characterizing a specific bias, whilst taking precautions to control other biases in each decomposition.

\begin{itemize}
\item \emph{Stellar Light Distribution}: Results from the \texttt{SMOOTHING} catalog (Section \ref{SMOOTHcat}) were used to characterize the biases from stellar light distribution (SLD) schemes used to create the synthetic images (Section \ref{sec:smooth}). We showed that there are no systematic biases from the choice of SLD scheme on integrated magnitude or half-light radius in a representative sample of galaxies (RIGs). However, the decomposition results for a few diffuse galaxies (i.e. those that have large radii for their stellar mass) show that they are systematically fainter and smaller for \emph{all} SLD schemes relative to the corresponding measurements derived directly from the synthetic images -- with systematic offsets as large as +3.0 magnitudes and -0.6 dex in half-light radius. The typical scatter in comparisons with alternative SLD schemes is roughly $\pm0.05$ mag in integrated magnitude and $\pm0.03$ dex in half-light radius -- though much of this is attributed to the large scatter from the diffuse systems. The scatter in the bulge-to-total fractions in comparisons with the fiducial scheme was roughly $\pm0.05$ for the alternative adaptive scheme and $\pm0.2$ in comparisons with constant SLD schemes. A systematic trend of reduced (B/T) measurements were identified in comparisons of fixed kernel radius SLD schemes relative to adaptive schemes. Decomposition results for galaxies using SLD schemes with fixed kernel radii had reduced (B/T) by up to 0.6 relative to adaptive schemes. Detailed structural estimates may therefore be affected by SLD schemes and the spatial resolution that they afford.


\item \emph{Camera Angle}: Results from the \texttt{CAMERAS} catalog (Section \ref{CAMScat}) were used to show that decomposition results from the RIGs in different projections yield largely consistent results for the majority of the sample (Section \ref{sec:projection}). Magnitude differences from decompositions of galaxies at different camera angles were typically $<0.1$ mag and half-light radius differences $<0.15$ dex. Diffuse galaxies generally had larger variations up to 2.0 mag in integrated magnitude and 0.6 dex in half-light radius. Bulge-to-total fraction differences appeared to depend on the median (B/T) across all camera angles. Galaxies that appear to be bulge- or disc-dominated from their median estimates have small variation in (B/T). But, galaxies with median $0.2<(B/T)_{50\%}<0.8$ show larger sensitivity to projection that appears to peak at $(B/T)_{50\%}\approx0.5$. It appears that when significant bulge \emph{and} disc components are present within a galaxy, then projection significantly affects photometric decompositions of these components.

\item \emph{Environment and Crowding}: Results from the \texttt{ASKA} catalog were used to show that the decomposition results are largely robust to biases from crowding, sky brightness, and PSF resolution (Section \ref{sec:crowding}). Statistical errors on integrated magnitude and half-light radius were $<0.05$ mag and $<0.025$ dex, respectively. However, diffuse galaxies have larger errors, up to 2 mag and 0.4 dex in magnitude and half-light radius. Statistical errors on the bulge-to-total fraction were related to the median (B/T) for the distribution of decompositions of each galaxy and were as large as 0.3. Median systematic errors on integrated magnitude and half-light radius were small, with $\Delta m_{r,50\%} < 0.05$ mag and $\Delta \log (rhl_{r,50\%}) < 0.05$ dex, for the majority of galaxies. Diffuse galaxies had large systematic offsets in integrated magnitude and half-light radius, with $\Delta m_{r,50\%}$ up to +2 mag and $\Delta \log (rhl_{r,50\%})$ down to -0.4 dex relative to the corresponding measurements derived directly from the synthetic images. 

\item \emph{Internal Segmentation}: Inspection of the images, models, residuals, and science masks enabled the identification of internal segmentation by locally discrete substructure in the surface brightness distributions of diffuse galaxies (Section \ref{sec:segmentation}). In each of the experiments, diffuse galaxies had the largest systematic and random errors. Using the decomposition results from previous experiments, we found that the deblending in diffuse galaxies with bright substructure was highly sensitive to all observational biases which were the root cause of their large systematic and random errors.

\end{itemize}

The results from our characterization of biases are summarized in Table \ref{table_biases} which shows the random and systematic errors that can be expected for key parameters. Error contributions from each realism effect are separated by row. Each pair of values indicates the errors for galaxies within $\log \mathrm{M}_{\star}/\mathrm{M}_{\odot}<11$ and $\mathrm{M}_{\odot}<11$ and $rhm>8$ kpc (second number) and those without (first number).

\begin{table*}	
	\centering
	\caption[Table of Biases]{Table of biases for parameter estimates from the bulge+disc decompositions: apparent magnitude, half-light radius, and (B/T). Values are computed as follows. (Row 1) SLD: Random errors on $m_r$ and $rhl_r$ are computed from the $16^{\mathrm{th}}-84^{\mathrm{th}}$ percentile interval of the estimates obtained from \texttt{fn16} scheme for all galaxies in the \texttt{SMOOTHING} catalog. Systematic errors are computed by taking the median of the differences between the estimates from \gimtwod{} and those computed directly from the synthetic images. The random error on (B/T) shown in the table is computed from the median of the difference between estimates from the \texttt{fc1kpc} and \texttt{fn16} SLD schemes. (Row 2) CAM: Random errors on $m_r$, $rhl_r$, and (B/T) are computed from the median of differences between the maximum and minimum estimates from each galaxy in the \texttt{CAMERAS} catalog. Systematic errors on $m_r$ and $rhl_r$ for each galaxy are computed by taking the median of differences between the estimates from \gimtwod{} and those computed directly from the synthetic images. The quoted values are the median systematic errors over all select galaxies. (Row 3) SKY: Random errors are computed $16^{\mathrm{th}}-84^{\mathrm{th}}$ percentile intervals of estimates for each parameter in the \texttt{ASKA} catalog over all insertions a galaxy. The quoted values are the median random errors over all select galaxies. Systematic errors on $m_r$ and $rhl_r$ are computed from the median of parameter estimates over all insertions of a galaxy. The quoted values are the median systematic errors over all select galaxies. Each table entry contains two values. The second is for the low-mass, diffuse galaxies in $\log \mathrm{M}_{\star}/\mathrm{M}_{\odot}<11$ and $rhm>8$ kpc in the RIG sample. The first is for all other galaxies in the RIG sample.}\label{table_biases}
	
	\vspace{5pt}
	$(A)$: $\notin$(M$_{\star}/$M$_{\odot}<10^{11}$ \& $rhm>8$ kpc), $\quad(B)$: $\in$(M$_{\star}/$M$_{\odot}<10^{11}$ \& $rhm>8$ kpc)
	\begin{tabular}{lcccccr} 
		\hline

		Bias & \shortstack{$\Delta m_{r,sys}$ (mag) \\ $A \quad\quad B$} &  \shortstack{$\Delta m_{r,rand}$ (mag)\\  $A \quad\quad B$} & \shortstack{ $\Delta \log rhl_{r,sys}$ \\ $A \quad\quad B$ } & \shortstack{$\Delta \log rhl_{r,rand}$ \\ $A \quad\quad B$} & \shortstack{$\Delta B/T_{r,rand}$ \\ $A \quad\quad B$} \\

		\hline
		SLD  &    0.029$\quad$0.324   &    0.134$\quad$ 1.300    &     -0.015$\quad$ -0.111     &     0.071$\quad$ 0.328     &     0.243$\quad$ 0.073     \\
		CAM &    0.042$\quad$0.456   &    0.098$\quad$ 0.448    &     -0.024$\quad$ -0.080     &     0.100$\quad$ 0.134     &     0.144$\quad$ 0.014     \\
		SKY  &    0.071$\quad$0.563    &    0.065$\quad$ 0.450    &     -0.030$\quad$ -0.120     &     0.028$\quad$ 0.106     &     0.085$\quad$ 0.032     \\		
		\hline
	\end{tabular}

\end{table*}

Having characterized the biases in our experiments and catalogs, we performed decompositions of each camera angle of every galaxy in the Illustris synthetic image catalog of \cite{2015MNRAS.447.2753T} to enable comparison with observations. We showed that roughly $30\%$ of galaxies in Illustris were affected by internal segmentation (Section \ref{size_flux}) and we have quantified the effect of this segmentation on estimates of size and flux using the \texttt{DISTINCT} catalog. Internal segmentation systematically reduced estimates of size (up to $\Delta \log rhl_r \approx -0.4$ dex) and flux (up to $\Delta m_r\approx +2.0$ mag fainter) for galaxies in Illustris and was a consistent bias in each experiment (i.e. no choice of SLD scheme reduced its effects in the decomposition results). However, decomposition results for integrated magnitude and half-light radius in the \texttt{DISTINCT} catalog for the majority of galaxies were consistent with the properties derived from the synthetic images before the addition of observational biases.

In the next paper in our series, \cite{cbottrell2017}, we employ our decompositions in comparisons with the properties derived from real galaxies in the SDSS. In particular, we focus on comparisons that use our estimates of integrated magnitude, half-light radius, and bulge-to-total fractions -- for which the systematic biases and statistical uncertainties are now quantitatively characterized.

\section*{Acknowledgements}

We thank the reviewer for their valuable and constructive feedback which greatly contributed to the quality of this paper. We thank Greg Snyder for useful discussions and input. PT acknowledges support for Program number HST-HF2-51384.001-A was provided by NASA through a grant from the Space Telescope Science Institute, which is operated by the Association of Universities for Research in Astronomy, Incorporated, under NASA contract NAS5-26555. This research made use of a University of Victoria computing facility funded by grants from the Canadian Foundation for Innovation and the British Columbia Knowledge and Development Fund. We thank the system administrators of this facility for their gracious support. Funding for the Sloan Digital Sky Survey IV has been provided by
the Alfred P. Sloan Foundation, the U.S. Department of Energy Office of
Science, and the Participating Institutions. SDSS-IV acknowledges
support and resources from the Center for High-Performance Computing at
the University of Utah. The SDSS web site is www.sdss.org. SDSS-IV is managed by the Astrophysical Research Consortium for the 
Participating Institutions of the SDSS Collaboration including the 
Brazilian Participation Group, the Carnegie Institution for Science, 
Carnegie Mellon University, the Chilean Participation Group, the French Participation Group, Harvard-Smithsonian Center for Astrophysics, 
Instituto de Astrof\'isica de Canarias, The Johns Hopkins University, 
Kavli Institute for the Physics and Mathematics of the Universe (IPMU) / 
University of Tokyo, Lawrence Berkeley National Laboratory, 
Leibniz Institut f\"ur Astrophysik Potsdam (AIP),  
Max-Planck-Institut f\"ur Astronomie (MPIA Heidelberg), 
Max-Planck-Institut f\"ur Astrophysik (MPA Garching), 
Max-Planck-Institut f\"ur Extraterrestrische Physik (MPE), 
National Astronomical Observatory of China, New Mexico State University, 
New York University, University of Notre Dame, 
Observat\'ario Nacional / MCTI, The Ohio State University, 
Pennsylvania State University, Shanghai Astronomical Observatory, 
United Kingdom Participation Group,
Universidad Nacional Aut\'onoma de M\'exico, University of Arizona, 
University of Colorado Boulder, University of Oxford, University of Portsmouth, 
University of Utah, University of Virginia, University of Washington, University of Wisconsin, 
Vanderbilt University, and Yale University.




\bibliographystyle{mnras}
\bibliography{Bib/references_full.bib} 




\appendix

\section{Catalog Parameters}
\label{sec:parameter_descriptions}
The decomposition catalogs: \texttt{SMOOTHING, CAMERAS, ASKA,} and \texttt{DISTINCT} are described by the same sets of parameters. Each decomposition catalog is split into two tables: ``n4" for the bulge+disc and ``pS" for the pure Sersic decompositions. Tables \ref{tab:illustris_n4}-\ref{tab:illustris_pS} describe the parameters accessible from these catalogs -- found in the online supplement to this paper. Additional information, such as the location into which a galaxy is placed in the SDSS, can also be found in the catalogs.

\begin{table*}
	\centering
	\caption[Bulge+disc decomposition parameters]{Illustris galaxy structural parameters from the $n_b$=4, $n_d=1$ bulge + disk decompositions (online supplementary information)\label{tab:illustris_n4}}
	
	\vspace{5pt}
	\begin{tabular}{ll} 
		\hline
		Column Name & Description \\
		\hline
simulID & Unique catalog ID combination of ObjID, subhaloID, camera (includes non-number characters) \\
subhaloID & Illustris galaxy subhalo ID \\
camera & Camera angle orientation, [0,1,2,3] \\
smooth & Stellar light distribution scheme \\
objID & Matched SDSS galaxy object ID (placement procedure) \\ 
run & SDSS run \\
rerun & SDSS rerun\\
camcol & SDSS camera column\\
field & SDSS field (provides unique SDSS corrected image ID with run, rerun, camcol)\\
colc\_g\_sim & pixel column in SDSS $g$-band corrected image\\
rowc\_g\_sim & pixel row in SDSS $g$-band corrected image\\
colc\_r\_sim & pixel column in SDSS $r$-band corrected image\\
rowc\_r\_sim & pixel row in SDSS $r$-band corrected image\\
g2dmag$\_$g\_sim & $g$-band apparent magnitude inferred from the total flux in the synthetic image \\
g2dmag$\_$r\_sim & $r$-band apparent magnitude inferred from the total flux in the synthetic image  \\
g2dmag$\_$g & $g$-band apparent magnitude of GIM2D output B+D model \\
g2dmag$\_$r & $r$-band apparent magnitude of GIM2D output B+D model \\
bt\_g & $g$-band bulge-to-total fraction\\ 
bt\_r & $r$-band bulge-to-total fraction\\ 
rhalf\_g & $g$-band galaxy semi-major axis, half-light radius in SDSS pixels\\
rhalf\_r & $r$-band galaxy semi-major axis, half-light radius in SDSS pixels\\
ocrhalf\_g & $g$-band galaxy circular half-light radius in SDSS pixels\\
ocrhalf\_r & $r$-band galaxy circular half-light radius in SDSS pixels\\
re & Bulge semi-major effective radius in SDSS pixels \\ 
e & Bulge ellipticity ($e \equiv  1 - b/a$, e = 0 for a circular bulge)\\ 
phib & Bulge position angle in degrees (measured clockwise from the $+y$ axis of SDSS images)\\ 
rd & Exponential disk scale length in SDSS pixels \\ 
incd & Disk inclination angle in degrees ($i \equiv$ 0 for a face-on disk) \\ 
phid & Disk position angle in degrees (measured clockwise from the $+y$ axis of SDSS images)\\
dx\_g & B+D model center offset from column position given by colc\_g\_sim on SDSS corrected $g$-band image (SDSS pixels)\\
dy\_g & B+D model center offset from row position given by rowc\_g\_sim on SDSS corrected $g$-band image (SDSS pixels)\\
dx\_r & B+D model center offset from column position given by colc\_r\_sim on SDSS corrected $r$-band image (SDSS pixels)\\
dy\_r & B+D model center offset from row position given by rowc\_r\_sim on SDSS corrected $r$-band image (SDSS pixels)\\
msk\_nflags & Number of unique flagged sources in the area subtended by the synthetic image in science image \\
msk\_bflags & Number of unique flagged sources contiguous to pixels belonging to the source with the same flag as the central pixel \\
\hline

	\end{tabular}

\end{table*}

\begin{table*}
	\centering
	\caption[Pure Sersic decomposition parameters]{Illustris galaxy structural parameters from pure Sersic decompositions (online supplementary information)\label{tab:illustris_pS}}
	
	\vspace{5pt}
	\begin{tabular}{ll} 
		\hline
		Column Name & Description \\
		\hline
simulID & Unique catalog ID combination of ObjID, subhaloID, camera (includes non-number characters)\\
subhaloID & Illustris galaxy subhalo ID \\
camera & Camera angle orientation, [0,1,2,3] \\
smooth & Stellar light distribution scheme \\
objID & Matched SDSS galaxy object ID (placement procedure) \\ 
run & SDSS run \\
rerun & SDSS rerun\\
camcol & SDSS camera column\\
field & SDSS field (provides unique SDSS corrected image ID with run, rerun, camcol)\\
colc\_g\_sim & pixel column in SDSS $g$-band corrected image\\
rowc\_g\_sim & pixel row in SDSS $g$-band corrected image\\
colc\_r\_sim & pixel column in SDSS $r$-band corrected image\\
rowc\_r\_sim & pixel row in SDSS $r$-band corrected image\\
g2dmag$\_$g\_sim & $g$-band apparent magnitude inferred from the total flux in the synthetic image \\
g2dmag$\_$r\_sim & $r$-band apparent magnitude inferred from the total flux in the synthetic image  \\
g2dmag$\_$g & $g$-band apparent magnitude of GIM2D output pure Sersic model \\
g2dmag$\_$r & $r$-band apparent magnitude of GIM2D output pure Sersic model \\
n & Galaxy Sersic index \\
rhalf\_g & $g$-band galaxy semi-major axis, half-light radius in SDSS pixels\\
rhalf\_r & $r$-band galaxy semi-major axis, half-light radius in SDSS pixels\\
ocrhalf\_g & $g$-band galaxy circular half-light radius in SDSS pixels\\
ocrhalf\_r & $r$-band galaxy circular half-light radius in SDSS pixels\\
re & Galaxy semi-major effective radius in SDSS pixels \\ 
e & Galaxy ellipticity ($e \equiv  1 - b/a$, e = 0 for a circular bulge)\\ 
phib & Galaxy position angle in degrees (measured clockwise from the $+y$ axis of SDSS images)\\ 
dx\_g & B+D model center offset from column position given by colc\_g\_sim on SDSS corrected $g$-band image (SDSS pixels)\\
dy\_g & B+D model center offset from row position given by rowc\_g\_sim on SDSS corrected $g$-band image (SDSS pixels)\\
dx\_r & B+D model center offset from column position given by colc\_r\_sim on SDSS corrected $r$-band image (SDSS pixels)\\
dy\_r & B+D model center offset from row position given by rowc\_r\_sim on SDSS corrected $r$-band image (SDSS pixels)\\
msk\_nflags & Number of unique flagged sources in the area subtended by the synthetic image in science image \\
msk\_bflags & Number of unique flagged sources contiguous to pixels belonging to the source with the same flag as the central pixel \\
\hline

	\end{tabular}

\end{table*}

\section{\textsc{Source Extractor} \& Internal Segmentation}
\label{sec:SExpars}
The goal of this paper is to apply \emph{the same} analysis used for real galaxy images to galaxies from a cosmological simulation, rather than optimize the parameters from \textsc{SExtractor} and \textsc{gim2d} to handle the synthetic images with realism. The key to the comparison is that the methodology is identical -- providing a direct diagnostic of the distinguishing properties of real and simulated galaxies. In this paper, we have shown that internal segmentation is the dominant bias for the synthetic images with realism. Meanwhile, internal segmentation is not identified as a significant systematic in analyses of real galaxies in the SDSS (\Seleven). The capacity of an experiment to identify such fundamental distinctions becomes limited when distinct methodologies are used for the real and simulated data sets. Therefore, we did not explore alternative deblending parameters in the main body of this paper. Nonetheless, it is possible to create a deblending scheme in which the systematics from internal segmentation in the synthetic images are reduced. However, in order to facilitate a fair comparison with observations, an alternative deblending scheme would need to be calibrated for realism (such as crowding -- to which the deblending \emph{must} be sensitive) in both observational and synthetic galaxy images. In this Appendix, we demonstrate that alternative parameters for deblending in \textsc{SExtractor} can enable improved accuracy modelling of galaxy properties for both B+D and $pS$ decompositions and removal of the systematics from internal segmentation.


\subsection{Deblending parameters}
\textsc{SExtractor} uses two parameters to determine whether a source above a pre-defined detection threshold is a single object or contains contributions from multiple closely-separated unique sources. The first, \texttt{DEBLEND\_NTHRESH}, defines a number of isophotes that extend from zero (in a sky-subtracted image) and the pixel containing the maximum flux in the object. Allocation of flux to a unique source occurs if two conditions are met for local maxima in the flux distribution: (1) the local minimum in the trough that separates each local maximum is in a lower isophotal layer than both local maxima; (2) the total flux contained in \emph{both} sources that defined by the first criterion contain some set fraction, \texttt{DEBLEND\_MINCONT}, of the total flux of the combined source. The fiducial deblending parameters used by \Seleven{} are: (\texttt{DEBLEND\_NTHRESH}=32, \texttt{DEBLEND\_MINCONT}=0.00005) -- which are calibrated for dealing with strongly crowded sources in the SDSS among other science metrics. The two parameters can be modified to achieve an arbitrary level of deblending in sources with non-monotonic surface brightness distributions (i.e. contain local maxima that are unique from the global). However, selecting a set of parameters that \emph{both} achieves reduced segmentation in the synthetic galaxy surface brightness distributions \emph{and} accurately handles the realism of crowding is a non-trivial challenge and is beyond the scope of what are trying to accomplish. Instead, we will use the more ``accurate'' model parameters derived using an alternative set of deblending parameters to solidify the claims we make about internal segmentation in a comparison with the fiducial set.

\subsection{Comparison with Alternative Deblending Methods}
We use the RIG sample in the experiment. The fiducial methodology is employed on each RIG in the same, uncrowded location in the SDSS. The reasons for this are two-fold: (1) we want to examine the differences that choices in the deblending parameters have on the bias from internal segmentation alone; (2) we have not attempted to calibrate either scheme for the synthetic images in crowded fields.
We select an alternative set of deblending parameters simply by increasing the fiducial \texttt{DEBLEND\_MINCONT}=0.00005 to \texttt{DEBLEND\_MINCONT}=0.05. The change in \texttt{DEBLEND\_MINCONT} has the effect of significantly increasing the strictness of the criteria for separating locally bright features from a source (requiring at least 5\% of the total flux to be contained in the local feature). The methodology employing the alternative deblending parameters is then applied to each RIG in the same location in the SDSS in which the fiducial methodology is employed. The experiment guarantees that a change in the segmentation map and measured parameters for a given galaxy is solely sensitive to the internal segmentation that arises from the choice of deblending parameters.

\begin{figure*}
	\includegraphics[width=0.49\linewidth]{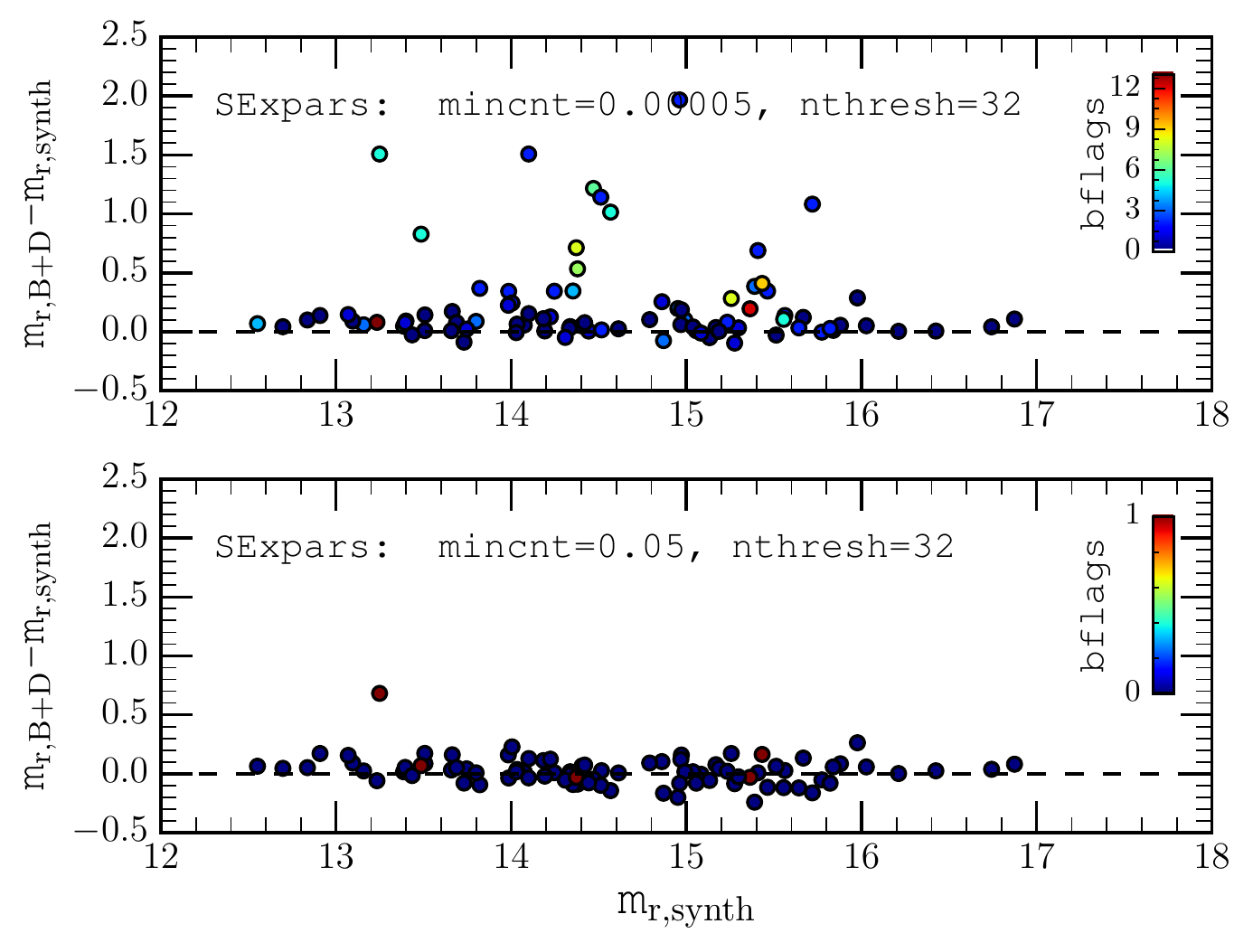}
	\includegraphics[width=0.49\linewidth]{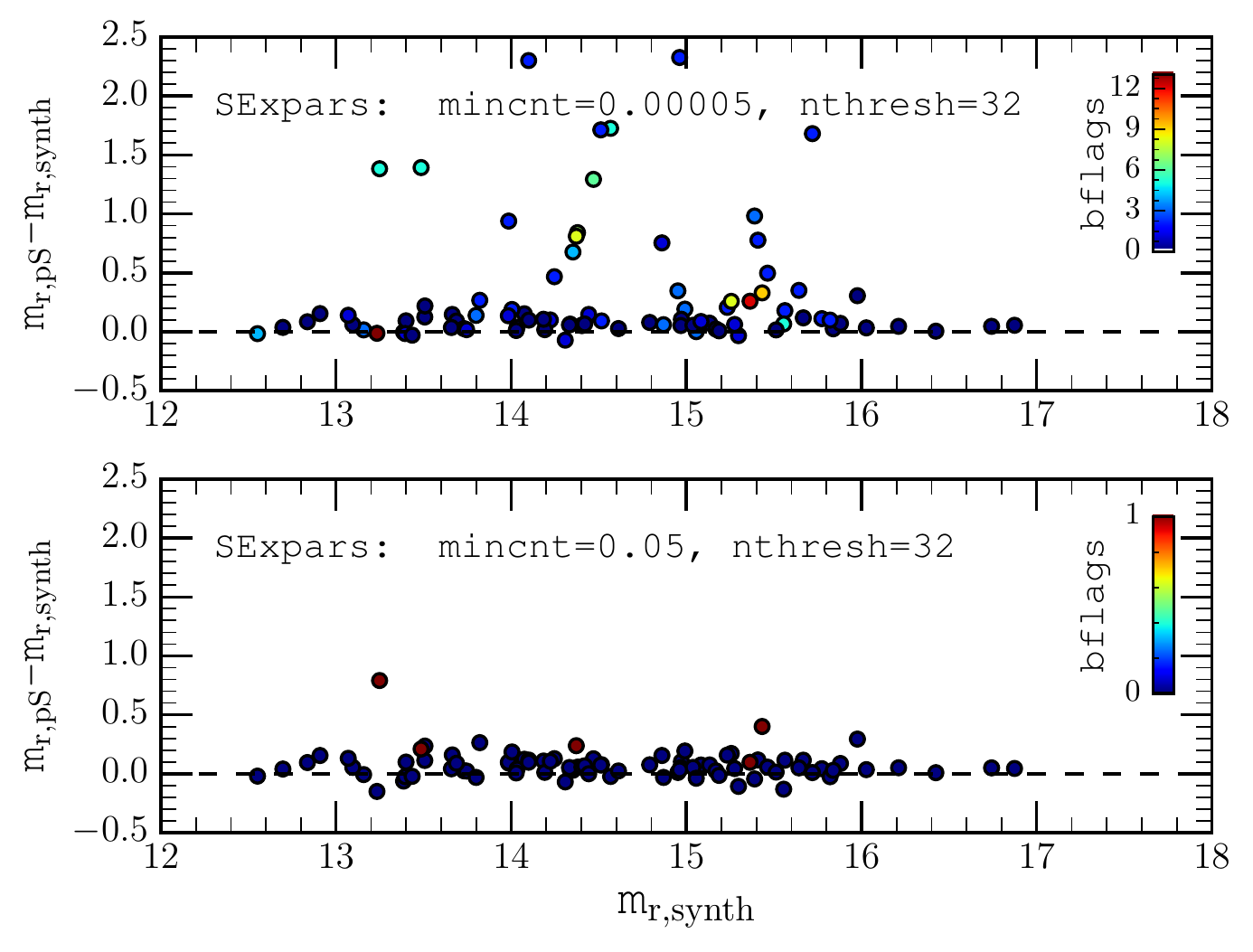}
    \caption[MINCNT comparisons: magnitude]{Comparisons of B+D (left panels) and $pS$ (right panels) model integrated magnitudes  with the input image magnitudes for the fiducial (upper panels) and alternative (lower panels) \textsc{SExtractor} deblending parameters. Points are coloured by \emph{bflags}. Note the difference in scale in the \emph{bflags} in the upper and lower panels -- corresponding to the significant reduction of internal segmentation using the alternative deblending scheme. Further note that the systematic offsets for galaxies with high \emph{bflags} in the upper panels are all but eliminated in the lower panels.}
    \label{fig:mincnt_mag}
\end{figure*}

\begin{figure*}
	\includegraphics[width=0.49\linewidth]{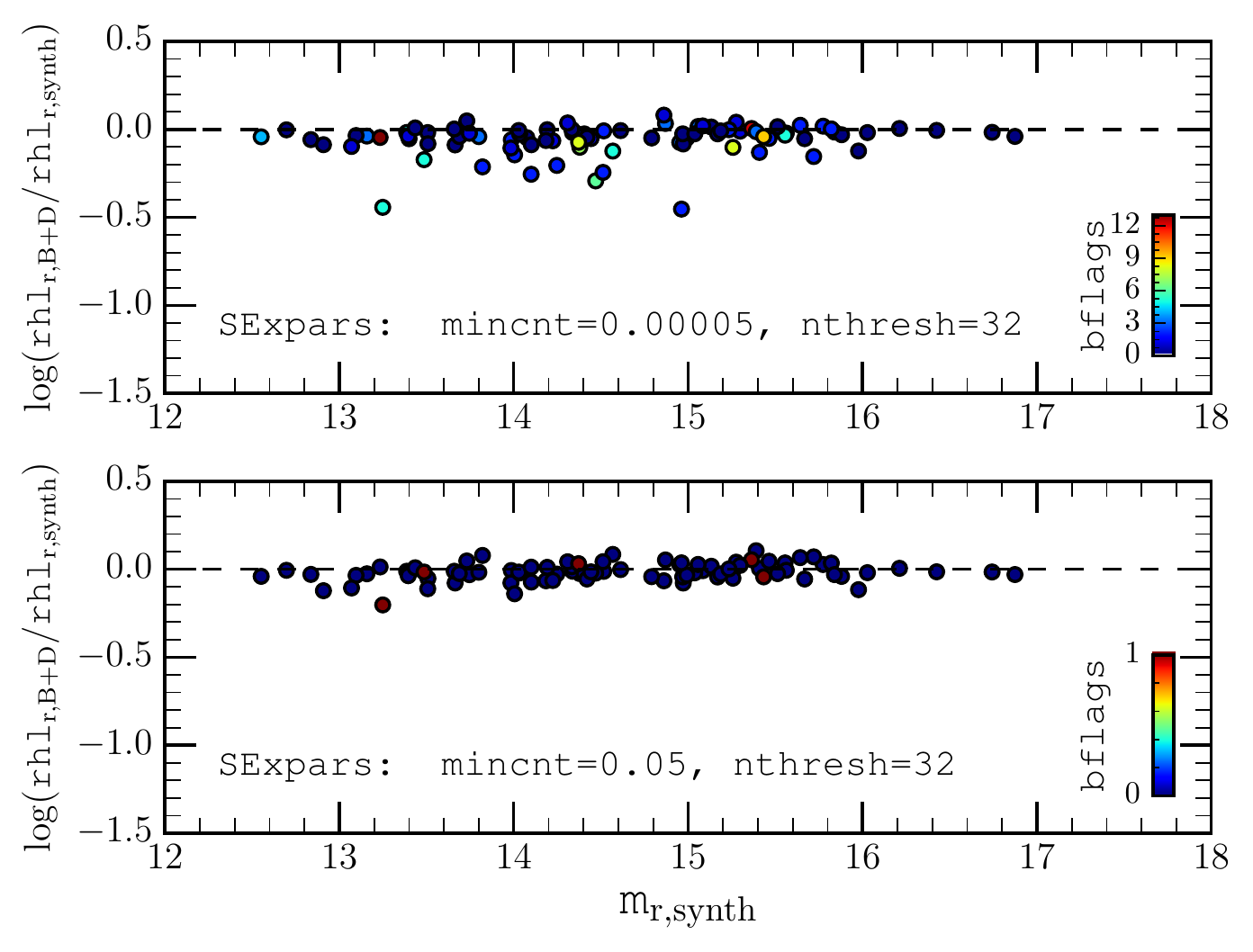}
	\includegraphics[width=0.49\linewidth]{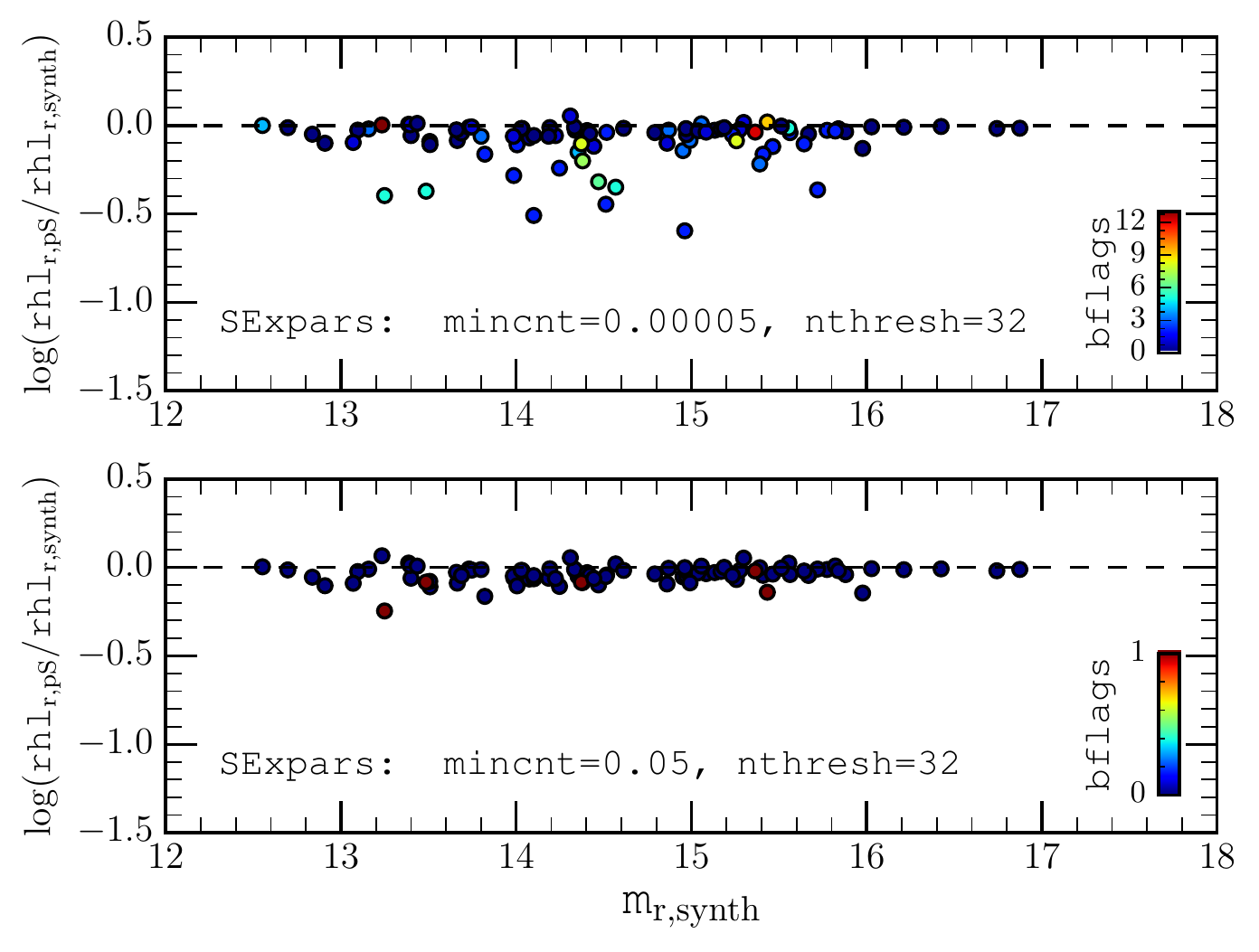}
    \caption[MINCNT comparisons: magnitude]{Comparisons of B+D (left panels) and $pS$ (right panels) model half-light radii with the input image half-light radii for the fiducial (upper panels) and alternative (lower panels) \textsc{SExtractor} deblending parameters. Note that the negatively offset galaxies in the upper panels mostly lie along the dashed lines in the lower panels.}
    \label{fig:mincnt_rhl}
\end{figure*}

\begin{figure*}
	\centering
	\includegraphics[width=0.9\linewidth]{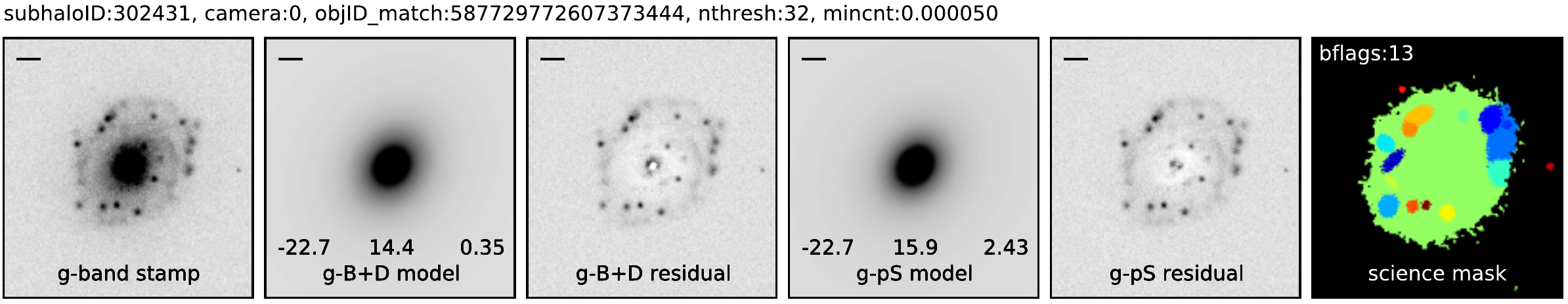}
	\includegraphics[width=0.9\linewidth]{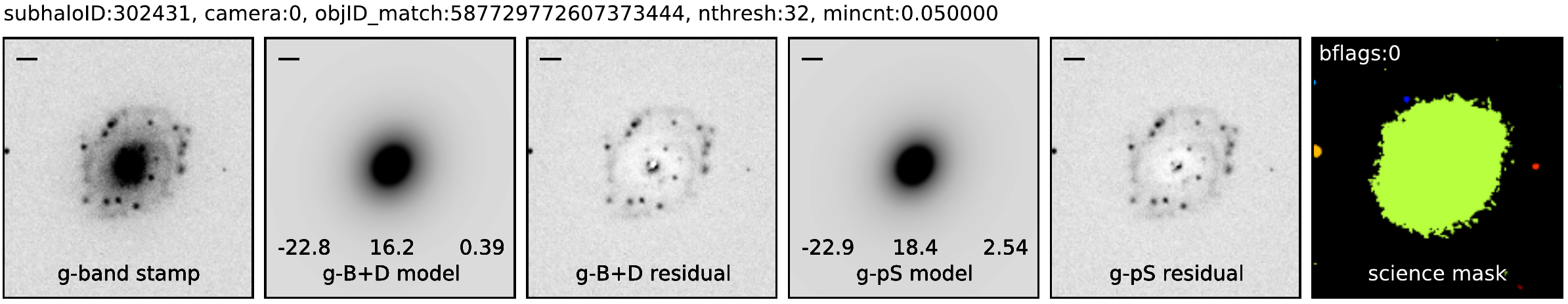}
	\includegraphics[width=0.9\linewidth]{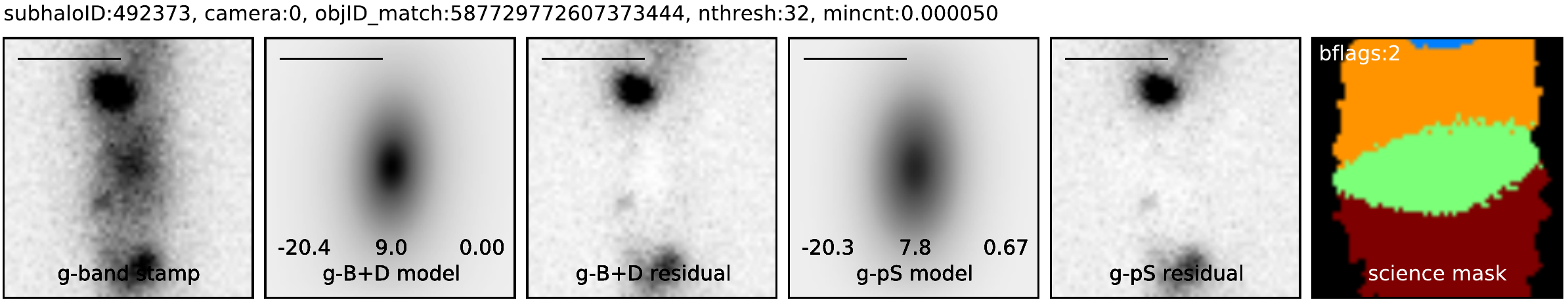}
	\includegraphics[width=0.9\linewidth]{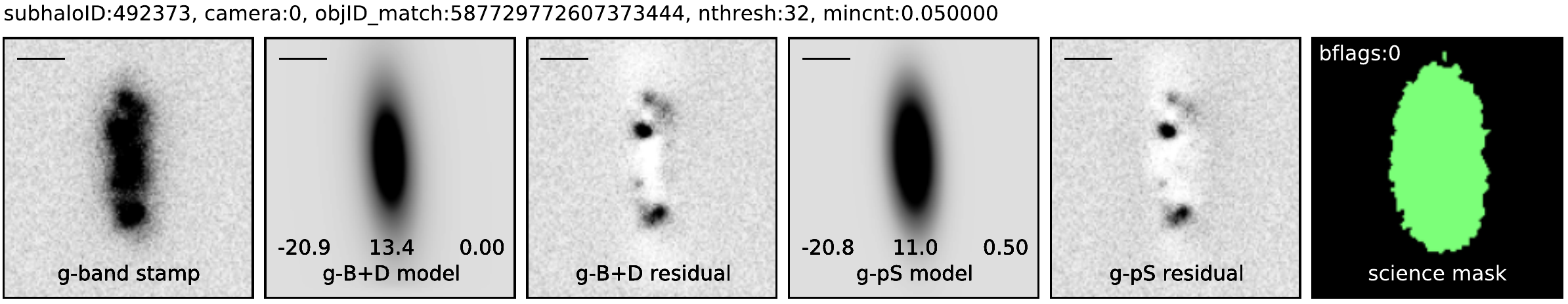}
    \caption[MINCNT images]{Similar to Figure \ref{fig:mosaic}, each row shows the $g$-band science image, B+D model, B+D residual, $pS$ model, $pS$ residual, and science mask for a galaxy with a specific set of segmentation parameters. The first and second rows show the results for the same galaxy with the fiducial (first row) and alternative (second row) deblending parameters. Note the masking of substructure in the fiducial deblending with respect to the alternative scheme. Note also that the best-fitting parameters for the galaxy in the first and second rows are less affected by the choice of deblending scheme because the clumpy features contain a small fraction of the total light from the galaxy. The third and fourth rows show a galaxy with larger discrepancy in the model parameters between the fiducial (third row) and alternative (fourth row) deblending schemes. While the galaxy is less segmented (as quantified by \emph{bflags} in the fiducial scheme), the modelling is restricted to a single discrete feature in the surface brightness distribution of the galaxy due to internal segmentation using the fiducial parameters. In this case, the choice of deblending parameters strongly affects the model parameters. Magnitudes are higher and half-light radii are correspondingly lower. Image sizes in each row may vary as the size of the image is set by the area corresponding to pixels flagged for modelling. The 10 kpc scale at the top left of each image can be used as an indicator of relative image size. }
    \label{fig:mincnt_images}
\end{figure*}

Figures \ref{fig:mincnt_mag} and \ref{fig:mincnt_rhl} compare the integrated magnitudes and circular aperture half-light radii from the B+D and $pS$ models to the corresponding input image properties for the fiducial (upper panels) and alternative (lower panels) deblending schemes. Points are coloured by \emph{bflags} for each set of deblending parameters. The large systematic offsets in magnitude (positive offsets) and half-light radius (negative offsets) seen for several RIGs with non-zero \emph{bflags} in the fiducial deblending are largely removed by using the alternative scheme. The result demonstrates that more accurate photometric modelling is possible for the synthetic galaxy images by modifying the \textsc{SExtractor} deblending parameters. The improvements to model accuracy that can be accomplished using an alternative deblending scheme are visually demonstrated in Figure \ref{fig:mincnt_images}. However, the cost of this improved accuracy in the modelling of the galaxy surface brightness distributions is a limited capacity for comparison with observations and a general ignorance of how robust these model parameters are to observationally consistent realism such as crowding.

\begin{figure*}
	\includegraphics[width=0.32\linewidth]{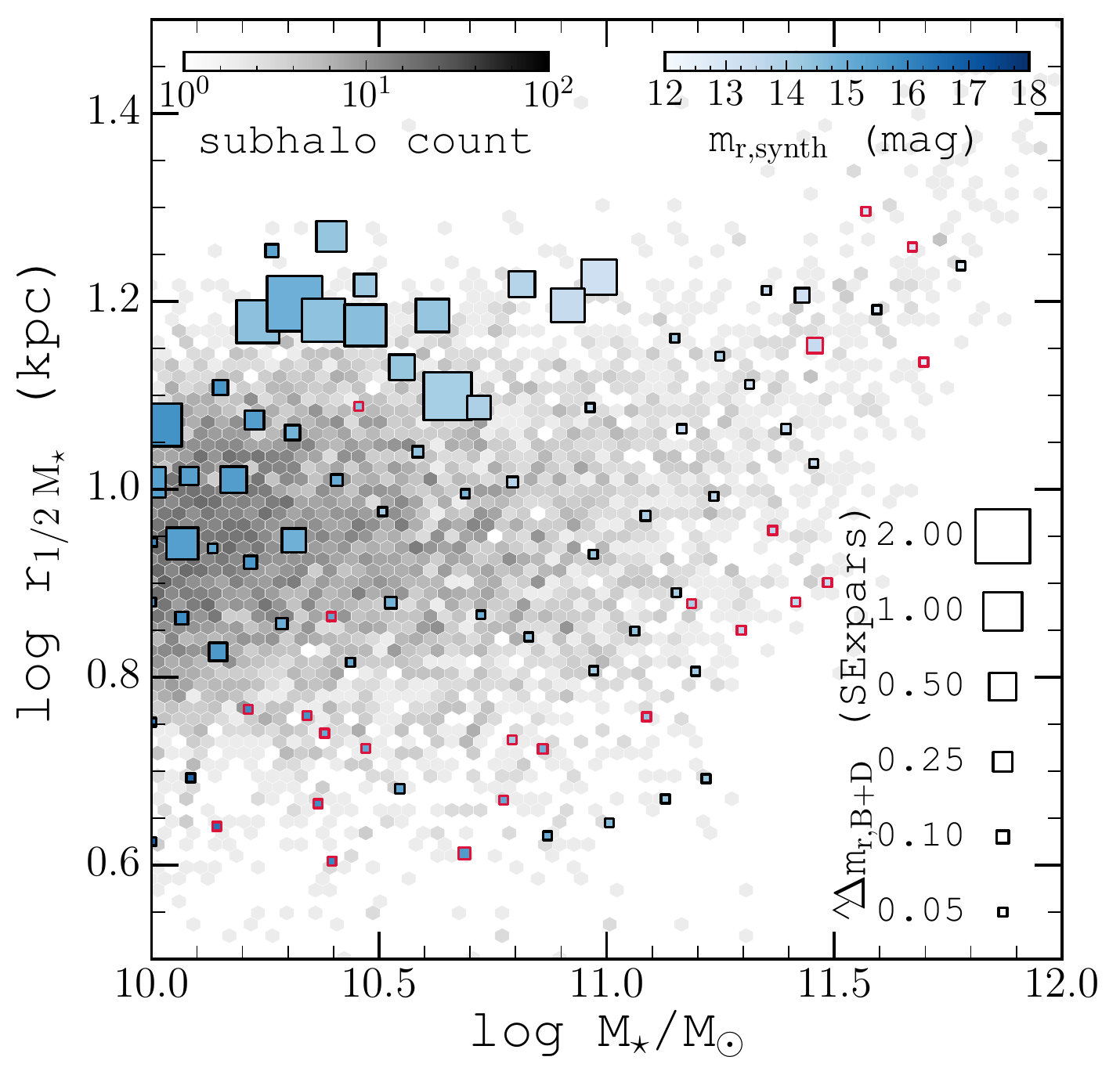}
	\includegraphics[width=0.32\linewidth]{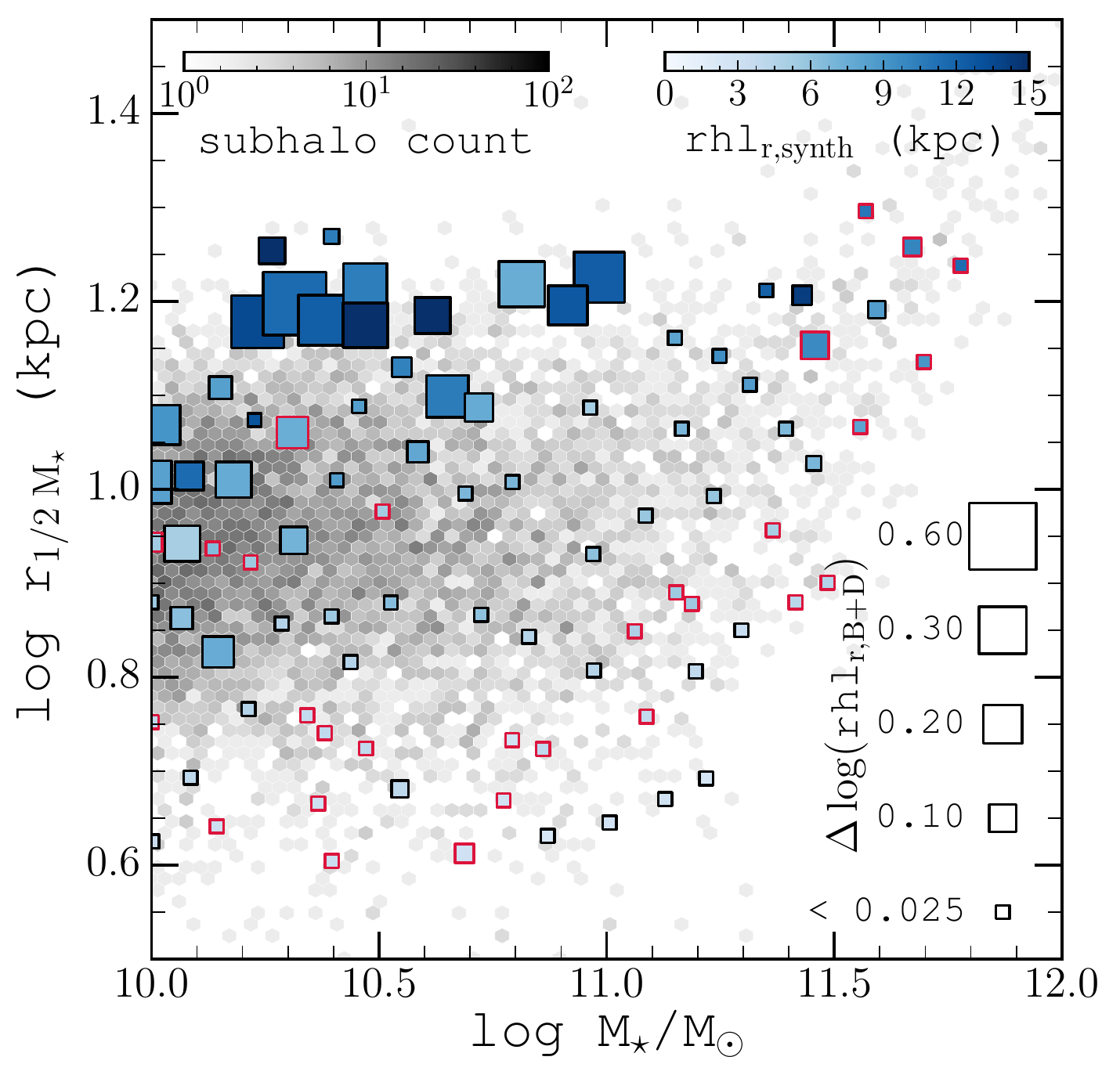}
	\includegraphics[width=0.32\linewidth]{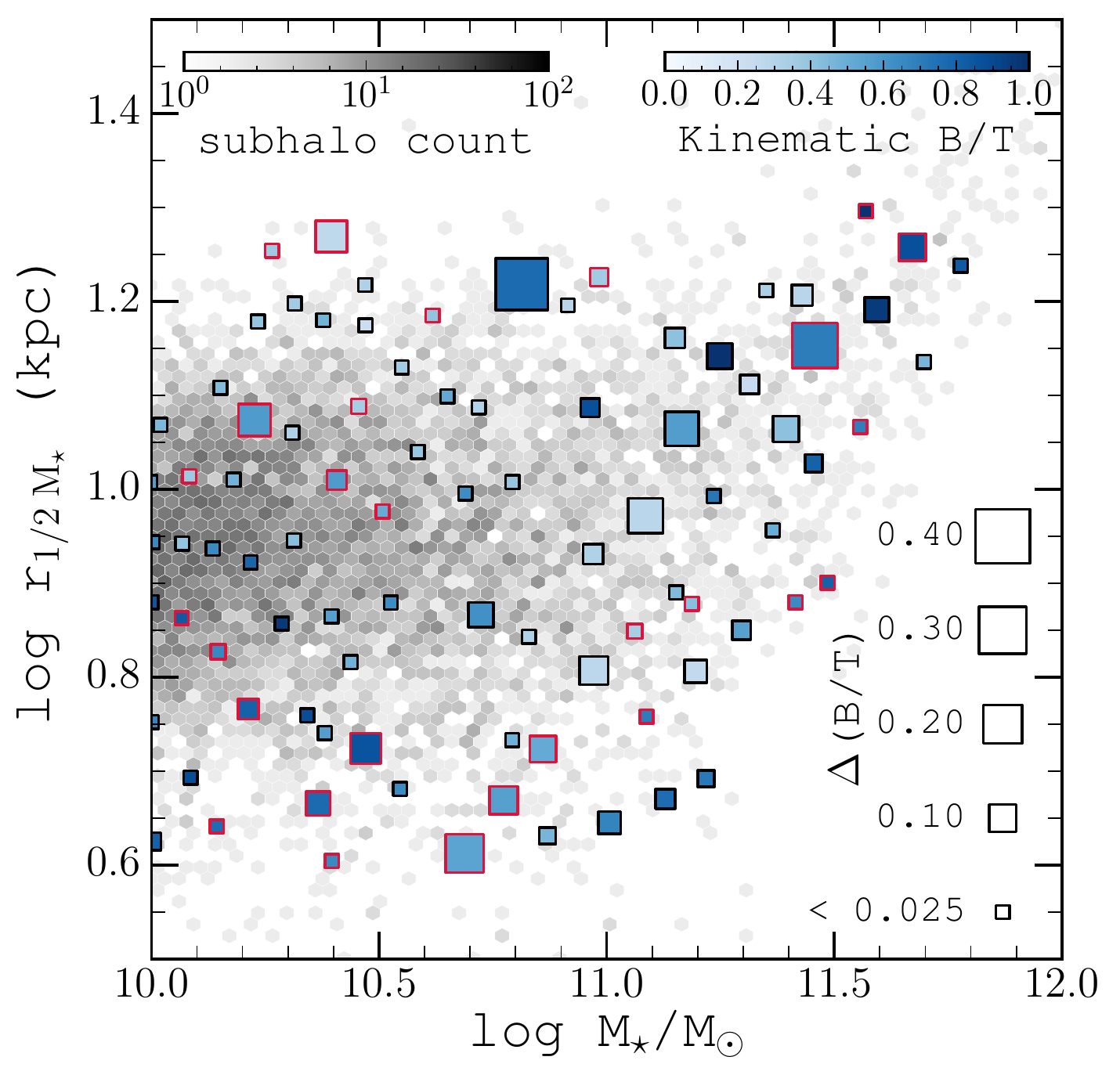}
	\includegraphics[width=0.32\linewidth]{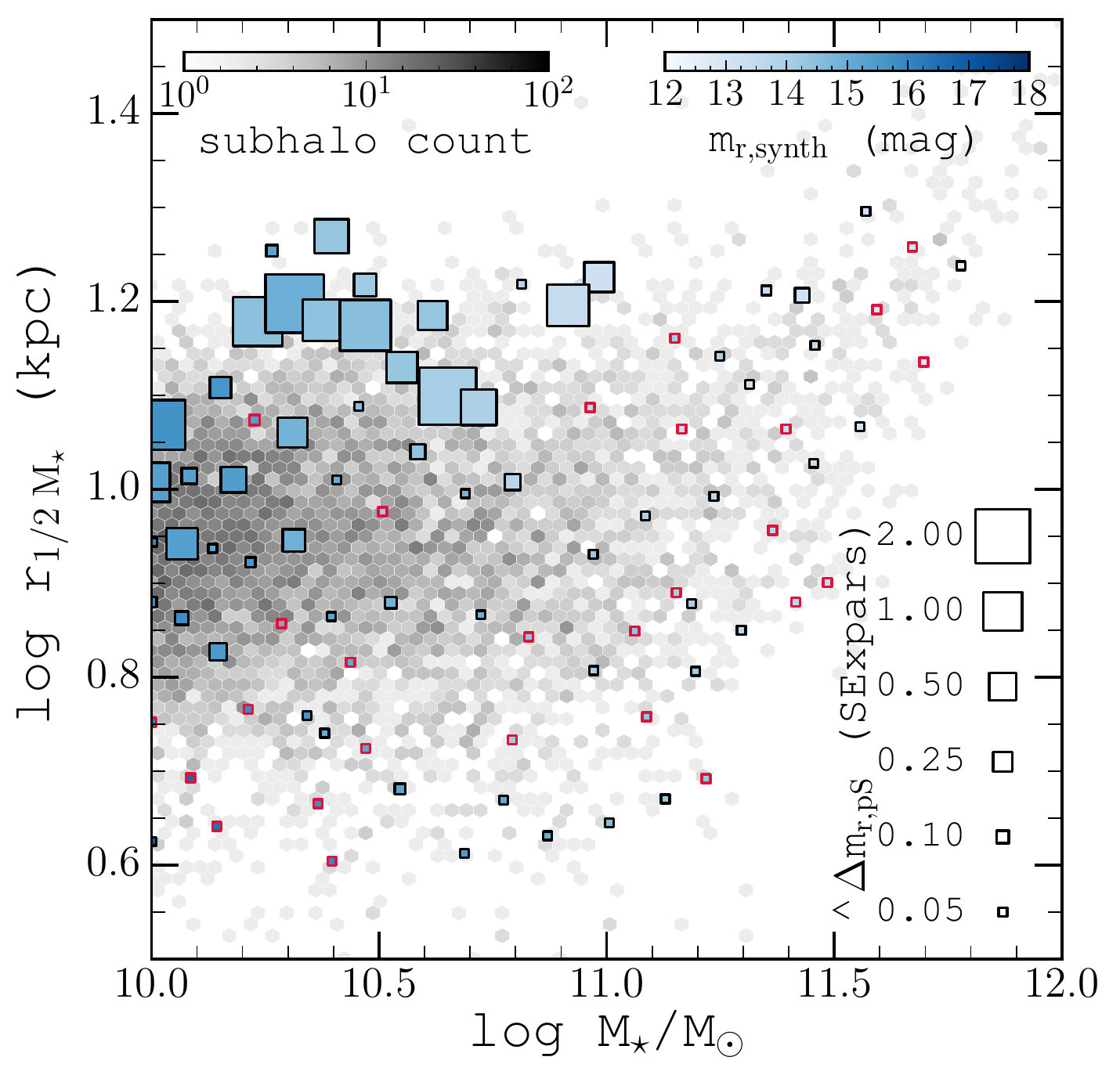}
	\includegraphics[width=0.32\linewidth]{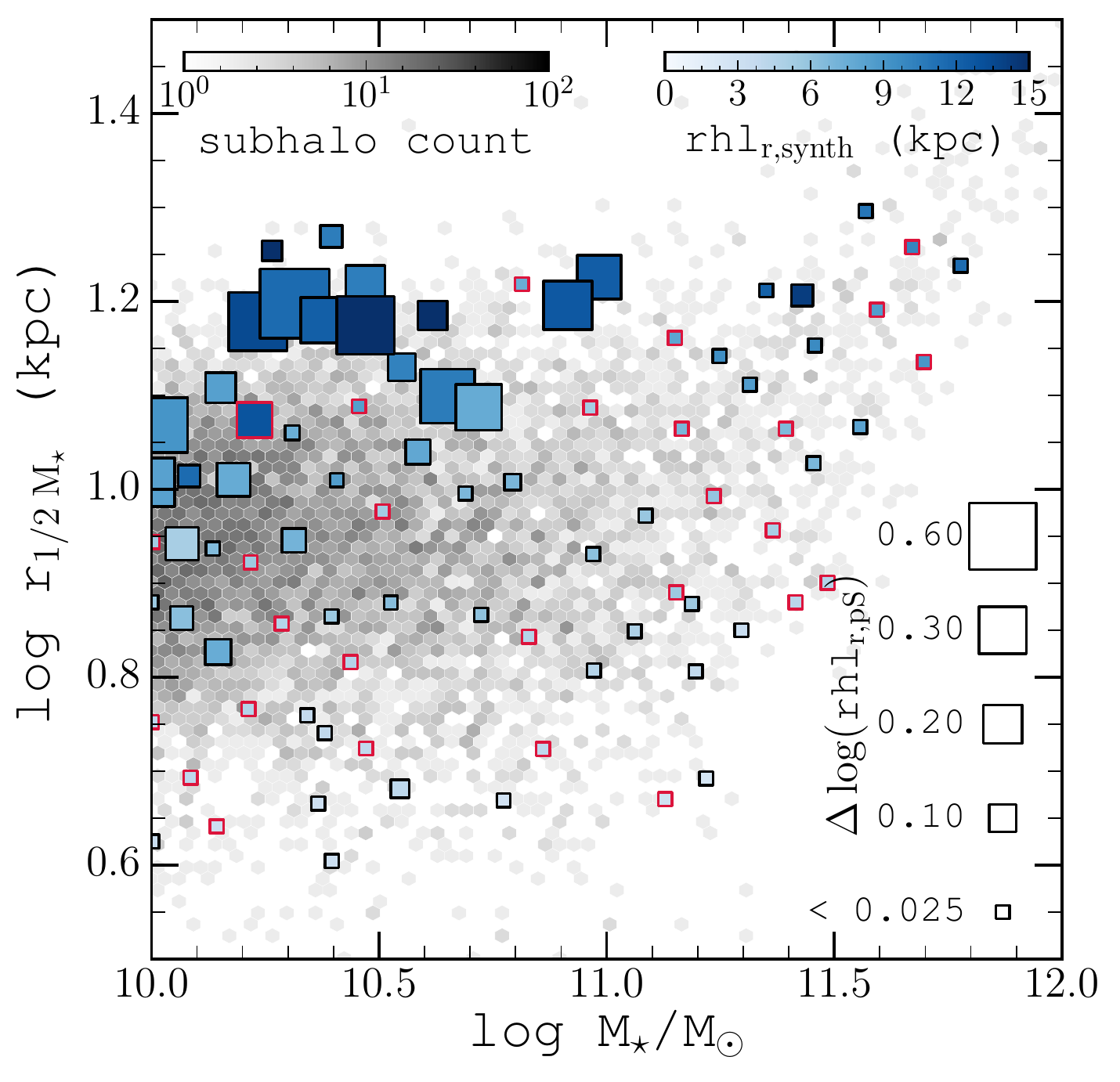}
	\includegraphics[width=0.32\linewidth]{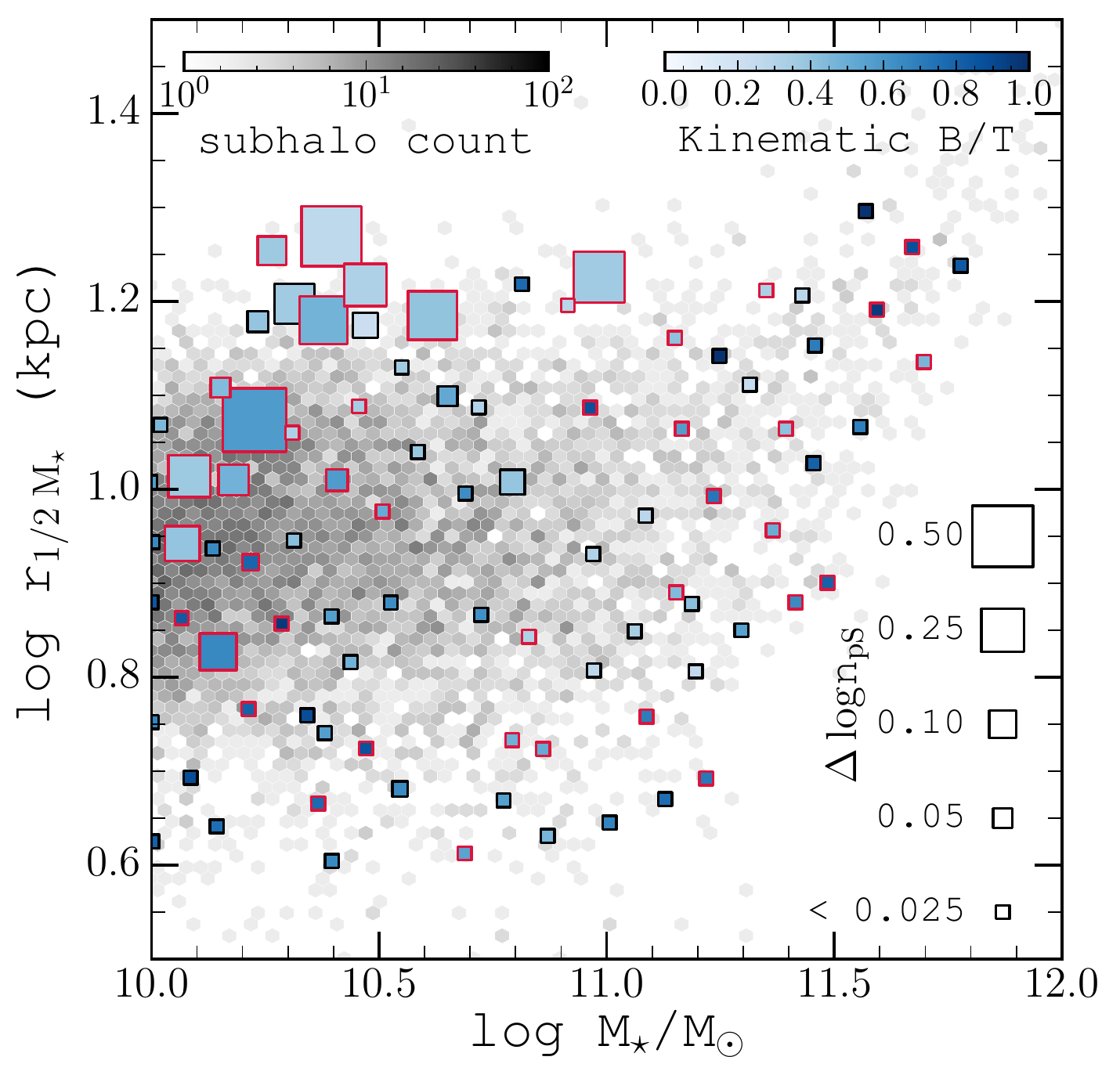}
    \caption[SExpar comparisons: all]{Offsets in B+D (upper panels) and $pS$ (lower panels) model parameters from the fiducial and alternative deblending analyses. In each case, the best-fitting parameters from the alternative deblending analysis are subtracted from those derived using the fiducial analysis. Positive and negative offsets are indicated by black and red borders on the coloured squares. Only the half-light radius border colours are reversed for visual impression (so as to match the magnitudes). Otherwise same as Figure \ref{fig:crowd_systematic_errors}. The rightmost panels show the differences in (B/T) fractions and Sersic indices. In these panels, colours indicate the kinematic (B/T) for each RIG computed from the angular momentum distribution of each galaxy assuming symmetry of the bulge about $j_z/j(E)=0$ and a rotationally supported kinematic disc. $j(E)$ is the maximum angular momentum of the stellar particles with positions between -50 and +50 of a particle in question in a rank-ordered list by binding energy for all stellar particles in the halo. Kinematic (B/T) is thus defined as $2\times N_{\star}\left(j_z/j(E)<0\right)/N_{\star}$. }
    \label{fig:SExpar_errors}
\end{figure*}

Figure \ref{fig:SExpar_errors} confirms that the high-\emph{bflags} galaxies from the fiducial scheme are indeed the same diffuse systems referenced throughout this paper. Each panel shows the RIGs projected on the stellar half-mass radius and total stellar mass plane. The face-colour of each square is represented by the variable described by the colourbar in the upper right of each panel. The size of a square represents the difference between the measured properties of a RIG in the analyses using fiducial and alternative deblending schemes (i.e. $\Delta m_r = m_{r,\mathrm{fid}}-m_{r,\mathrm{alt}}$, $\Delta\log rhl = \log(rhl_{r,\mathrm{fid}}/rhl_{r,\mathrm{alt}})$, etc.). Black and red borders on each square are used to denote positive and negative offsets, respectively, in the fiducial model estimates from those derived using the alternative deblending parameters (except for half-light radius, which is reversed to match border colours in magnitude). Upper panels show the properties derived from the B+D decompositions and lower panels show the results from the $pS$ decompositions. The large offsets in magnitude and half-light radius occur for galaxies with low intrinsic magnitudes and simultaneously large intrinsic sizes, as expected. Furthermore, the offsets in the B+D and $pS$ model magnitudes are remarkably similar. The similarity in the B+D and $pS$ model offsets demonstrates that while both optimization models are affected by internal segmentation, the bias is consistent in each case. 

The rightmost panels of Figure \ref{fig:SExpar_errors} show the differences in (B/T) fractions and Sersic indices. The colour of each square in these panels represents the kinematic (B/T) computed from the angular momenta of the stellar particles (see Figure caption). The errors in the Sersic indices corroborate with the errors in the magnitudes and sizes. Galaxies in which only a part of the substructure is modelled show the largest offsets in Sersic index. Alternatively, (B/T) offsets do not follow the trend seen for the other model parameters. The offsets also do not follow the kinematic (B/T) as one would expect if kinematic (B/T) traces the average photometric (B/T) over all environments seen in in Figure \ref{fig:crowd_random_errors} (with which the (B/T) offsets shown here \emph{are} correlated). We explore kinematic and photometric (B/T) fractions in detail in a forthcoming paper \citep{cbottrell2017}.

\section{Representative Illustris Galaxies (RIGs): Images}
\label{sec:RIG_images}

\begin{figure*}
	\center\includegraphics[width=\linewidth]{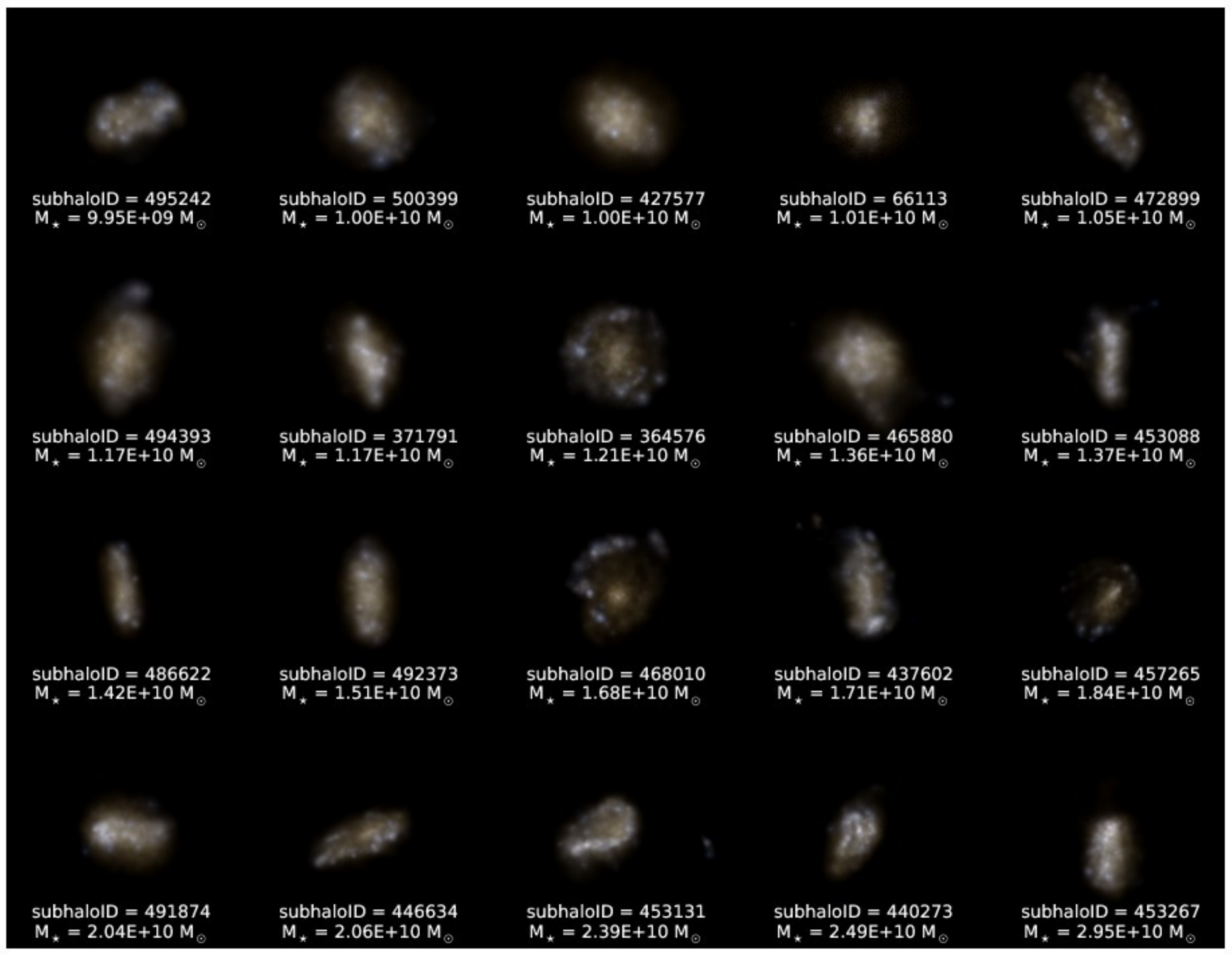}
    \caption[RIGS 1-20]{Representative Illustris Galaxy (RIG) Sample synthetic images ordered by total stellar mass. The synthetic images are composites of the SDSS $gri$ colours.}
    \label{fig:RIGS_1}
\end{figure*}

Figures \ref{fig:RIGS_1}-\ref{fig:RIGS_5} show the representative Illustris galaxies (RIGs). The galaxies uniformly span the distribution of Illustris galaxies in total stellar mass and half-mass radius.

\begin{figure*}
	\center\includegraphics[width=\linewidth]{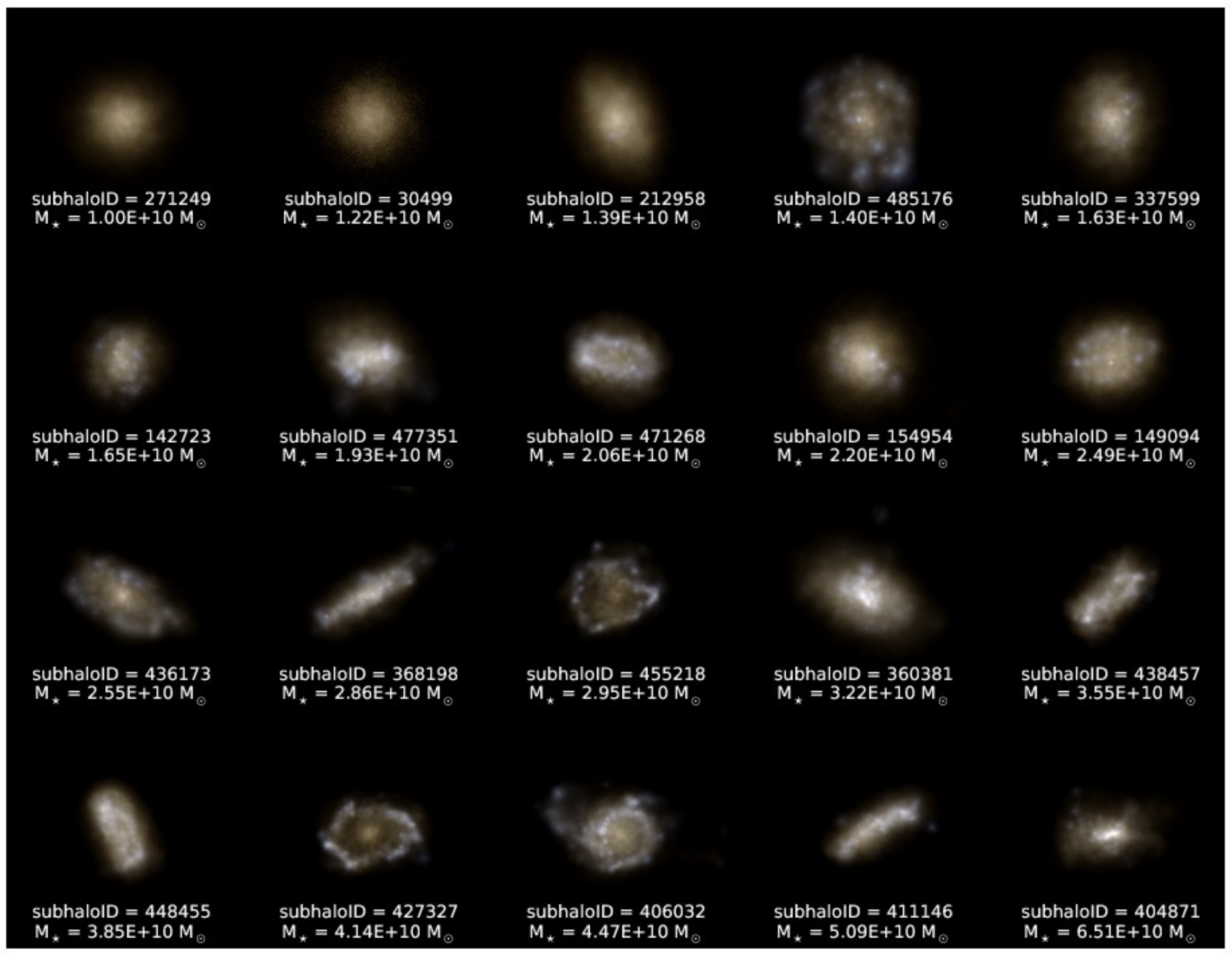}
    \caption[RIGS 21-40]{Representative Illustris Galaxy (RIG) Sample synthetic images ordered by total stellar mass. The synthetic images are composites of the SDSS $gri$ colours.}
    \label{fig:RIGS_2}
\end{figure*}
\begin{figure*}
	\center\includegraphics[width=\linewidth]{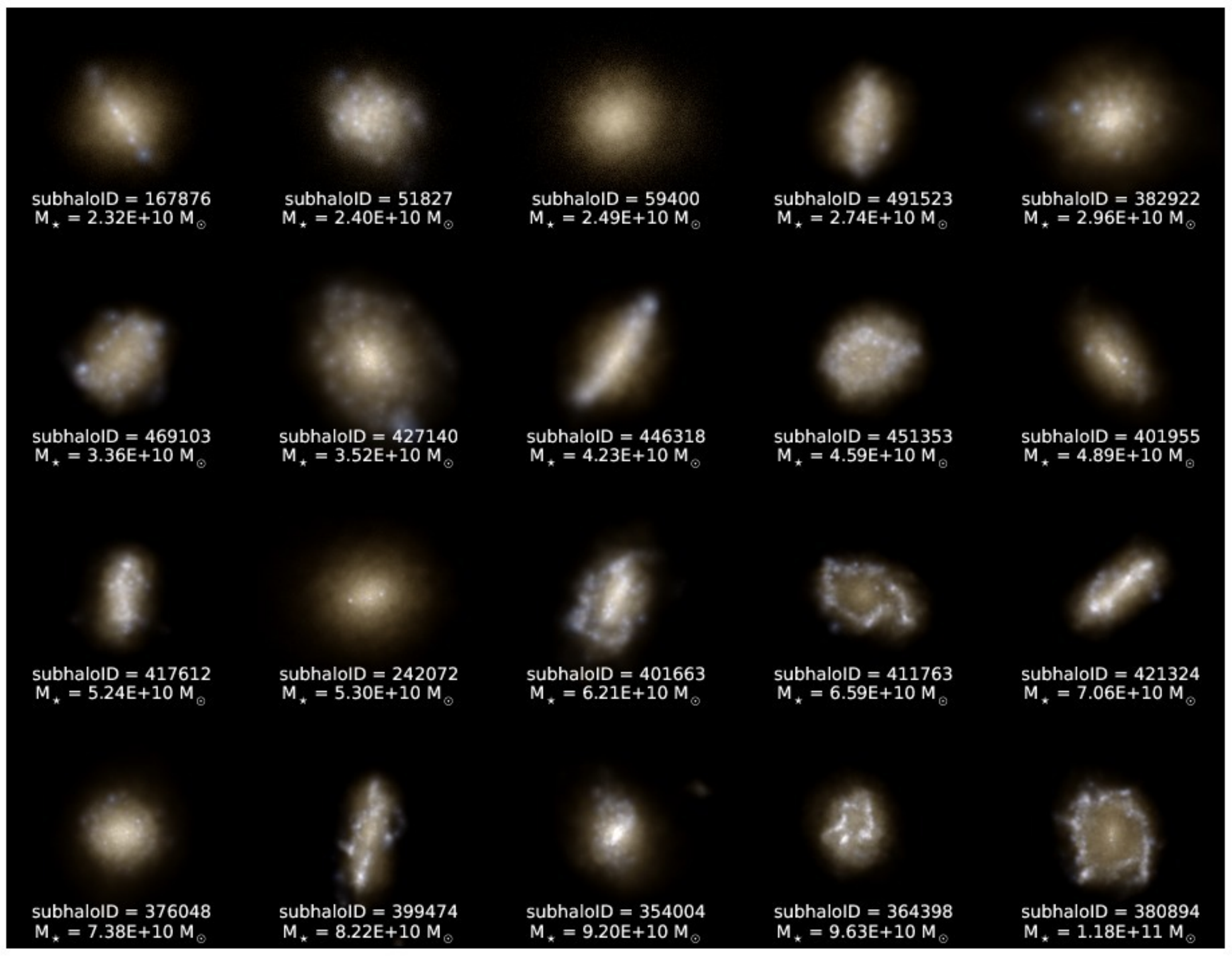}
    \caption[RIGS 41-60]{Representative Illustris Galaxy (RIG) Sample synthetic images ordered by total stellar mass. The synthetic images are composites of the SDSS $gri$ colours.}
    \label{fig:RIGS_3}
\end{figure*}
\begin{figure*}
	\center\includegraphics[width=\linewidth]{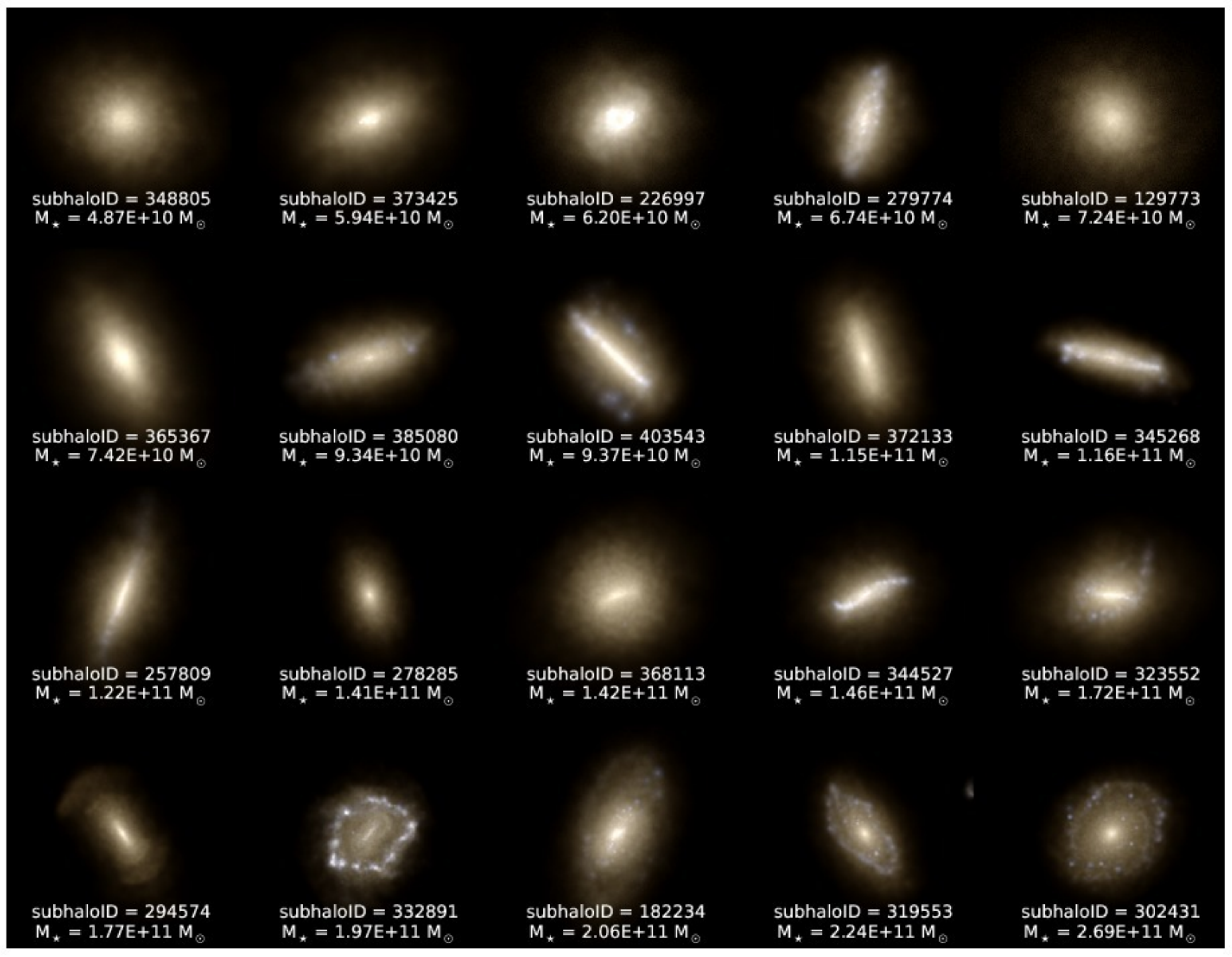}
    \caption[RIGS 61-80]{Representative Illustris Galaxy (RIG) Sample synthetic images ordered by total stellar mass. The synthetic images are composites of the SDSS $gri$ colours.}
    \label{fig:RIGS_2}
\end{figure*}
\begin{figure*}
	\center\includegraphics[width=\linewidth]{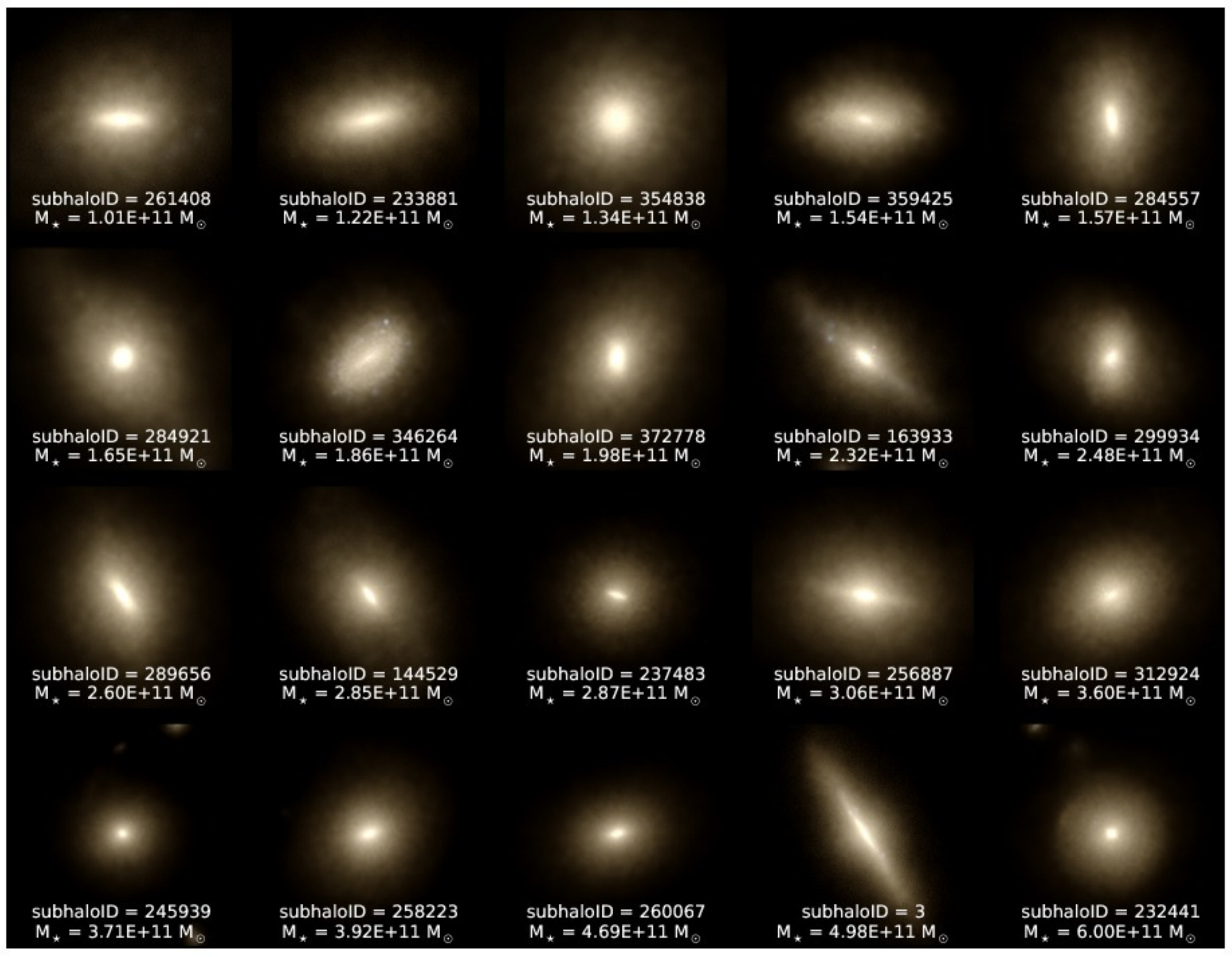}
    \caption[RIGS 81-100]{Representative Illustris Galaxy (RIG) Sample synthetic images ordered by total stellar mass. The synthetic images are composites of the SDSS $gri$ colours.}
    \label{fig:RIGS_5}
\end{figure*}


\bsp	
\label{lastpage}
\end{document}